\documentclass[structabstract]{aa}
\usepackage{txfonts}
\usepackage{graphicx}
\usepackage{natbib}
\usepackage{longtable}
\usepackage{subeqnarray}
\usepackage{cases}
\usepackage{supertabular}

\begin{document}

\title{Kinetic temperature of massive star forming molecular clumps measured with formaldehyde}
%\thanks{Appendices are only available in electronic form at http://www.aanda.org.}}

\author{X. D. Tang\inst{1,2,3}
\and C. Henkel\inst{1,4}
\and K. M. Menten\inst{1}
\and X. W. Zheng\inst{5}
\and J. Esimbek\inst{2,3}
\and J. J. Zhou\inst{2,3}
\and C. C. Yeh\inst{1}
\and C. K\"{o}nig\inst{1}
\and Y. Yuan\inst{2,3}
\and Y. X. He\inst{2,3}
\and D. L. Li\inst{2,3}}

\institute{ Max-Planck-Institut f\"{u}r Radioastronomie, Auf dem H\"{u}gel 69, 53121 Bonn, Germany\\
\email{xdtang@mpifr-bonn.mpg.de}
\and Xinjiang Astronomical Observatory, Chinese Academy of Sciences, 830011 Urumqi, China\\
\and Key Laboratory of Radio Astronomy, Chinese Academy of Sciences, 830011 Urumqi, China\\
\and Astronomy Department, King Abdulaziz University, PO Box 80203, 21589 Jeddah, Saudi Arabia\\
\and Department of Astronomy, Nanjing University, 210093 Nanjing, China\\}

%\date{Received / Accepted}

\abstract
{For a general understanding of the physics involved in the star
formation process, measurements of physical parameters such as
temperature and density are indispensable. The chemical and
physical properties of dense clumps of molecular clouds are
strongly affected by the kinetic temperature. Therefore, this
parameter is essential for a better understanding of the
interstellar medium. Formaldehyde, a molecule which traces
the entire dense molecular gas, appears to be the most
reliable tracer to directly measure the gas kinetic temperature.}
{We aim to determine the kinetic temperature with spectral
lines from formaldehyde and to compare the results with those
obtained from ammonia lines for a large number of massive clumps.}
{Three 218 GHz transitions ($J$$_{K_AK_C}$ = 3$_{03}$-2$_{02}$,
3$_{22}$-2$_{21}$, and 3$_{21}$-2$_{20}$) of para-H$_2$CO were
observed with the 15m James Clerk Maxwell Telescope (JCMT)
toward 30 massive clumps of the
Galactic disk at various stages of high-mass star formation.
Using the RADEX non-LTE model, we derive the gas kinetic
temperature modeling the measured
para-H$_2$CO 3$_{22}$-2$_{21}$/3$_{03}$-2$_{02}$ and
3$_{21}$-2$_{20}$/3$_{03}$-2$_{02}$ ratios.}
{The gas kinetic temperatures derived from the
para-H$_2$CO (3$_{21}$-2$_{20}$/3$_{03}$-2$_{02}$) line
ratios range from 30 to 61 K with an average of 46 $\pm$ 9 K.
A comparison of kinetic temperature derived from
para-H$_2$CO, NH$_3$, and the dust emission indicates
that in many cases para-H$_2$CO traces a similar kinetic
temperature to the NH$_3$ (2,2)/(1,1) transitions and the
dust associated with the HII regions. Distinctly higher
temperatures are probed by para-H$_2$CO in the clumps
associated with outflows/shocks. Kinetic temperatures
obtained from para-H$_2$CO trace turbulence to a higher
degree than NH$_3$ (2,2)/(1,1) in the massive clumps.
The non-thermal velocity dispersions of para-H$_2$CO
lines are positively correlated with the gas kinetic
temperature. The massive clumps are significantly
influenced by supersonic non-thermal motions.}
{}
\maketitle

\section{Introduction}
Ammonia (NH$_3$) is frequently
used as the standard molecular cloud thermometer
\citep{Ho1983,Walmsley1983,Danby1988,Mangum2013b}.
However, its abundance can vary strongly in different environments
(e.g., 10$^{-5}$ in hot cores; \citealt{Mauersberger1987} and
10$^{-8}$ in dark clouds; \citealt{Benson1983}) and is extremely
affected by a high UV flux. Species like CH$_3$C$_2$H and CH$_3$CN,
also sensitive to kinetic temperature, are not widespread enough
(e.g., \citealt{Gusten1985,Bally1987,Nummelin1998}). Therefore,
these molecules are of limited use as reliable probes to trace
the gas kinetic temperature \citep{Mangum1993b}.

Formaldehyde (H$_2$CO) is a ubiquitous molecule in molecular clouds
\citep{Downes1980,Cohen1981,Bieging1982,Cohen1983,Zylka1992,Mangum2008,
Ao2013,Mangum2013a,Tang2013,Ginsburg2015,Ginsburg2016}.
It is thought to be formed on the surface of dust grains by successive
hydrogenation of CO \citep{Watanabe2002,Woon2002,Hidaka2004},
it is released into the gas phase by shocks or UV heating, and is
destroyed by photodissociation. Unlike for NH$_3$, the fractional
abundance of H$_2$CO does not vary substantially and is similar
even when comparing e.g., the hot core with the compact ridge
of the well-studied Orion-KL nebula \citep{Mangum1990,Mangum1993a,Caselli1993,Johnstone2003}.

%%%%%%%%%%%%%%%%%%%%%%%%%%%%%%%%%%%%%%
%\centering
\begin{table*}[t]
\footnotesize
\caption{Source parameters.}
%{\label{kstars} Table 1. Source Positions}
\label{table:source}
\centering
\begin{tabular}
%{p{2cm}p{2cm}p{2cm}p{1.4cm}p{2cm}}
{c c c c c c c ccc}
\hline\hline % inserts double horizontal lines
Sources & RA(J2000) & DEC(J2000) & $N$(NH$_3$) & $N$(H$_2$) &
$S$$_{870\mu m}$ & $T_{\rm kin}$(NH$_3$) & $T_{\rm dust}$(HiGal) & Distance& Association \\
& $^{h}$ {} $^{m}$ {} $^{s}$ & \degr {} \arcmin {} \arcsec &
10$^{15}$ cm$^{-2}$ & 10$^{22}$ cm$^{-2}$ & Jy & K & K & kpc &\\
\hline % inserts single horizontal line
G5.89-0.39 &18:00:30.34 &-24:04:00.21 & 2.29 & 87.19 & 41.64& 39.7$\pm$3.2  &54.1$\pm$4.1  & 2.47  & HII \\
G5.90-0.44 &18:00:43.60 &-24:04:51.09 & 1.95 & 34.36 & 14.74& 31.5$\pm$2.6  &33.3$\pm$6.5  & 2.45  & HII \\
G5.97-1.36 &18:04:22.14 &-24:28:28.28 & 3.16 & 5.80  & 2.72 & 37.9$\pm$26.3 &30.0          & 13.8  &  \\
G6.91-0.22 &18:02:05.32 &-23:05:18.03 & 3.98 & 18.53 & 4.65 & 15.6$\pm$0.8  &12.2$\pm$0.8  & 3.86  &  \\
G9.04-0.52 &18:07:42.14 &-21:23:00.84 & 3.16 & 11.74 & 2.70 & 14.4$\pm$0.7  &19.6$\pm$4.5  & 12.21 &  IRDC \\
G9.21-0.20 &18:06:52.03 &-21:04:27.35 & 3.72 & 18.04 & 5.37 & 18.6$\pm$1.0  &27.0$\pm$2.4  & 11.95 &  IRDC \\
G9.88-0.75 &18:10:18.87 &-20:45:24.26 & 2.88 & 23.84 & 8.44 & 23.1$\pm$1.4  &26.1$\pm$3.9  & 3.72  &  IRDC \\
G10.99-0.08&18:10:06.50 &-19:27:46.24 & 4.68 & 14.74 & 2.87 & 12.6$\pm$0.5  &11.4$\pm$0.6  & 3.67  & IRDC \\
G11.92-0.61&18:13:58.04 &-18:54:24.91 & 2.82 & 32.66 & 10.93& 21.4$\pm$1.1  &23.3$\pm$0.1  & 3.91  & EGO \\
G12.43-1.11&18:16:51.70 &-18:41:26.52 & 1.05 & 19.30 & 7.91 & 29.0$\pm$3.0  &...           & 12.5  &  HII \\
G12.68-0.18&18:13:53.79 &-18:01:52.60 & 6.92 & 26.61 & 11.51& 32.1$\pm$6.4  &27.7$\pm$3.9  & 4.85  &  IRDC \\
G12.91-0.26&18:14:39.34 &-17:52:04.53 & 3.55 & 50.29 & 20.17& 28.0$\pm$1.7  &28.1$\pm$4.2  & 3.85  & EGO \\
G13.28-0.30&18:15:39.85 &-17:34:34.89 & 2.95 & 10.76 & 2.29 & 13.5$\pm$0.6  &11.1$\pm$0.3  & 4.04  & IRDC \\
G14.20-0.19&18:16:58.91 &-16:42:08.26 & 3.39 & 22.53 & 7.30 & 20.6$\pm$0.8  &23.1$\pm$0.5  & 3.83  &  IRDC, EGO? \\
G14.33-0.64&18:18:54.52 &-16:47:41.24 & 2.57 & 67.36 & 25.52& 25.5$\pm$1.4  &27.3$\pm$7.5  & 2.55  & EGO \\
G15.66-0.50&18:20:59.44 &-15:33:14.52 & 2.82 & 14.34 & 5.30 & 24.6$\pm$1.0  &26.3          & 0.08  &  IRDC\\
G17.10+1.02&18:18:14.80 &-13:34:14.51 & 3.16 & 2.05  & 1.13 & 60.6$\pm$62.1 &28.1$\pm$1.1  & 2.05  &  \\
G18.21-0.34&18:25:21.56 &-13:13:39.55 & 2.29 & 14.00 & 3.90 & 17.3$\pm$1.5  &10.5$\pm$0.2  & 12.34 &  IRDC\\
G19.01-0.03&18:25:44.48 &-12:22:42.20 & 2.14 & 17.24 & 5.35 & 19.5$\pm$1.1  &26.4$\pm$4.3  & 4.37  &  HII, EGO?\\
G22.55-0.52&18:34:12.70 &-09:28:21.49 & 2.69 & 11.58 & 3.09 & 16.6$\pm$0.7  &24.7$\pm$1    & 10.84 &  IRDC \\
G28.61-0.03&18:43:37.56 &-03:51:37.98 & 2.19 & ...   & 1.05 & 41.2$\pm$16.0 &29.3$\pm$5    & 11.81 &  \\
G28.86+0.07&18:43:46.02 &-03:35:29.79 & 3.63 & ...   & 7.56 & 48.4$\pm$6.8  &33.2$\pm$4.3  & 6.28  & HII \\
G30.24+0.57&18:44:29.16 &-02:08:22.85 & 3.02 & 2.23  & 1.04 & 37.6$\pm$14.6 &28.1$\pm$4.2  & 13.36 &  \\
G30.70-0.07&18:47:36.16 &-02:00:58.22 & 8.13 & 73.44 & 28.37& 26.3$\pm$2.2  &40.0          & 5.64  &  IRDC \\
G31.40-0.26&18:49:33.05 &-01:29:03.02 & 2.40 & 21.14 & 9.58 & 35.2$\pm$2.9  &29.8$\pm$3.4  & 5.41  & HII \\
G31.70-0.49&18:50:56.88 &-01:19:17.41 & 1.62 & 6.60  & 1.16 & 11.7$\pm$0.8  &13.7$\pm$4.3  & 4.99  & IRDC \\
G34.37-0.66&18:56:25.80 &+00:58:41.43 & 9.33 & 7.47  & 1.12 & 10.5$\pm$0.9  &33.0$\pm$6.5  & 0.84  & IRDC \\
G35.03+0.35&18:54:00.63 &+02:01:21.85 & 1.86 & 19.49 & 8.62 & 33.5$\pm$4.5  &39.2          & 3.49  & EGO \\
G35.19-0.74&18:58:11.96 &+01:40:23.87 & 2.04 & 27.21 & 10.73& 27.2$\pm$1.5  &32.1$\pm$10.3 & 2.31  & HII, EGO \\
G37.87-0.40&19:01:53.62 &+04:12:51.45 & 1.00 & ...   & 10.08& 33.9$\pm$8.2  &36.7$\pm$4.6  & 9.38  &  HII \\

\hline %inserts single line
\end{tabular}
\tablefoot{Parameters related to NH$_3$, $N$(H$_2$), and
$S$$_{870\mu m}$ are selected from \cite{Wienen2012}.
Distances are the kinematic distances presented by \cite{Wienen2015}.
Last column: HII = HII region, IRDC = infrared dark cloud,
EGO = extended green object.}
\end{table*}

Since the relative populations of the $K_{\rm a}$ ladders of
H$_2$CO are governed by collisions, line ratios involving
different $K_{\rm a}$ ladders are good tracers of the kinetic
temperature \citep{Mangum1993b,Muhle2007}. Particularly useful
are the three transitions of
para-H$_2$CO ($J$$_{K_AK_C}$ = 3$_{03}$-2$_{02}$, 3$_{22}$-2$_{21}$,
and 3$_{21}$-2$_{20}$), which can be measured simultaneously
at $\sim$ 218 GHz with a bandwidth of 1 GHz and whose relative
strengths (para-H$_2$CO 3$_{22}$-2$_{21}$/3$_{03}$-2$_{02}$
and 3$_{21}$-2$_{20}$/3$_{03}$-2$_{02}$) provide a sensitive
thermometer, possibly the best of the very few that are
available for the analysis of dense molecular gas. In the
case of optically thin emission, the line ratios are sensitive
to gas kinetic temperatures $\lesssim$ 50 K with a
small measurement uncertainty \citep{Mangum1993b}, which is
similar to the kinetic temperature range that the
NH$_3$ (2,2)/(1,1) ratio is sensitive to \citep{Ho1983,Mangum1992,Mangum2013a}.

Measurements of the dense molecular ridge in NGC2024 with
the para-H$_2$CO (3$_{03}$-2$_{02}$ and 3$_{22}$-2$_{21}$)
transitions show that the derived kinetic temperatures are
warmer ($T_{\rm kin}$(H$_2$CO) $\sim$ 45 -- 85 K;
\citealt{Watanabe2008}) than those traced by NH$_3$ (2,2)/(1,1)
($T_{\rm kin}$(NH$_3$) $\sim$ 27 -- 55 K; \citealt{Schulz1991}).
Using the three transitions of para-H$_2$CO at $\sim$ 218 GHz
to measure the kinetic temperature of the starburst galaxy
M82 shows that the derived kinetic temperature
($T_{\rm kin}$(H$_2$CO) $\sim$ 200 K; \citealt{Muhle2007})
is significantly higher than the temperature deduced from
the NH$_3$ (1,1)-(3,3) lines ($T_{\rm kin}$(NH$_3$) $\sim$ 60 K;
\citealt{Weiss2001}) and the dust temperature
($T_{\rm dust}$ $\sim$ 48 K; \citealt{Colbert1999}). It
is the higher $T_{\rm kin}$ value from H$_2$CO which is
representative for the bulk of the molecular gas in M82
\citep{Muhle2007}. \cite{Ao2013} and \cite{Ginsburg2016}
used the same para-H$_2$CO transitions to measure the kinetic
temperature of the dense molecular clouds near the Galactic
center. They found that these H$_2$CO-derived gas kinetic
temperatures (average $\sim$ 65 K) are uniformly higher
than the NH$_3$ (2,2)/(1,1) temperatures and the dust
temperatures of 14 -- 30 K. Overall, para-H$_2$CO, a molecule
which traces the entire dense molecular gas without much bias
because of a lack of drastic changes in abundance, appears
to be the best long-sought tracer of kinetic temperature of
the dense molecular gas at various stages of star formation.

The APEX Telescope Large Area Survey of the GALaxy (ATLASGAL)
\citep{Schuller2009}, using the Large APEX Bolometer Camera
(LABOCA) at 870 $\mu$m \citep{Siringo2009}, presents observations
in a Galactic longitude range of $\pm$60$^\circ$ and latitude
range of $\pm$1.5$^\circ$. This introduces a global view of
star formation at submm wavelengths and identifies a large
number of massive clumps forming high-mass stars at various
stages in the inner Galaxy \citep{Contreras2013,Urquhart2014,Csengeri2014}.
In this paper, we aim to measure the kinetic temperature
with three transitions of
para-H$_2$CO (J$_{K_AK_C}$ = 3$_{03}$-2$_{02}$, 3$_{22}$-2$_{21}$,
and 3$_{21}$-2$_{20}$) toward the massive clumps selected
from the ATLASGAL survey. Our main goals are the following:
(a) determining to what degree the kinetic temperatures obtained
from NH$_3$ and from para-H$_2$CO differ from each other,
(b) seeking a correlation between the temperature of
the gas and that of the dust, and (c) searching for a
correlation between kinetic temperature and line width as
is expected in the case of conversion of turbulent
energy into heat. In Sections 2 and 3, we introduce our
observations of the para-H$_2$CO triplet and the data
reduction, and describe the main results. The comparison
of kinetic temperatures derived from para-H$_2$CO, NH$_3$,
and dust is discussed in Section 4. Our main conclusions
are summarized in Section 5.

\setcounter{table}{1}
\begin{table*}
\centering
%\begin{minipage}{180mm}
\caption{Para-H$_2$CO spectral parameters.}
\label{table:H2CO}
\begin{tabular}{cccccc}
\hline\hline
Sources & Transition & $\int$$T$$_{\rm mb}$d$v$  & $V_{\rm lsr}$ & FWHM & $T$$_{\rm mb}$\\
&  & K km s$^{-1}$ & km s$^{-1}$ & km s$^{-1}$ & K \\
\hline
            & 3$_{03}$ -- 2$_{02}$ & 44.80(2.27) & 10.40(0.18) & 7.47(0.46) & 5.63\\
G5.89-0.39  & 3$_{22}$ -- 2$_{21}$ & 11.01(0.23) & 9.40(0.06)  & 5.54(0.14) & 1.87\\
            & 3$_{21}$ -- 2$_{20}$ & 11.47(0.23) & 9.39(0.05)  & 5.61(0.14) & 1.91\\
\hline
            & 3$_{03}$ -- 2$_{02}$ & 7.96(0.67)  & 10.39(0.20) & 4.78(0.47) & 1.57\\
G5.90-0.44  & 3$_{22}$ -- 2$_{21}$ & 0.61(0.13)  & 9.59(0.18)  & 1.94(0.50) & 0.30\\
            & 3$_{21}$ -- 2$_{20}$ & 1.41(0.21)  & 9.85(0.34)  & 4.83(0.98) & 0.27\\
\hline
            & 3$_{03}$ -- 2$_{02}$ & 3.93(0.11)  & 14.18(0.03) & 2.19(0.08) & 1.69\\
G5.97-1.36  & 3$_{22}$ -- 2$_{21}$ & ...         & ...         & ...        & ... \\
            & 3$_{21}$ -- 2$_{20}$ & ...         & ...         & ...        & ... \\
\hline
            & 3$_{03}$ -- 2$_{02}$ & 2.99(0.20)  & 21.26(0.16) & 5.14(0.39) & 0.54\\
G6.91-0.22  & 3$_{22}$ -- 2$_{21}$ & ...         & ...         & ...        & ... \\
            & 3$_{21}$ -- 2$_{20}$ & ...         & ...         & ...        & ... \\
\hline
            & 3$_{03}$ -- 2$_{02}$ & 0.80(0.13)  & 37.41(0.17) & 2.28(0.56) & 0.33\\
G9.04-0.52  & 3$_{22}$ -- 2$_{21}$ & ...         & ...         & ...        & ... \\
            & 3$_{21}$ -- 2$_{20}$ & ...         & ...         & ...        & ... \\
\hline
            & 3$_{03}$ -- 2$_{02}$ & 1.00(0.14)  & 42.44(0.17) & 2.38(0.37) & 0.40\\
G9.21-0.20  & 3$_{22}$ -- 2$_{21}$ & ...         & ...         & ...        & ... \\
            & 3$_{21}$ -- 2$_{20}$ & ...         & ...         & ...        & ... \\
\hline
            & 3$_{03}$ -- 2$_{02}$ & 6.43(0.27)  & 28.47(0.07) & 3.82(0.22) & 1.59\\
G9.88-0.75  & 3$_{22}$ -- 2$_{21}$ & ...         & ...         & ...        & ... \\
            & 3$_{21}$ -- 2$_{20}$ & ...         & ...         & ...        & ... \\
\hline
            & 3$_{03}$ -- 2$_{02}$ & 3.76(0.20)  & 35.99(0.09) & 3.49(0.23) & 1.01\\
G11.92-0.61 & 3$_{22}$ -- 2$_{21}$ & ...         & ...         & ...        & ... \\
            & 3$_{21}$ -- 2$_{20}$ & ...         & ...         & ...        & ... \\
\hline
            & 3$_{03}$ -- 2$_{02}$ & 7.40(0.17)  & 40.26(0.03) & 3.35(0.10) & 2.07\\
G12.43-1.11 & 3$_{22}$ -- 2$_{21}$ & ...         & ...         & ...        & ... \\
            & 3$_{21}$ -- 2$_{20}$ & ...         & ...         & ...        & ... \\
\hline
            & 3$_{03}$ -- 2$_{02}$ & 4.31(0.21)  & 54.85(0.13) & 5.84(0.41) & 0.70\\
G12.68-0.18 & 3$_{22}$ -- 2$_{21}$ & ...         & ...         & ...        & ... \\
            & 3$_{21}$ -- 2$_{20}$ & ...         & ...         & ...        & ... \\
\hline
            & 3$_{03}$ -- 2$_{02}$ & 12.53(0.40) & 37.24(0.08) & 5.02(0.21) & 2.34\\
G12.91-0.26 & 3$_{22}$ -- 2$_{21}$ & 2.17(0.19)  & 37.76(0.13) & 3.24(0.30) & 0.63\\
            & 3$_{21}$ -- 2$_{20}$ & 3.51(0.37)  & 36.60(0.33) & 5.94(0.75) & 0.56\\
\hline
            & 3$_{03}$ -- 2$_{02}$ & 7.99(0.34)  & 40.00(0.10) & 4.66(0.23) & 1.61\\
G14.20-0.19 & 3$_{22}$ -- 2$_{21}$ & 1.63(0.23)  & 39.46(0.31) & 4.71(0.87) & 0.33\\
            & 3$_{21}$ -- 2$_{20}$ & 1.61(0.20)  & 40.55(0.33) & 4.63(0.52) & 0.33\\
\hline
            & 3$_{03}$ -- 2$_{02}$ & 19.66(0.23) & 22.57(0.02) & 3.86(0.06) & 4.79\\
G14.33-0.64 & 3$_{22}$ -- 2$_{21}$ & 5.89(0.21)  & 22.28(0.06) & 3.82(0.17) & 1.44\\
            & 3$_{21}$ -- 2$_{20}$ & 4.50(0.19)  & 22.10(0.07) & 3.45(0.17) & 1.23\\
\hline
            & 3$_{03}$ -- 2$_{02}$ & 5.86(0.24)  & -5.14(0.10) & 5.07(0.29) & 1.09\\
G15.66-0.50 & 3$_{22}$ -- 2$_{21}$ & ...         & ...         & ...        & ... \\
            & 3$_{21}$ -- 2$_{20}$ & ...         & ...         & ...        & ... \\
\hline
            & 3$_{03}$ -- 2$_{02}$ & 1.06(0.16)  & 19.94(0.22) & 3.01(0.46) & 0.33\\
G17.10+1.02 & 3$_{22}$ -- 2$_{21}$ & ...         & ...         & ...        & ... \\
            & 3$_{21}$ -- 2$_{20}$ & ...         & ...         & ...        & ... \\
\hline
            & 3$_{03}$ -- 2$_{02}$ & 3.39(0.20)  & 46.09(0.18) & 6.01(0.45) & 0.53\\
G18.21-0.34 & 3$_{22}$ -- 2$_{21}$ & ...         & ...         & ...        & ... \\
            & 3$_{21}$ -- 2$_{20}$ & ...         & ...         & ...        & ... \\
\hline
            & 3$_{03}$ -- 2$_{02}$ & 5.86(0.21)  & 59.67(0.09) & 5.47(0.27) & 1.00\\
G19.01-0.03 & 3$_{22}$ -- 2$_{21}$ & ...         & ...         & ...        & ... \\
            & 3$_{21}$ -- 2$_{20}$ & ...         & ...         & ...        & ... \\
\hline
            & 3$_{03}$ -- 2$_{02}$ & 2.77(0.21)  & 75.66(0.17) & 4.62(0.42) & 0.56\\
G22.55-0.52 & 3$_{22}$ -- 2$_{21}$ & ...         & ...         & ...        & ... \\
            & 3$_{21}$ -- 2$_{20}$ & ...         & ...         & ...        & ... \\
\hline
            & 3$_{03}$ -- 2$_{02}$ & 1.51(0.19)  & 47.92(0.35) & 5.70(0.71) & 0.26\\
G28.61-0.03 & 3$_{22}$ -- 2$_{21}$ & ...         & ...         & ...        & ... \\
            & 3$_{21}$ -- 2$_{20}$ & ...         & ...         & ...        & ... \\
\hline
            & 3$_{03}$ -- 2$_{02}$ & 2.07(0.26)  & 98.43(0.16) & 3.30(0.47) & 0.59\\
G28.86+0.07 &                      & 4.69(0.26)  & 103.30(0.11)& 4.11(0.26) & 1.07\\
\hline
\end{tabular}
%\end{minipage}
\end{table*}
%%%%%%%%%%%%%%%%%%%%%
\setcounter{table}{1}
\begin{table*}
\centering
%\begin{minipage}{180mm}
\caption{continued}
\begin{tabular}{cccccc}
\hline\hline
Sources & Transition & $\int$$T$$_{\rm mb}$d$v$  & $V_{\rm lsr}$ & FWHM & $T$$_{\rm mb}$\\
&  & K km s$^{-1}$ & km s$^{-1}$ & km s$^{-1}$ & K \\
\hline
G28.86+0.07 & 3$_{22}$ -- 2$_{21}$ & 0.96(0.17)  & 103.40(0.21)& 2.49(0.55) & 0.36\\
            & 3$_{21}$ -- 2$_{20}$ & 1.31(0.17)  & 103.30(0.18)& 2.70(0.44) & 0.46\\
\hline
            & 3$_{03}$ -- 2$_{02}$ & 9.21(0.23)  & 90.25(0.05) & 4.67(0.15) & 1.86\\
G30.70-0.07 & 3$_{22}$ -- 2$_{21}$ & 2.47(0.17)  & 89.92(0.14) & 4.34(0.35) & 0.53\\
            & 3$_{21}$ -- 2$_{20}$ & 1.86(0.19)  & 90.43(0.16) & 3.37(0.38) & 0.51\\
\hline
            & 3$_{03}$ -- 2$_{02}$ & 11.81(0.17) & 86.70(0.03) & 4.86(0.08) & 2.29\\
G31.40-0.26 & 3$_{22}$ -- 2$_{21}$ & 1.74(0.16)  & 87.17(0.19) & 3.85(0.36) & 0.43\\
            & 3$_{21}$ -- 2$_{20}$ & 2.00(0.20)  & 87.00(0.19) & 3.82(0.47) & 0.49\\
\hline
            & 3$_{03}$ -- 2$_{02}$ & 9.53(0.21)  & 52.70(0.06) & 5.16(0.12) & 1.73\\
G35.03+0.35 & 3$_{22}$ -- 2$_{21}$ & 2.29(0.26)  & 52.55(0.38) & 6.64(0.93) & 0.33\\
            & 3$_{21}$ -- 2$_{20}$ & 2.23(0.21)  & 52.92(0.23) & 4.76(0.47) & 0.44\\
\hline
            & 3$_{03}$ -- 2$_{02}$ & 16.43(0.29) & 36.42(0.05) & 6.68(0.14) & 2.31\\
G35.19-0.74 & 3$_{22}$ -- 2$_{21}$ & 2.31(0.21)  & 36.20(0.16) & 4.01(0.54) & 0.54\\
            & 3$_{21}$ -- 2$_{20}$ & 1.49(0.17)  & 36.12(0.18) & 2.98(0.39) & 0.47\\
\hline
            & 3$_{03}$ -- 2$_{02}$ & 10.76(0.94) & 60.93(0.27) & 6.48(0.68) & 1.56\\
G37.87-0.40 & 3$_{22}$ -- 2$_{21}$ & ...         & ...         & ...        & ... \\
            & 3$_{21}$ -- 2$_{20}$ & ...         & ...         & ...        & ... \\
\hline
\end{tabular}
\end{table*}																													
\section{Selection of targets, observations, and data reduction}
We have selected 30 massive clumps of the Galactic disk
at various stages of high-mass star
formation and with strong NH$_3$ emission
from the ATLASGAL survey (see Table \ref{table:source}).
At Effelsberg, the NH$_3$ $(J,K)$ = (1,1), (2,2), and (3,3)
lines have been measured by Wienen et al. (2012) who determined
kinetic temperatures, based on the NH$_3$ (2,2)/(1,1) ratio,
ranging from 11 to 61 K. The sample of high-mass star forming
clouds at various evolutionary stages contains infrared dark
clouds (IRDCs), clouds hosting extended green objects (EGOs)
that are thought to trace outflows and are generally thought
to be in an early stage of massive star formation
\citep{Cyganowski2008,Cyganowski2011,Chen2013a,Chen2013b},
and clouds associated with HII regions based on the SIMBAD
Astronomical Database\footnote{%
  \tiny
http://simbad.u-strasbg.fr/simbad/}.

Sources observed are listed in Table \ref{table:source}.
Our observations were carried out in 2015 April, July, and
October with the 15m James Clerk Maxwell Telescope telescope (JCMT)
on Mauna Kea. The beam size is $\sim$ 23$''$ and the main-beam
efficiency is $\eta_{\rm mb}$ = $T_a^*/T_{\rm mb}$ $\simeq$ 0.7
at 218 GHz\footnote{%
  \tiny
http://www.eaobservatory.org/jcmt/instrumentation/heterodyne/rxa}.
The para-H$_2$CO J$_{K_AK_C}$ = 3$_{03}$-2$_{02}$, 3$_{22}$-2$_{21}$,
and 3$_{21}$-2$_{20}$ transitions have rest frequencies of 218.222,
218.475, and 218.760 GHz, respectively, which are measured
simultaneously by employing the ACSIS digital autocorrelation
spectrometer with the special backend  configuration
RxA$_-$H$_2$CO$_-$250$\times$3 allowing for three windows,
each with a bandwidth of 250 MHz\footnote{%
  \tiny
http://www.eaobservatory.org/jcmt/instrumentation/heterodyne/acsis/}.
This provides a velocity resolution of 0.084\,km\,s$^{-1}$ for
para-H$_2$CO (3$_{03}$-2$_{02}$ and 3$_{22}$-2$_{21}$) and
0.042\,km\,s$^{-1}$ for para-H$_2$CO (3$_{21}$-2$_{20}$);
CH$_3$OH (4$_{22}$-3$_{12}$) at 218.440 GHz is also observed
together with para-H$_2$CO (3$_{22}$-2$_{21}$).

%\subsection{Data reduction and exhibition}
Data reduction for spectral lines was performed using Starlink\footnote{%
  \tiny
http://starlink.eao.hawaii.edu/starlink} and GILDAS\footnote{%
  \tiny
http://www.iram.fr/IRAMFR/GILDAS}. To enhance signal-to-noise
ratios (S/N) in individual channels, we smoothed contiguous channels
to a velocity resolution $\sim$0.33 km s$^{-1}$.

\section{Results}
Of the 30 massive clumps observed (see Table \ref{table:source}),
25 are detected in the para-H$_2$CO (3$_{03}$-2$_{02}$) line.
Among the 25 para-H$_2$CO (3$_{03}$-2$_{02}$) detections,
10 also show the para-H$_2$CO (3$_{22}$-2$_{21}$) and
(3$_{21}$-2$_{20}$) lines, while 18 also exhibit emission from
the CH$_3$OH (4$_{22}$-3$_{12}$) line (218.440 GHz), which is
well separated from the para-H$_2$CO (3$_{22}$-2$_{21}$) transition in all cases.
The para-H$_2$CO and the CH$_3$OH line spectra are presented in
Figures \ref{figure:H2CO-spectrum} and \ref{figure:CH3OH-spectrum}.
Line parameters are listed in Tables \ref{table:H2CO} and \ref{table:CH3OH},
where integrated intensity ($\int$$T$$_{\rm mb}$d$v$), local
standard of rest velocity ($V_{\rm lsr}$), line width (FWHM), and
peak antenna brightness temperature ($T$$_{\rm mb}$) were obtained
from Gaussian fits. Five sources show no H$_2$CO and CH$_3$OH,
namely G10.99-0.08, G13.28-0.3, G30.24+0.57, G31.70-0.49,
and G34.37-0.66. G30.24+0.57 has a low H$_2$ column density of
2.2 $\times$ 10$^{22}$ cm$^{-2}$ \citep{Wienen2012}. G10.99-0.08,
G13.28-0.3, G31.70-0.49, and G34.37-0.66 are associated with infrared
dark clouds, which have a low kinetic temperature in the range of
10.5 -- 13.5 K. This suggests that para-H$_2$CO and CH$_3$OH are
excited with difficulty in these low-density and/or low-temperature
regions. Nevertheless, high detection rates of H$_2$CO ($\sim$83\%)
and CH$_3$OH ($\sim$60\%) indicate that the two molecular species
are commonly formed in the massive clumps of our sample.

\subsection{H$_2$CO column density}
To determine the para-H$_2$CO column densities and gas kinetic
temperatures, we use the RADEX non-LTE model \citep{van der Tak2007}
offline code\footnote{%
  \tiny
http://var.sron.nl/radex/radex.php} with collision rates from
\cite{Wiesenfeld2013}.
The RADEX code needs five input parameters: background temperature,
kinetic temperature, H$_2$ density, para-H$_2$CO column density,
and line width. For the background temperature, we adopt 2.73 K.
Model grids for the para-H$_2$CO lines encompass 40 densities
($n$(H$_2$) = 10$^3$ -- 10$^8$ cm$^{-3}$), 40 para-H$_2$CO column
densities ($N$(para-H$_2$CO) = 10$^{12}$ -- 10$^{16}$ cm$^{-2}$),
and 40 temperatures ranging from 10 to 110 K.
For the line width, we use the observed line width value.

%%%%%%%%%%%%%%%%%%%Fig.1-S870um-NH3-H2CO-CH3OH-intensities%%%%%%%%
\begin{figure}[t]
\vspace*{0.2mm}
\begin{center}
\includegraphics[width=8.5cm]{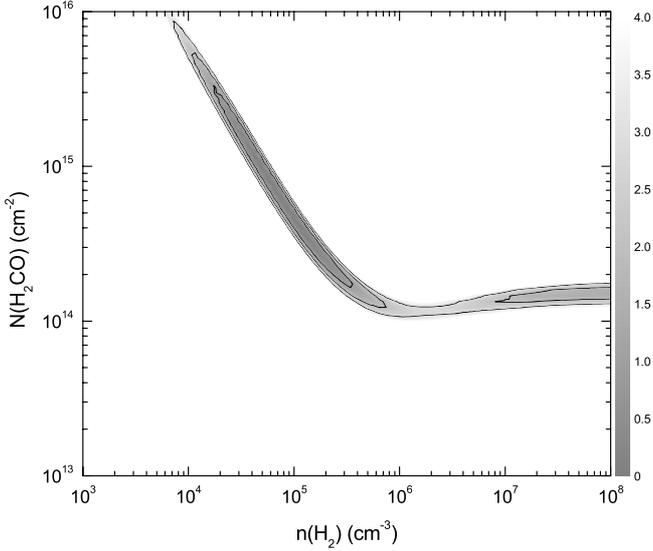}
\end{center}
\caption{Example of RADEX Non-LTE modeling of the $N$(H$_2$CO)--$n$(H$_2$)
relation for a kinetic temperature of 40 K. The source is G5.89-0.39.
Contours enclose regions of low reduced $\chi$$^2_{\rm red}$ based on
3$_{03}$-2$_{02}$ and 3$_{22}$-2$_{21}$  para-H$_2$CO brightness temperatures.
Contours are $\chi$$^2_{\rm red}$ = 3, 2, and 1.}
\label{figure:NH2CO}
\end{figure}

%%%%%%%%%%%%%%%%%%%%%%%%%%%%%%%%%%%%%%%%%%%%%%%%%%%%%%%%%%%%%%%%%%%%%%%%%%%%%%%%%%%%
\begin{table}[h]
\small
\caption{CH$_3$OH (4$_{22}$-3$_{12}$) spectral parameters.}
%{\label{kstars} Table 1. Source Positions}
\label{table:CH3OH}
\centering
\begin{tabular}
%{p{2cm}p{2cm}p{2cm}p{1.4cm}p{2cm}}
{c c c c c}
\hline\hline % inserts double horizontal lines
Sources &  $\int$$T$$_{\rm mb}$d$v$  & $V_{\rm lsr}$ & FWHM & $T$$_{\rm mb}$\\
&  K km s$^{-1}$ & km s$^{-1}$ & km s$^{-1}$ & K \\
\hline % inserts single horizontal line
G5.89-0.39  & 7.66(0.21) & 9.52(0.06)   & 4.75(0.15) & 1.51  \\
G5.90-0.44  & ...        & 9.89(0.16)   & ...        & ...   \\
G5.97-1.36  & 0.33(0.09) & 22.63(0.12)  & 1.01(0.27) & 0.31  \\
G6.91-0.22  & 1.16(0.17) & 21.00(0.27)  & 3.96(0.69) & 0.27  \\
G9.21-0.20  & 0.59(0.13) & 42.11(0.16)  & 1.43(0.40) & 0.39  \\
G9.88-0.75  & 0.93(0.17) & 28.18(0.29)  & 3.33(0.80) & 0.26  \\
G11.92-0.61 & 1.81(0.49) & 35.78(0.31)  & 4.69(2.31) & 0.36  \\
G12.43-1.11 & 1.79(0.24) & 39.52(0.23)  & 3.58(0.62) & 0.47  \\
G12.68-0.18 & 1.66(0.31) & 54.49(0.52)  & 5.79(1.68) & 0.27  \\
G12.91-0.26 & 6.51(0.39) & 37.70(0.16)  & 5.43(0.40) & 1.13  \\
G14.20-0.19 & 2.83(0.30) & 39.94(0.25)  & 5.14(0.66) & 0.51  \\
G14.33-0.64 & 8.54(0.36) & 22.01(0.07)  & 3.51(0.19) & 2.29  \\
G19.01-0.03 & 1.87(0.21) & 59.93(0.24)  & 4.64(0.70) & 0.39  \\
G28.86+0.07 & 2.80(0.29) & 103.10(0.19) & 3.88(0.50) & 0.67  \\
G30.70-0.07 & 6.49(0.27) & 90.06(0.10)  & 4.63(0.24) & 1.31  \\
G31.40-0.26 & 3.94(0.24) & 87.12(0.14)  & 4.71(0.35) & 0.79  \\
G35.03+0.35 & 1.91(0.27) & 52.99(0.36)  & 4.97(0.77) & 0.36  \\
G35.19-0.74 & 3.64(0.34) & 35.96(0.22)  & 4.96(0.60) & 0.69  \\
\hline %inserts single line
\end{tabular}
\end{table}

We ran RADEX to obtain beam averaged para-H$_2$CO column densities and calculated
the behavior of the $\chi$$^2_{\rm red}$ value of the observed
3$_{03}$-2$_{02}$ and 3$_{22}$-2$_{21}$ (or 3$_{21}$-2$_{20}$)
para-H$_2$CO line brightness temperatures (see Figure \ref{figure:NH2CO}).
The value of $\chi$$^2_{\rm red}$ is defined as\\
\begin{equation}
\chi^2_{\rm red} = \Sigma_{\rm i} \frac{(T_{\rm R(obs)_i}-T_{\rm R(mod)_i})^2}{\sigma_{T_{\rm R(obs)_i}}^2},
\end{equation}
where $T_{\rm R(obs)_i}$ and $T_{\rm R(mod)_i}$ represent the
observed main beam brightness temperatures ($T_{\rm mb}$) and
RADEX non-LTE modeled brightness temperatures, and
$\sigma$$_{T_{\rm R(obs)_i}}^2$ represents the uncertainty in
$T_{\rm R(obs)_i}$ including the rms noise in the spectra and
the absolute temperature calibration uncertainty. One degree
of freedom is used to the fit $\chi$$^2_{\rm red}$ value.
The reduced $\chi$$^2_{\rm red}$ value depends of course on
$T_{\rm kin}$, but also to a lesser degree on H$_2$ density and
para-H$_2$CO column density. To provide a feeling of the related
uncertainties, we take as an example source G5.89-0.39
(see Figure \ref{figure:NH2CO}), which is a typical case.
This figure shows that $\chi$$^2_{\rm red}$ depends on the
H$_2$ density and para-H$_2$CO column density at low densities
($n$(H$_2$) $<$ 10$^6$ cm$^{-3}$), while the kinetic temperature
is kept constant at $\sim$ 40 K (which is close to the actual
temperature, see below). For higher densities, $\chi$$^2_{\rm red}$
decreases only slowly with H$_2$ density. The entire plot provides
a lower limit to the column density at $N$(H$_2$CO) $\sim$ 10$^{14}$ cm$^{-2}$.

The H$_2$ density of the ATLASGAL clumps is $\sim$ 10$^5$ cm$^{-3}$
\citep{Beuther2002,Motte2003,Wienen2012}. As can be seen in
Figure \ref{figure:NH2CO}, our characteristic source G5.89-0.39 also
shows the lowest $\chi$$^2_{\rm red}$ values near this density, so we
adopt an H$_2$ volume density of $n$(H$_2$) = 10$^5$ cm$^{-3}$.
The results are listed in Table \ref{table:NH2CO-Tkin}. Including all
sources, the $N$(para-H$_2$CO) range is 0.4 -- 47 $\times$ 10$^{13}$ cm$^{-2}$
with an average of 6.5 $\times$ 10$^{13}$ cm$^{-2}$, which agrees
with the results from other star forming regions and from
protostellar cores \citep{Mangum1993b,Hurt1996,Watanabe2008}.
At densities $n$(H$_2$) = 10$^5$ cm$^{-3}$, the fractional abundance
$N$(para-H$_2$CO)/$N$(H$_2$) becomes 0.4 -- 5.4 $\times$ 10$^{-10}$,
where $N$(H$_2$) is derived from the 870 $\mu$m continuum
emission \citep{Wienen2012}.

\begin{figure*}[t]
\vspace*{0.2mm}
\begin{center}
\includegraphics[width=5.8cm]{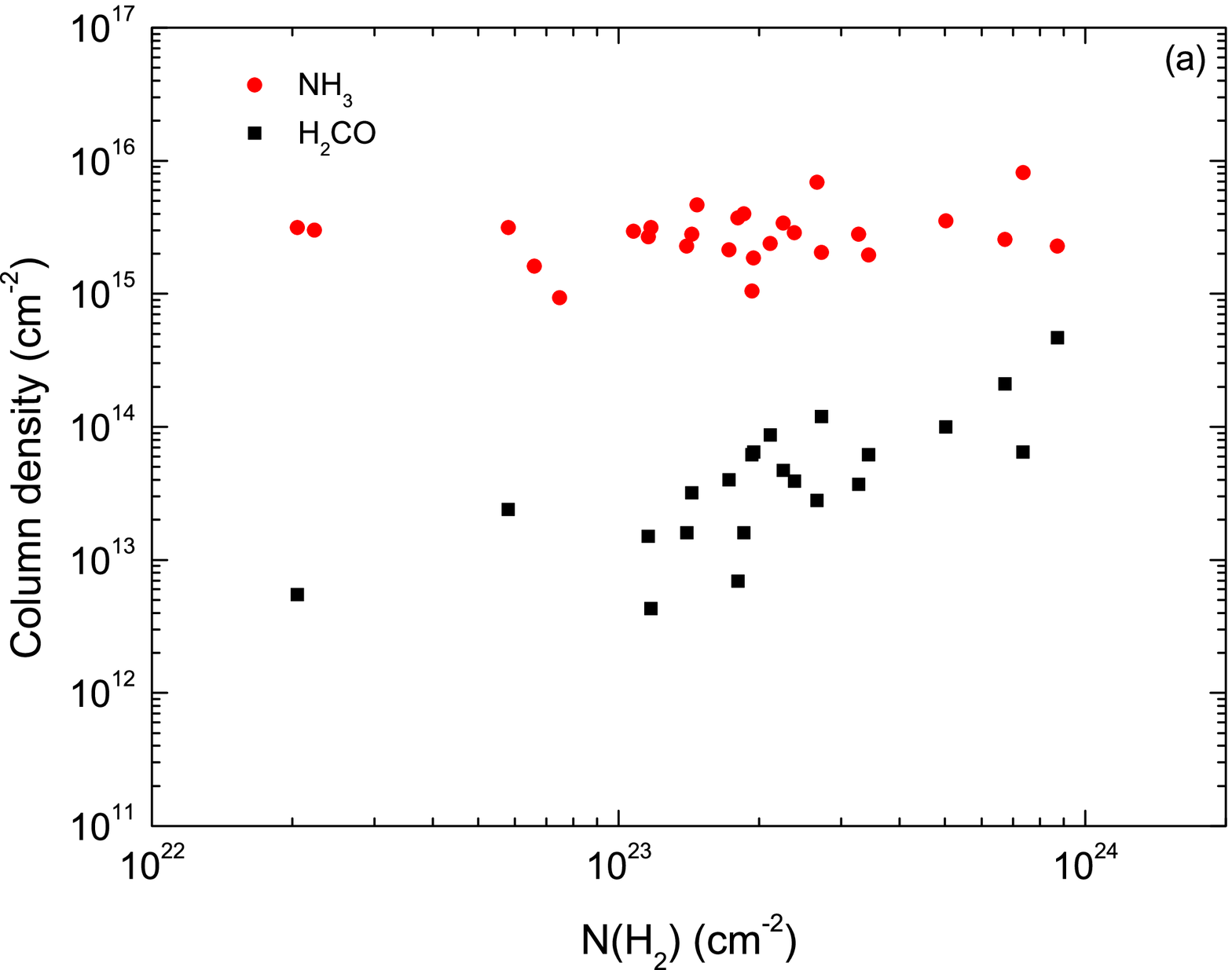}
\includegraphics[width=5.8cm]{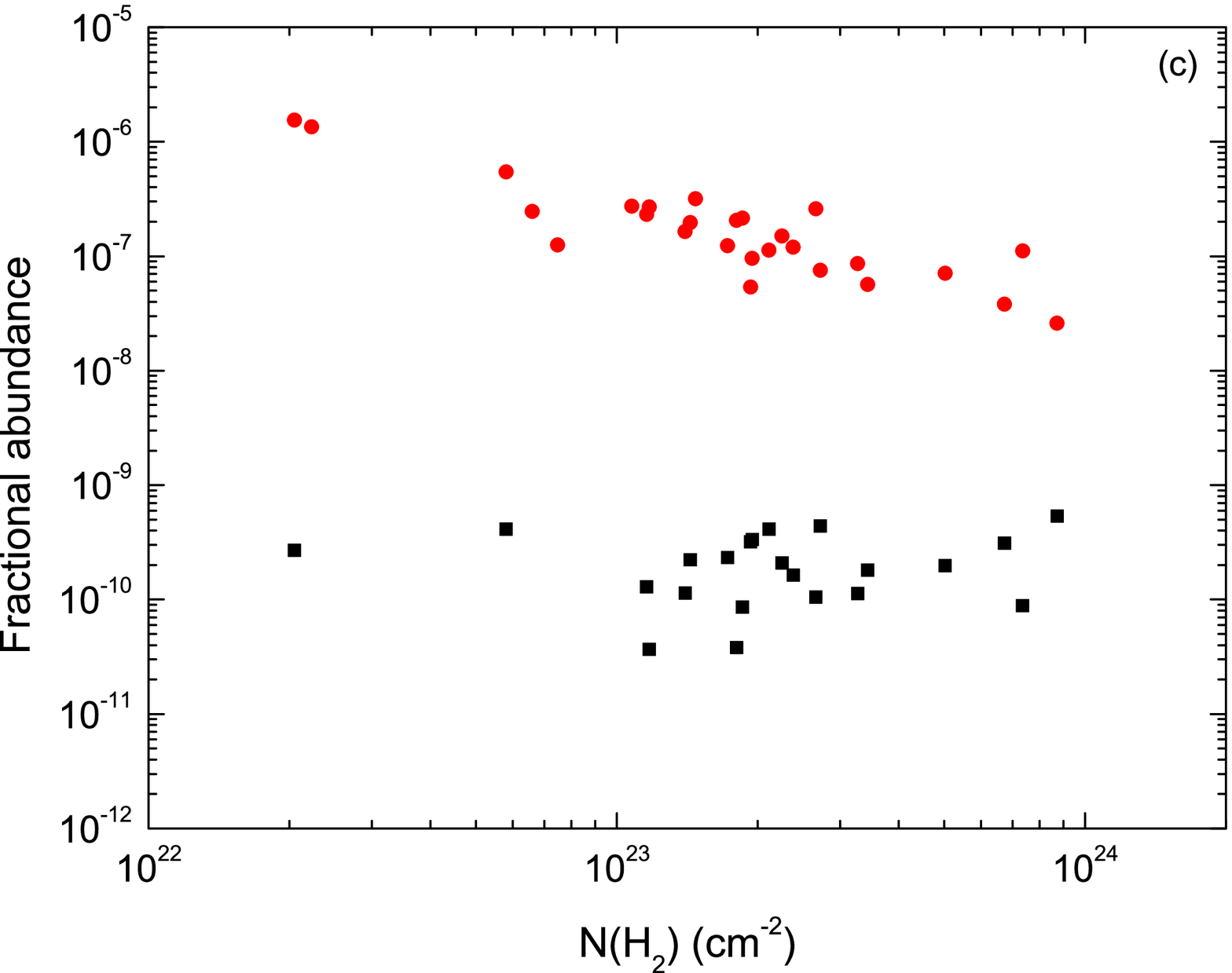}
\includegraphics[width=5.8cm]{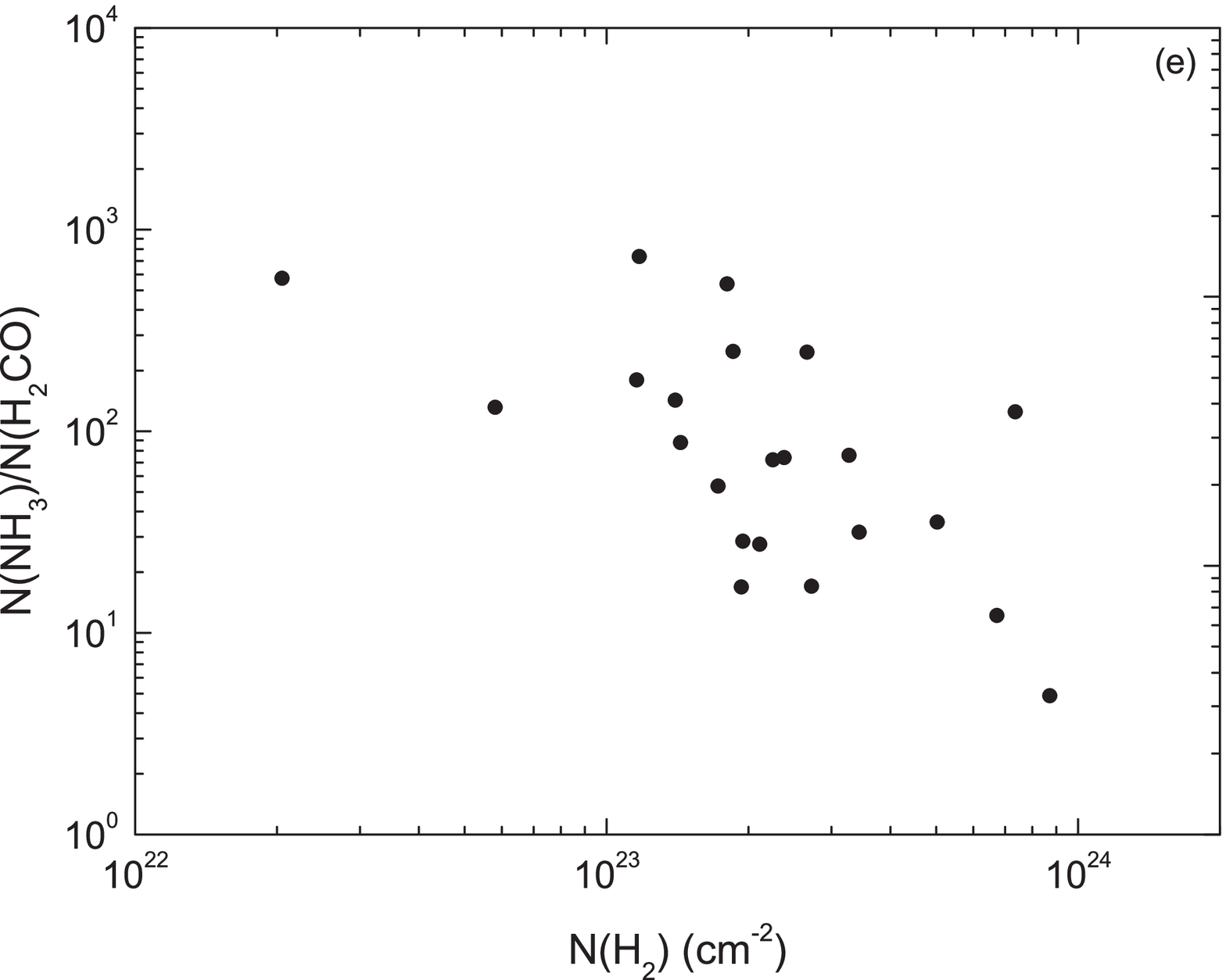}
\includegraphics[width=5.8cm]{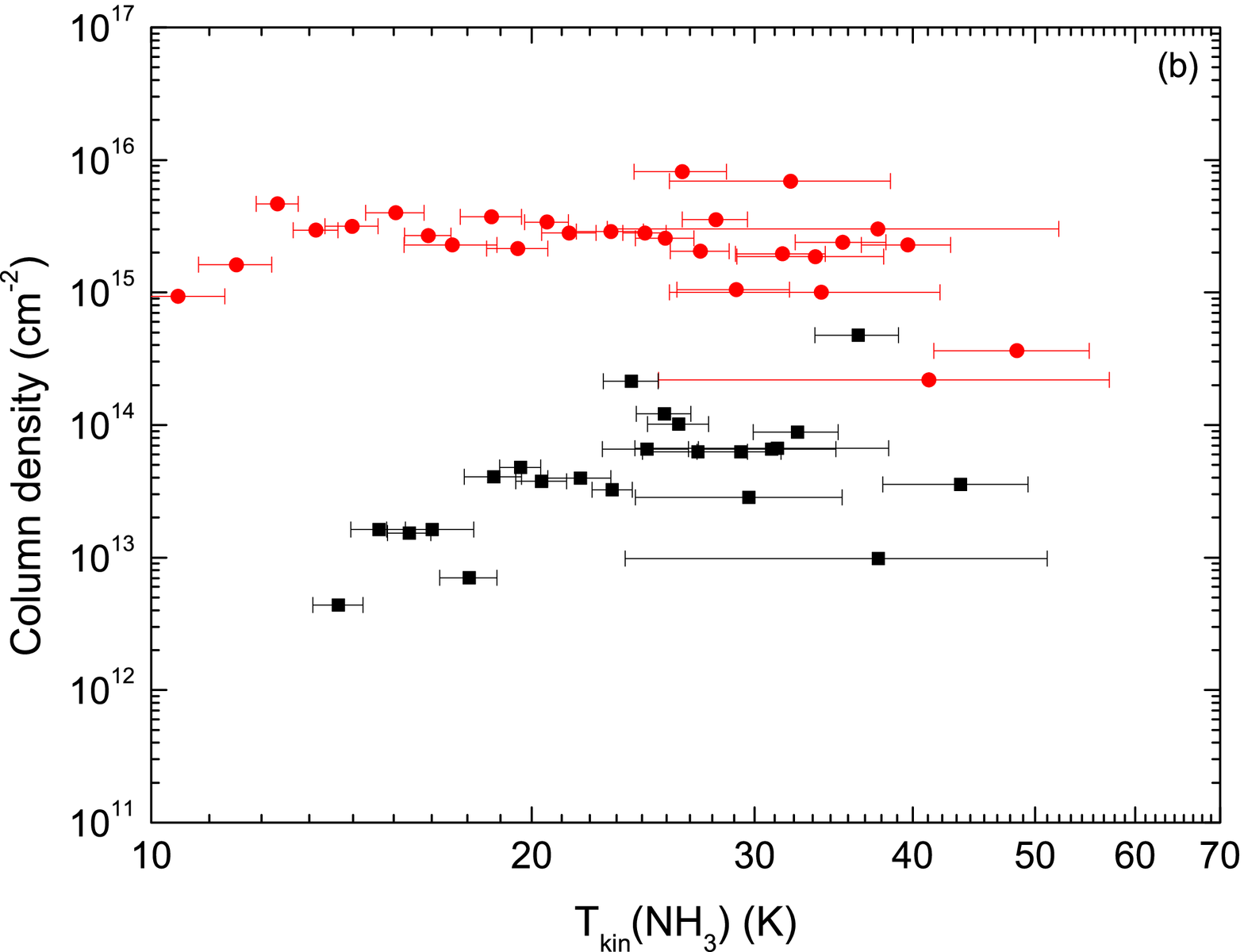}
\includegraphics[width=5.8cm]{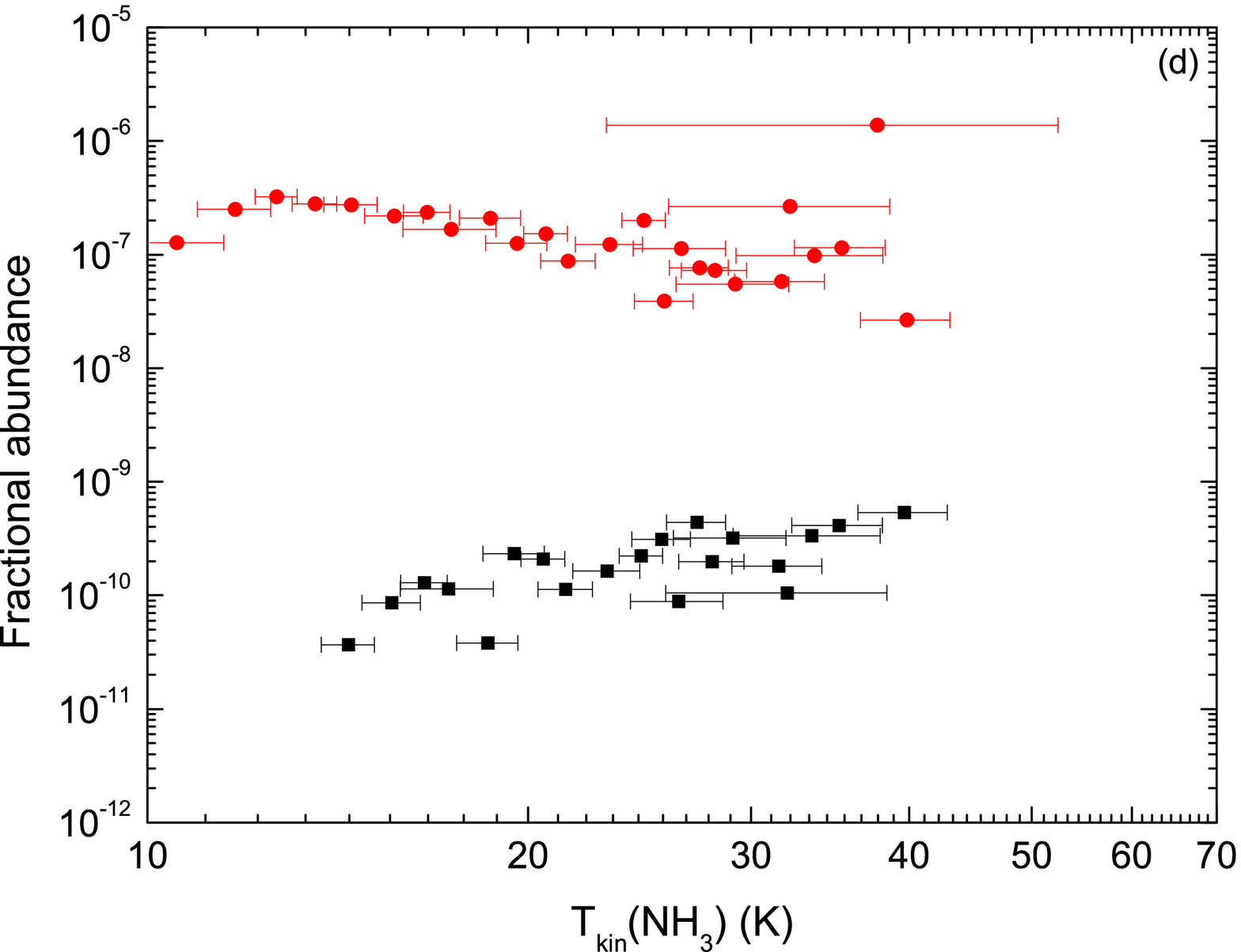}
\includegraphics[width=5.8cm]{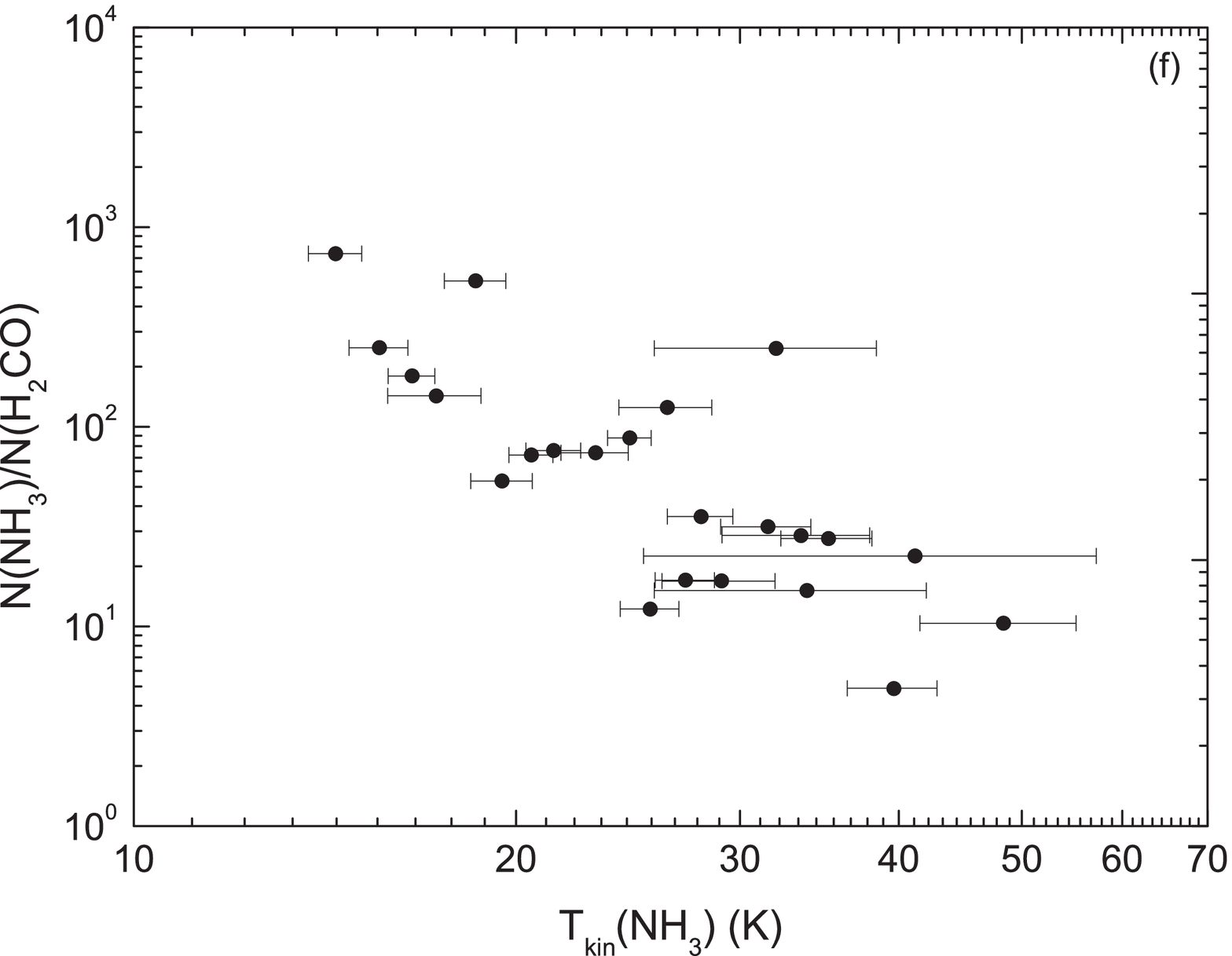}
\end{center}
\caption{Column densities of $N$(para-NH$_3$) and $N$(para-H$_2$CO)
(derived at density 10$^5$ cm$^{-3}$) (a, b), fractional abundance
of $N$(para-NH$_3$)/$N$(H$_2$) and $N$(para-H$_2$CO)/$N$(H$_2$) (c, d),
and $N$(para-NH$_3$)/$N$(para-H$_2$CO) (e, f) vs. column density
$N$(H$_2$) and kinetic temperature $T_{\rm kin}$(NH$_3$).}
\label{figure:N(H2CO)-N(NH3)-N(H2)-Tk}
\end{figure*}

The column densities of para-NH$_3$ derived from the (1,1) and (2,2)
lines \citep{Wienen2012}, those of para-H$_2$CO (derived at density
10$^5$ cm$^{-3}$), and the fractional abundances of
$N$(para-H$_2$CO)/$N$(H$_2$), $N$(para-NH$_3$)/$N$(H$_2$), and
$N$(para-NH$_3$)/$N$(para-H$_2$CO) with corresponding H$_2$ column density and
kinetic temperature $T_{\rm kin}$(NH$_3$) are shown in
Figure \ref{figure:N(H2CO)-N(NH3)-N(H2)-Tk}. The para-NH$_3$ column densities
range from 10$^{15}$ to 10$^{16}$ cm$^{-2}$ and show no correlation with
the H$_2$ column density and gas kinetic temperature in the massive
clumps (see Figure \ref{figure:N(H2CO)-N(NH3)-N(H2)-Tk} (a, b)). Variations
in the fractional abundance of $N$(para-NH$_3$)/$N$(H$_2$) amount to nearly
two orders of magnitude (2.6 $\times$ 10$^{-8}$ -- 1.5 $\times$ 10$^{-6}$).
The $N$(para-NH$_3$)/$N$(H$_2$) ratio decreases with increasing H$_2$ column density and
kinetic temperature (see Figure \ref{figure:N(H2CO)-N(NH3)-N(H2)-Tk} (c, d)).
The para-H$_2$CO column density increases proportionally with the H$_2$
column density and gas kinetic temperature
(see Figure \ref{figure:N(H2CO)-N(NH3)-N(H2)-Tk} (a, b)). The fractional
abundance of $N$(para-H$_2$CO)/$N$(H$_2$) remains stable with increasing
H$_2$ column density and kinetic temperature
(see Figure \ref{figure:N(H2CO)-N(NH3)-N(H2)-Tk} (c, d)). Nevertheless,
the scatter amounts to 0.4 -- 5.4, i.e., by more than a factor of 10.
The relative abundances $N$(para-NH$_3$)/$N$(para-H$_2$CO) range from
4.9 $\times$ 10$^0$ to 7.4 $\times$ 10$^2$ and decrease with H$_2$
column density and kinetic temperature
(see Figure \ref{figure:N(H2CO)-N(NH3)-N(H2)-Tk} (e, f)).
The stable fractional para-H$_2$CO abundances as a function of $N$(H$_2$)
and $T$$_{\rm kin}$ (see Figure \ref{figure:N(H2CO)-N(NH3)-N(H2)-Tk} (c, d)
indicates that H$_2$CO is a more reliable tracer of the H$_2$ column density than NH$_3$.

%%%%%%%%%%%%%%%%%%%Fig.1-S870um-NH3-H2CO-CH3OH-intensities%%%%%%%%

%%%%%%%%%%%%%%%%%%%Fig.1-S870um-NH3-H2CO-CH3OH-intensities%%%%%%%%
\begin{figure*}[t]
\vspace*{0.2mm}
\begin{center}
\includegraphics[width=8.5cm]{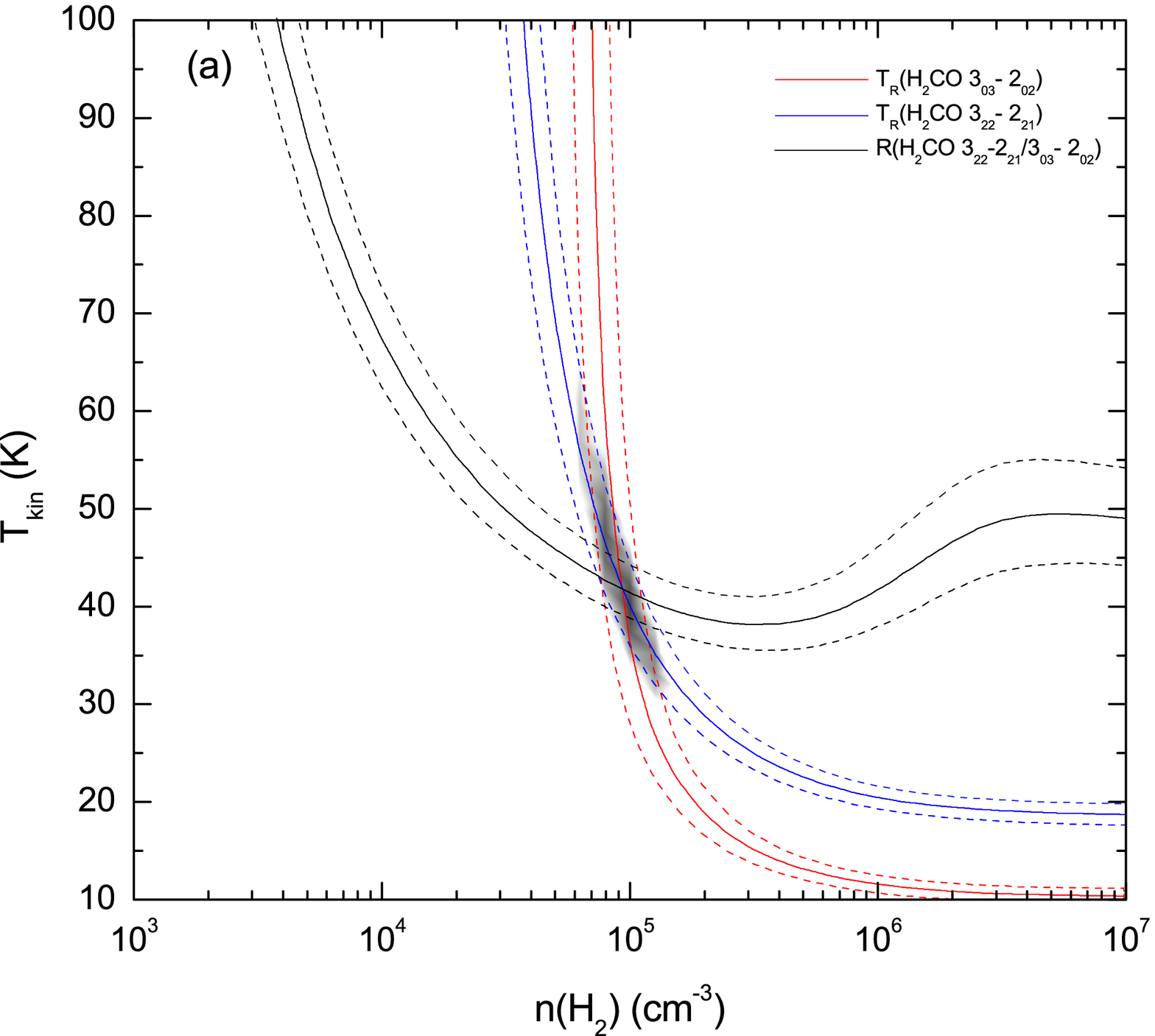}
\includegraphics[width=8.5cm]{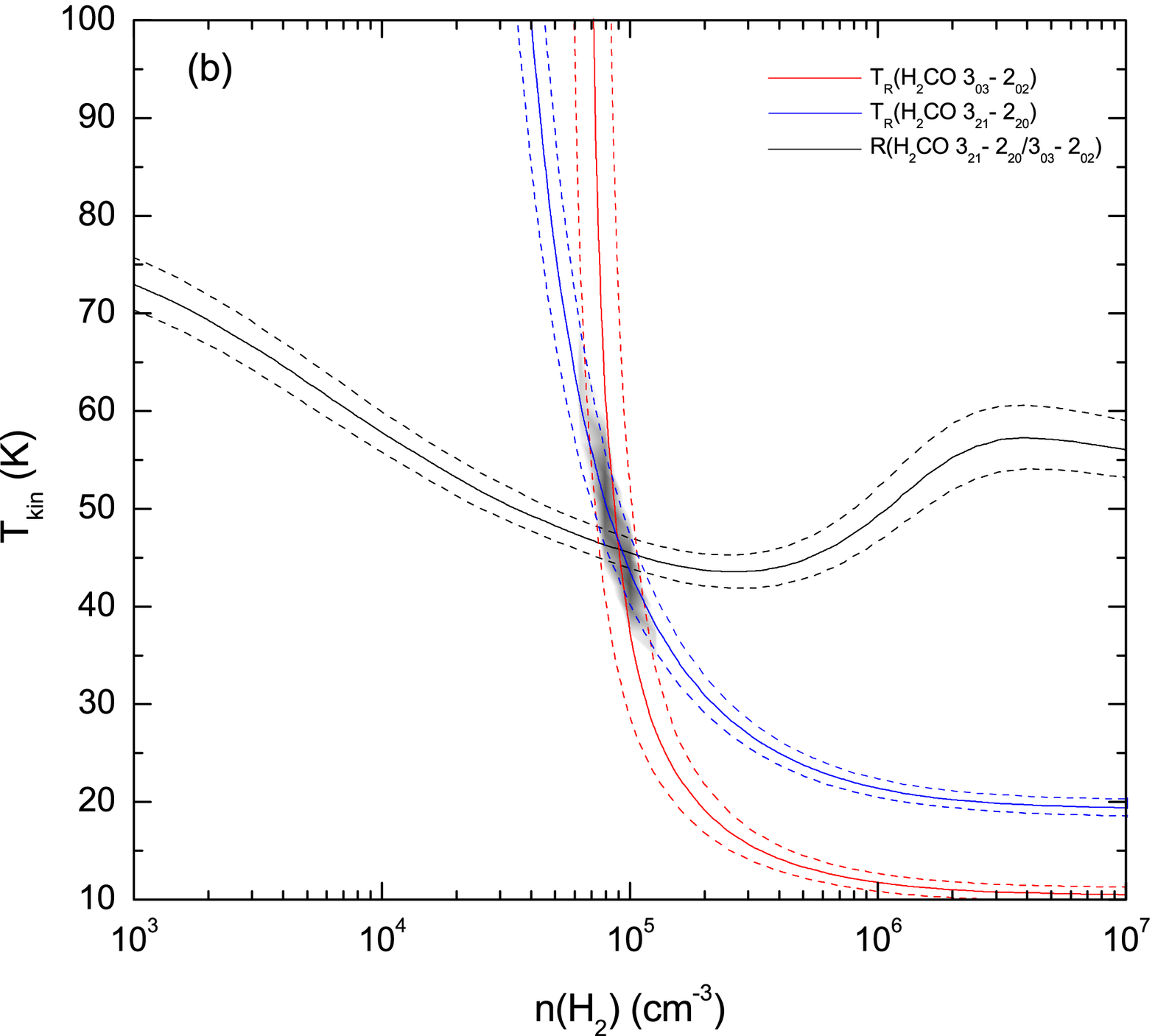}
\end{center}
\caption{Example of RADEX non-LTE modeling of the para-H$_2$CO kinetic
temperature for G5.89-0.39.  Para-H$_2$CO 3$_{03}$-2$_{02}$ (red solid
and dotted lines represent observed values and uncertainties),
3$_{22}$-2$_{21}$ and 3$_{21}$-2$_{20}$ (a and b, blue solid and dotted
lines) line brightness temperatures and
para-H$_2$CO 3$_{22}$-2$_{21}$/3$_{03}$-2$_{02}$ and
3$_{21}$-2$_{20}$/3$_{03}$-2$_{02}$ ratios (black solid and dotted lines).
The gray region is characterized by $\chi$$^2_{\rm red}$ ($<$ 1.5)
with density $n$(H$_2$) and kinetic temperature $T_{\rm kin}$ for a
para-H$_2$CO column density 4.7 $\times$ 10$^{14}$ cm$^{-2}$.}
\label{figure:Tkin}
\end{figure*}

We also derive averaged column densities and fractional abundances
of para-NH$_3$ and para-H$_2$CO in the subsamples consisting of HII
regions, EGOs, and IRDCs. For NH$_3$, the average column densities
$N$(para-NH$_3$) are 1.84 ($\pm$1.00) $\times$ 10$^{15}$, 2.32
($\pm$0.45) $\times$ 10$^{15}$, and 3.55 ($\pm$2.00) $\times$ 10$^{15}$ cm$^{-2}$,
with the errors representing the standard deviations of the mean.
The fractional abundances $N$(para-NH$_3$)/$N$(H$_2$) are
0.74 ($\pm$0.25) $\times$ 10$^{-7}$, 0.74 ($\pm$0.37) $\times$ 10$^{-7}$,
and 2.06 ($\pm$0.67) $\times$ 10$^{-7}$ in HII regions, EGOs, and IRDCs,
respectively. For H$_2$CO, the average column densities $N$(para-H$_2$CO)
are 1.16 ($\pm$1.35) $\times$ 10$^{14}$, 1.04 ($\pm$0.93) $\times$ 10$^{14}$,
and 2.81 ($\pm$2.00) $\times$ 10$^{13}$ cm$^{-2}$. Fractional abundances
$N$(para-H$_2$CO)/$N$(H$_2$) are 3.32 ($\pm$1.36) $\times$ 10$^{-10}$,
2.53 ($\pm$1.21) $\times$ 10$^{-10}$, and 1.23 ($\pm$0.66) $\times$ 10$^{-10}$
in HIIs, EGOs, and IRDCs. Average variations of fractional abundances
of $N$(para-H$_2$CO)/$N$(H$_2$) in different stages of star formation
amount to nearly a factor of 3, which is similar to the amount of
change seen in the fractional abundance $N$(para-NH$_3$)/$N$(H$_2$).
Therefore, we confirm that H$_2$CO can be widely used as a probe to
trace the dense gas without drastic changes in abundance during various
stages of star formation.

%%%%%%%%%%%%%%%%%%%%%%%%%%%%%%%%%%%%%%%%%%%%%%%%%%%%%%%%%%%%%%%%%%%%%%%%%%%%%%%%%%%%
\begin{table*}[t]
%\small
%\tiny
%\scriptsize
\caption{Para-H$_2$CO column densities and kinetic temperature.}
\centering
\begin{tabular}
%{p{2cm}p{2cm}p{2cm}p{1.4cm}p{2cm}}
{c c cccccc}
\hline\hline % inserts double horizontal lines
& \multicolumn{1}{c}{$N$(para-H$_2$CO)} & & \multicolumn{3}{c}{Kinetic temperature}\\
\cline{2-2} \cline{4-6}
Sources & $n$(H$_2$)=$10^5$ cm$^{-3}$ & & 3$_{22}$-2$_{21}$/3$_{03}$-2$_{02}$
& 3$_{21}$-2$_{20}$/3$_{03}$-2$_{02}$ & $T_{\rm LTE}$\\
        & cm$^{-2}$ & & K & K & K\\
\hline % inserts single horizontal line
G5.89-0.39  &4.7$\times10^{14}$ &&42$^{+5}_{-5}$ &45$^{+3}_{-2}$   & 58  \\
G5.90-0.44  &6.2$\times10^{13}$ &&28$^{+6}_{-5}$ &36$^{+4}_{-6}$   & 24  \\
G5.97-1.36  &2.4$\times10^{13}$ &&   ...         &    ...          & ... \\
G6.91-0.22  &1.6$\times10^{13}$ &&   ...         &    ...          & ... \\
G9.04-0.52  &4.3$\times10^{12}$ &&   ...         &    ...          & ... \\
G9.21-0.20  &6.9$\times10^{12}$ &&   ...         &    ...          & ... \\
G9.88-0.75  &3.9$\times10^{13}$ &&   ...         &    ...          & ... \\
G11.92-0.61 &3.7$\times10^{13}$ &&   ...         &    ...          & ... \\
G12.43-1.11 &6.2$\times10^{13}$ &&   ...         &    ...          & ... \\
G12.68-0.18 &2.8$\times10^{13}$ &&   ...         &    ...          & ... \\
G12.91-0.26 &1.0$\times10^{14}$ &&45$^{+5}_{-3}$ &47$^{+9}_{-7}$   & 40  \\
G14.20-0.19 &4.7$\times10^{13}$ &&53$^{+9}_{-8}$ &61$^{+14}_{-12}$ & 58  \\
G14.33-0.64 &1.6$\times10^{14}$ &&53$^{+4}_{-4}$ &51$^{+6}_{-4}$   & 76  \\
G15.66-0.50 &3.2$\times10^{13}$ &&   ...         &    ...          & ... \\
G17.10+1.02 &5.5$\times10^{12}$ &&   ...         &    ...          & ... \\
G18.21-0.34 &1.6$\times10^{13}$ &&   ...         &    ...          & ... \\
G19.01-0.03 &4.0$\times10^{13}$ &&   ...         &    ...          & ... \\
G22.55-0.52 &1.5$\times10^{13}$ &&   ...         &    ...          & ... \\
G28.61-0.03 &9.7$\times10^{12}$ &&   ...         &    ...          & ... \\
G28.86+0.07 &3.5$\times10^{13}$ &&41$^{+8}_{-6}$ &51$^{+12}_{-10}$ & 47  \\
G30.70-0.07 &6.5$\times10^{13}$ &&41$^{+5}_{-5}$ &40$^{+3}_{-2}$   & 65  \\
G31.40-0.26 &8.7$\times10^{13}$ &&41$^{+5}_{-3}$ &49$^{+5}_{-4}$   & 36  \\
G35.03+0.35 &6.5$\times10^{13}$ &&51$^{+7}_{-6}$ &48$^{+6}_{-6}$   & 56  \\
G35.19-0.74 &1.2$\times10^{14}$ &&37$^{+4}_{-3}$ &30$^{+3}_{-3}$   & 34  \\
G37.87-0.40 &6.6$\times10^{13}$ &&  ...          &    ...          & ... \\
\hline %inserts single line
\end{tabular}
\label{table:NH2CO-Tkin}
\end{table*}

\subsection{Kinetic temperature}
The para-H$_2$CO (3$_{03}$-2$_{02}$) line is the strongest of the three 218
GHz para-H$_2$CO transitions. In order to avoid small uncertain values in
the denominator, we used the para-H$_2$CO 3$_{22}$-2$_{21}$/3$_{03}$-2$_{02}$
and 3$_{21}$-2$_{20}$/3$_{03}$-2$_{02}$ ratios to derive the kinetic temperature.
The two ratios trace the kinetic temperature with an uncertainty of
$\lesssim$ 25\% below 50 K \citep{Mangum1993b}.
An example is presented to show how the parameters are constrained by the
reduced $\chi$$^2_{\rm red}$ value, line brightness, and line ratio distribution
of para-H$_2$CO in the $T_{\rm kin}$-$n$(H$_2$) parameter space in Figure \ref{figure:Tkin}.
We used the column density derived at 10$^5$ cm$^{-3}$ to constrain the
kinetic temperature.

Our results are listed in Table \ref{table:NH2CO-Tkin}. The
para-H$_2$CO 3$_{22}$-2$_{21}$/3$_{03}$-2$_{02}$ line ratio is sensitive
to the gas density at spatial densities $n$(H$_2$) $<$ 10$^5$ cm$^{-3}$
(see Figure \ref{figure:Tkin}), so it seems that this
line ratio is not quite as good as 3$_{21}$-2$_{20}$/3$_{03}$-2$_{02}$ as a
thermometer to trace kinetic temperature in the low-density regions of a
molecular cloud. At high density $n$(H$_2$) $\gtrsim$ 10$^5$ cm$^{-3}$, the
two ratios (3$_{22}$-2$_{21}$/3$_{03}$-2$_{02}$ and 3$_{21}$-2$_{20}$/3$_{03}$-2$_{02}$)
show a similar behavior to kinetic temperature and spatial density.
The comparison of kinetic temperatures derived from both
para-H$_2$CO 3$_{22}$-2$_{21}$/3$_{03}$-2$_{02}$ and 3$_{21}$-2$_{20}$/3$_{03}$-2$_{02}$
ratios suggests that the two ratios trace similar temperatures at a
density of 10$^5$ cm$^{-3}$ (see Table \ref{table:NH2CO-Tkin}).
The para-H$_2$CO 3$_{22}$-2$_{21}$ and 3$_{21}$-2$_{20}$ transitions, have
similar energy above the ground state, $E_{\rm u}$ $\simeq$ 68 K, similar line
brightness (see Table \ref{table:H2CO}), and are often detected at the same time
(e.g., \citealt{Bergman2011,Wang2012,Lindberg2012,Ao2013,Immer2014,Trevino2014,Ginsburg2016});
therefore, para-H$_2$CO 3$_{22}$-2$_{21}$/3$_{03}$-2$_{02}$ and
3$_{21}$-2$_{20}$/3$_{03}$-2$_{02}$ ratios are both good thermometers to
determine kinetic temperature in dense regions ($n$(H$_2$)
$\gtrsim$ 10$^5$ cm$^{-3}$). However, at lower densities,
the 3$_{21}$-2$_{20}$/3$_{03}$-2$_{02}$ ratio should be preferred.

The para-H$_2$CO line intensity ratios 3$_{22}$-2$_{21}$/3$_{03}$-2$_{02}$
and 3$_{21}$-2$_{20}$/3$_{03}$-2$_{02}$ can provide a measurement of the
kinetic temperature of the gas in local thermodynamic equilibrium (LTE).
The kinetic temperature can be calculated from para-H$_2$CO
transitions assuming that the lines are optically thin, and originate
from a high-density region \citep{Mangum1993b}
\begin{equation}
T_{\rm kin} =  \frac{47.1}{ln(0.556\frac{I(3_{03}-2_{02})}{I(3_{22}-2_{21})})},
\end{equation}
where $I$(3$_{03}$-2$_{02}$)/$I$(3$_{22}$-2$_{21}$) is the para-H$_2$CO
integrated intensity ratio. The results of the kinetic temperature
calculations from the para-H$_2$CO 3$_{03}$-2$_{02}$/3$_{22}$-2$_{21}$
integrated intensity ratio are listed in Table \ref{table:NH2CO-Tkin}.
The kinetic temperatures derived from this method have an uncertainty of
$\lesssim$ 30\% \citep{Mangum1993b}. Considering this uncertainty, the
temperatures derived from LTE and the RADEX non-LTE model are consistent
(see Table \ref{table:NH2CO-Tkin}).

%%%%%%%%%%%%%%%%%%%Fig.1-S870um-NH3-H2CO-CH3OH-intensities%%%%%%%%
\begin{figure*}[t]
\vspace*{0.2mm}
\begin{center}
\includegraphics[width=14cm]{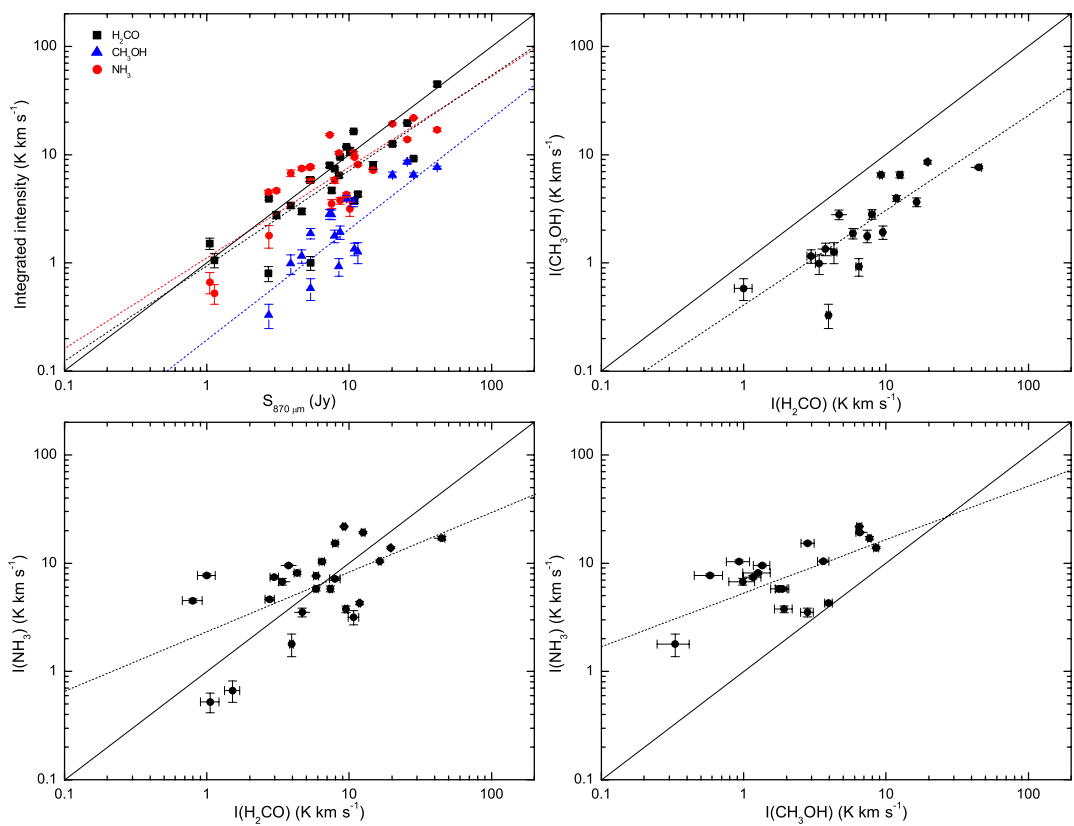}
\end{center}
\caption{Comparison of integrated intensities of para-H$_2$CO (3$_{03}$-2$_{02}$),
CH$_3$OH (4$_{22}$-3$_{12}$), NH$_3$ (1,1), and 870 $\mu$m continuum flux densities.
Dashed lines are the results from linear fits. Solid lines correspond to Y = X.
Gauss fitted peak temperatures and line widths of NH$_3$ (1,1) are from \cite{Wienen2012}.
Assuming Gaussian profiles, we plot integrated intensities calculated with
$\int$$T(v){\rm d}v$ = $\sqrt{\pi/4ln2}$$ \cdot$ $T_{\rm pk}$ $\cdot$ $\Delta$V$_{\rm FWHM}$.}
\label{figure:S870um-NH3-H2CO-CH3OH}
\end{figure*}

\section{Discussion}
\subsection{Comparison of H$_2$CO, CH$_3$OH, NH$_3$, and 870 $\mu$m emission}
We compare the integrated intensities of para-H$_2$CO (3$_{03}$-2$_{02}$),
CH$_3$OH (4$_{22}$-3$_{12}$), and NH$_3$ (1,1) with 870 $\mu$m emission
in Figure \ref{figure:S870um-NH3-H2CO-CH3OH}. It shows that the molecules
follow the 870 $\mu$m intensity distribution. The integrated intensities
of para-H$_2$CO (3$_{03}$-2$_{02}$), CH$_3$OH (4$_{22}$-3$_{12}$), and
NH$_3$ (1,1) are also compared in Figure \ref{figure:S870um-NH3-H2CO-CH3OH}.
There is a good correlation between para-H$_2$CO (3$_{03}$-2$_{02}$) and
CH$_3$OH integrated intensities (correlation coefficient $R^2$ $\sim$ 0.7).
Line widths of para-H$_2$CO (3$_{03}$-2$_{02}$) and CH$_3$OH also tend to
be similar (see Tables \ref{table:H2CO} and \ref{table:CH3OH}).
This suggests that the two molecules may trace similar regions and/or are
chemically linked in their parent massive clumps.

H$_2$CO and CH$_3$OH are thought to be formed by successive hydrogenation
of CO on grain surfaces: CO $\rightarrow$ HCO $\rightarrow$
H$_2$CO $\rightarrow$ CH$_3$O $\rightarrow$ CH$_3$OH \citep{Watanabe2002,Woon2002,Hidaka2004}.
Previous observations of para-H$_2$CO and CH$_3$OH in the Orion Bar
photon-dominated region (PDR) have suggested that para-H$_2$CO traces
the interclump material. CH$_3$OH is found mainly in the clumps, so that
the two species trace different environments \citep{Leurini2006,Leurini2010}.
Our result differs from what is found for the Orion Bar, but is consistent
with the majority of results where the two species are similarly distributed
as in e.g., W3, CrA, L1157, W33, and NGC 2264
\citep{Wang2012,Lindberg2012,Gomez2013,Immer2014,Cunningham2016}.
The likely reason is the different molecular environment. CH$_3$OH is
more easily photodissociated than H$_2$CO in the PDRs.

For the integrated intensities of $I$(para-H$_2$CO)-$I$(NH$_3$)
and $I$(CH$_3$OH)-$I$(NH$_3$), correlation coefficients ($R^2$)
are 0.4 and 0.3, respectively, so they are only weakly correlated.
Nearly all line widths of para-H$_2$CO (3$_{03}$-2$_{02}$) and
CH$_3$OH are greater than those of NH$_3$ (1,1) (see Tables
\ref{table:H2CO} and \ref{table:CH3OH}, and Tables 1 and 2 in
\cite{Wienen2012}). The weak correlation can be explained if
para-H$_2$CO and CH$_3$OH trace a higher density gas than NH$_3$ (1,1).

\subsection{Comparison of kinetic temperatures derived from H$_2$CO and NH$_3$}
\begin{figure*}[t]
\vspace*{0.2mm}
\begin{center}
\includegraphics[width=17cm]{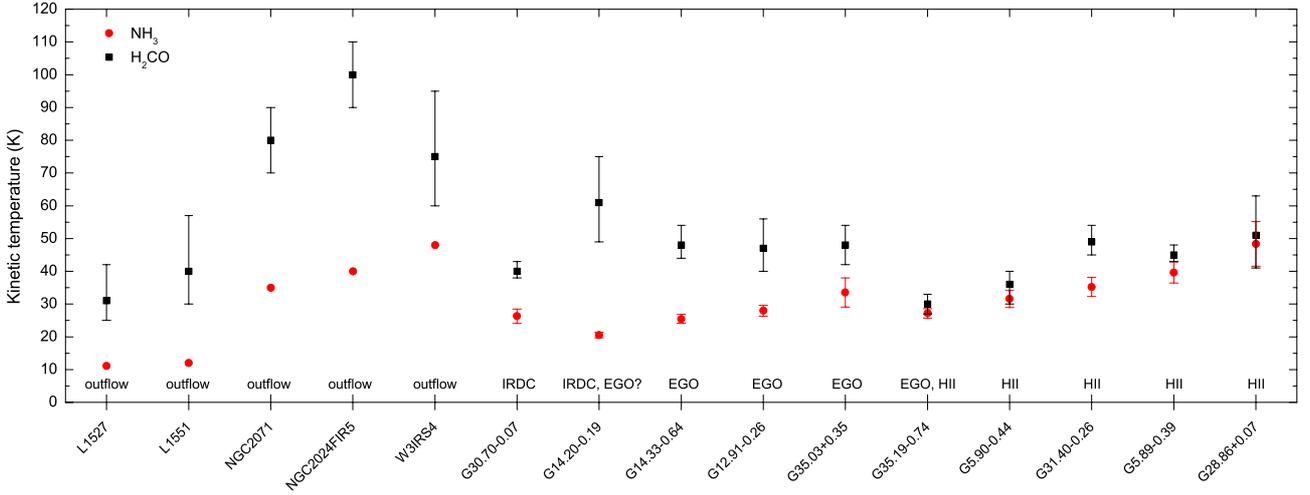}
\end{center}
\caption{Comparison of kinetic temperatures derived from
para-H$_2$CO 3$_{21}$-2$_{20}$/3$_{03}$-2$_{02}$ (black squares)
and NH$_3$ (2,2)/(1,1) (red points) ratios. Para-H$_2$CO and
NH$_3$ kinetic temperatures of L1551, L1527, NGC2071, NGC2024FIR5,
and W3IRS4 are selected from \cite{Takano1986}, \cite{Mangum1993b},
\cite{Moriarty-Schieven1995}, \cite{Jijina1999}, and \cite{Watanabe2008}.}
\label{figure:Tk-NH3-H2CO}
\end{figure*}

For our massive clump samples with kinetic temperatures derived
by para-H$_2$CO (3$_{21}$-2$_{20}$/3$_{03}$-2$_{02}$) and
NH$_3$ (2,2)/(1,1), the kinetic temperature ranges for para-H$_2$CO
from 30 to 61 K (average 46 $\pm$ 9 K), and for NH$_3$ from 21
to 48 K (average 32 $\pm$ 8 K), respectively.
The comparison of kinetic temperature derived from the
para-H$_2$CO (3$_{21}$-2$_{20}$/3$_{03}$-2$_{02}$) and the
NH$_3$ (2,2)/(1,1) line ratios is shown in Figure \ref{figure:Tk-NH3-H2CO}.
The kinetic temperatures derived from para-H$_2$CO and NH$_3$
agree in five sources, namely in G5.89-0.39, G5.90-0.44, G28.86+0.07,
G31.40-0.26, and G35.19-0.74. Higher kinetic temperatures
(difference $>$ 10 K) traced by para-H$_2$CO as compared to
NH$_3$ are found in G12.91-0.26, G14.20-0.19, G14.33-0.64, G30.70-0.07,
and G35.03+0.35. It seems that para-H$_2$CO traces a slightly higher
temperature than NH$_3$ (2,2)/(1,1) in the massive clumps.
The probable reason is that para-H$_2$CO may trace hotter and denser
regions, while the NH$_3$ (2,2)/(1,1) line ratio traces cooler and
more diffuse gas \citep{Ginsburg2016}. The different beam sizes for
para-H$_2$CO (JCMT beam $\sim$ 23$''$) and NH$_3$ (Effelsberg beam
$\sim$ 40$''$) data also have to be considered. The source sizes (FWHM)
derived from para-H$_2$CO range from 20$''$ to 31$''$ \citep{Csengeri2014},
which match the JCMT beam but are smaller than the Effelsberg beam.
The smaller JCMT beam size compared to Effelsberg might imply that
the para-H$_2$CO data focus more on the inner active cloud cores
than the NH$_3$ data do. Therefore, the determination of kinetic temperature
differences may be influenced, to a certain degree, by beam size.

%%%%%%%%%%%%%%%%%%%Fig.1-Tkin-NH3-H2CO-intensities%%%%%%%%

For an evaluation of whether beam size or other parameters play
the dominant role in revealing differences between
$T_{\rm kin}$(para-H$_2$CO) and $T_{\rm kin}$(NH$_3$ (2,2)/(1,1)),
we have to check observational results in a systematic way.
The NH$_3$ (1,1) and (2,2) transitions are sensitive to cold
(10 -- 40 K; \citealt{Ho1983,Mangum1992,Mangum2013a}) and dense
($\gtrsim$ 10$^4$ cm$^{-3}$; \citealt{Rohlfs2004}) gas.
Previous para-H$_2$CO (3$_{22}$-2$_{21}$/3$_{03}$-2$_{02}$)
and NH$_3$ (2,2)/(1,1) observations toward protostars, bipolar
flows, submillimeter clumps, far infrared sources, active star
formation sources, the Galactic center, Large Magellanic Clouds,
and starburst galaxies show significantly different gas kinetic
temperatures. Previous observed results with different telescopes
are listed in Table \ref{table:Pre-TH2CO-TNH3}. For L1527, L1551,
N159W, and W3IRS4, the kinetic temperature difference still is
$<$ 30 K. Larger differences ($>$ 45 K) are found in NGC2071 and
NGC2024FIR5. The most significant difference is in the starburst
galaxy M82 ($T_{\rm kin}$(NH$_3$) deduced from (1,1)-(3,3),
\citealt{Weiss2001}). Toward NGC2071 and M82, the NH$_3$ beam was
the larger one, but in the case of NGC2024FIR5 we find the opposite.
We also find the opposite in the case of N159W, $T_{\rm kin}$(NH$_3$) $<$ $T_{\rm kin}$(para-H$_2$CO)
in spite of a smaller ammonia beam. Therefore, the beam size
difference between our JCMT para-H$_2$CO and the Effelsberg
NH$_3$ data is likely not a dominant factor.

As shown in Figure \ref{figure:Tk-NH3-H2CO}, it seems that the differences
between $T_{\rm kin}$(para-H$_2$CO) and $T_{\rm kin}$(NH$_3$ (2,2)/(1,1)
vary with evolutionary stage of the respective massive star formation region.
The derived kinetic temperatures
from para-H$_2$CO are distinctly higher than those from
NH$_3$ (2,2)/(1,1) in the clumps associated with EGOs
(difference $>$ 14 K; G12.91-0.26, G14.20-0.19, G14.33-0.64, and G35.03+0.35).
Similar temperature differences have been found in L1527,
L1551, NGC2024FIR5, NGC2071, and W3IRS4, which are well-known
outflow objects. The derived kinetic temperatures
from NH$_3$ (2,2)/(1,1) may reflect an average temperature
of cooler and more diffuse gas.
The outflow/shock could heat the dense gas
traced by H$_2$CO. Therefore, in these cases, para-H$_2$CO
probes higher temperature gas which appears to be related to
gas excited by star formation activities (e.g., outflows, shocks).
The kinetic temperatures derived from para-H$_2$CO and
NH$_3$ (2,2)/(1,1) are in agreement in the sources associated
with HII regions (difference $<$ 14 K;
G5.89-0.39, G5.90-0.44, G28.86+0.07, and
G31.40-0.26). This indicates that temperature gradients
potentially probed by para-H$_2$CO and NH$_3$ (2,2)/(1,1) in different
parts of the clouds are small.
To conclude, para-H$_2$CO is a good
thermometer, like NH$_3$, to trace the gas kinetic temperature
($T_{\rm kin}$(gas) $\gtrsim$ 30 K) in the molecular environment
surrounding HII regions. Large differences in kinetic temperatures between
$T_{\rm kin}$(para-H$_2$CO) and $T_{\rm kin}$(NH$_3$ (2,2)/(1,1)
may indicate clouds in different evolutionary stages of massive star formation.

The kinetic temperatures based on para-H$_2$CO data disagree with
the values obtained from NH$_3$ (1,1) and (2,2), but agree with
the properties of the high-excitation component traced by CO in
the starburst galaxy M82 \citep{Muhle2007}. \cite{Ao2013} found
that the para-H$_2$CO kinetic temperatures are consistent with
the temperatures derived from high-$J$ NH$_3$ \citep{Mauersberger1986}
in the Galactic CMZ. Higher excited NH$_3$ lines commonly lead
to higher kinetic temperatures. Therefore, if higher NH$_3$
levels (e.g., NH$_3$ (2,2)/(4,4); \cite{Mangum2013a,Gong2015})
are also involved in measuring the kinetic temperatures, the
values derived from para-H$_2$CO and NH$_3$ might become consistent
in these sources where we have found a discrepancy. Thus
detailed comparisons of $T_{\rm kin}$ values deduced from
para-H$_2$CO and high-$J$ NH$_3$ transitions would be meaningful.

%%%%%%%%%%%%%%%%%%%%%%%%%%%%%%%%%%%%%%%%%%%%%%%%%%%%%%%%%%%%%%%%%%%%%%%%%%%%%%%%%%%%
\begin{table}[t]
\caption{Previous results of observed para-H$_2$CO and NH$_3$ temperatures.}
\centering
\begin{tabular}
%{p{2cm}p{2cm}p{2cm}p{1.4cm}p{2cm}}
{cccccccc}
\hline\hline % inserts double horizontal lines
%Sources &    & N(H$_2$CO) &  &\\  \cline{2-5}
& \multicolumn{2}{c}{NH$_3$} & & \multicolumn{2}{c}{para-H$_2$CO}&\\
\cline{2-3} \cline{5-6}
Sources & Beam & $T_{\rm kin}$ & & Beam & $T_{\rm kin}$ & Ref.\\
        & arcsec &  K && arcsec & K & \\
\hline % inserts single horizontal line
L1527       & 88 &   11  & & 28    & 31$^{+11}_{-6}$   & 1,2 \\
L1551       & 40 &   12  & & 28    & 40$^{+17}_{-10}$  & 1,3 \\
W3IRS4      & 40 &   48  & & 30    & 75$^{+20}_{-15}$  & 3,4 \\
NGC2071     & 88 &   35  & & 23,30 & 80$^{+10}_{-10}$  & 3,4,5 \\
NGC2024FIR5 & 3  &   40  & & 23    & 100$^{+10}_{-10}$ & 3,6 \\
N159W       & 9  &   16  & & 23    & 29$^{+5}_{-5}$    & 7,8 \\
M82         & 40 &   60  & & 23    & 200 & 9,10 \\
\hline %inserts single line
\end{tabular}
\label{table:Pre-TH2CO-TNH3}
\tablebib{(1) \cite{Moriarty-Schieven1995}; (2) \cite{Takano1986};
(3) \cite{Jijina1999}; (4) \cite{Mangum1993b}; (5) \cite{Mitchell2001};
(6) \cite{Watanabe2008}; (7) \cite{Heikkila1999}; (8) \cite{Ott2010};
(9) \cite{Muhle2007}; (10) \cite{Weiss2001}.}
\end{table}

%%%%%%%%%%%%%%%%%%%Fig.1-Tkin-322-321-intensities%%%%%%%%
\begin{figure}[t]
\vspace*{0.2mm}
\begin{center}
\includegraphics[width=8.5cm]{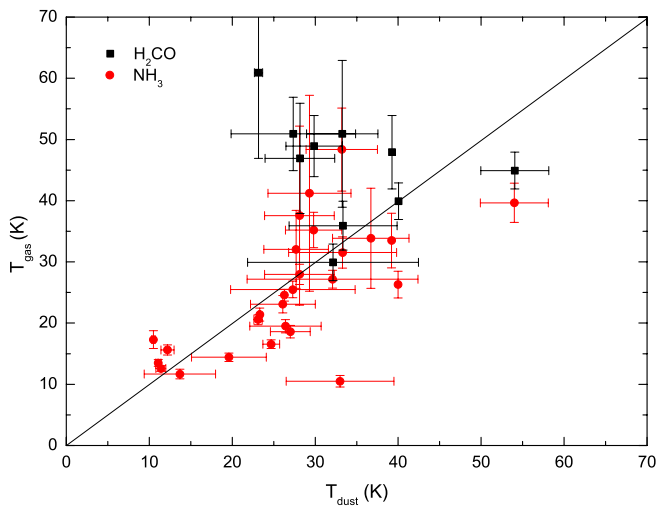}
\end{center}
\caption{Comparison of gas kinetic temperatures derived from
para-H$_2$CO 3$_{21}$-2$_{20}$/3$_{03}$-2$_{02}$ (black squares)
and NH$_3$ (2,2)/(1,1) (red points) ratios against the HiGal
dust temperatures. The straight line indicates  locations of
equal temperatures. Two sources (G5.97-1.36 and G17.10+1.02)
with particularly large $T_{\rm kin}$(NH$_3$) errors are not
shown here.}
\label{figure:Tdust-TH2CO-TNH3}
\end{figure}

%%%%%%%%%%%%%%%%%%%Fig.1-velocity-dispersion-intensities%%%%%%%%
\begin{figure}[t]
\vspace*{0.2mm}
\begin{center}
\includegraphics[width=8.5cm]{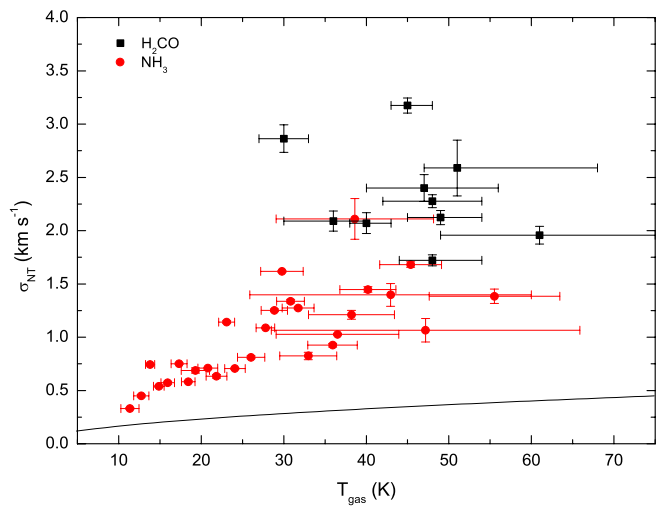}
\end{center}
\caption{Non-thermal velocity dispersion ($\sigma_{NT}$) vs.
gas kinetic temperature for para-H$_2$CO (black squares) and
NH$_3$ (red points). For the black squares, the gas kinetic
temperatures were derived from
para-H$_2$CO 3$_{21}$-2$_{20}$/3$_{03}$-2$_{02}$ line ratios.
For the red points, they were obtained from the NH$_3$ (2,2)/(1,1)
line ratios. The black line is the corresponding thermal
sound speed. Two sources (G5.97-1.36 and G17.10+1.02) with
particularly large $T_{\rm kin}$(NH$_3$) errors are not shown here.}
\label{figure:velocity-dispersion}
\end{figure}

\subsection{Comparison of kinetic temperatures derived from the gas and the dust}
The observed gas and dust temperatures do not agree in the Galactic
CMZ \citep{Gusten1981,Ao2013,Ott2014,Ginsburg2016,Immer2016}.
However, the temperatures derived from dust and gas are often in
agreement in the active dense clumps of Galactic disk clouds
\citep{Dunham2010,Giannetti2013,Battersby2014}. The dust temperatures
are obtained from SED fitting to Herschel HiGal data at 70, 160,
250, 350, and 500 $\mu$m and ATLASGAL data at 870 $\mu$m for our
samples, following the method described in K\"{o}nig et al. (submitted).
The results are listed in Table \ref{table:source}. The derived
kinetic temperature range is 11 -- 54 K (average 26 $\pm$ 8 K).
A comparison of gas kinetic temperature derived from
para-H$_2$CO (3$_{21}$-2$_{20}$/3$_{03}$-2$_{02}$) and NH$_3$ (2,2)/(1,1)
against HiGal dust temperatures is shown in Figure \ref{figure:Tdust-TH2CO-TNH3}.
For the temperatures derived from para-H$_2$CO, most sources
(G12.91-0.26, G14.20-0.19, G14.33-0.64, G28.86+0.07, G31.40-0.26, and G35.03+0.35)
show a higher temperature (difference $>$ 9 K) than the HiGal
dust temperature. The difference is due to the
dust that may trace an average temperature covering a rather wide range of densities
and is not as influenced by outflowing gas as para-H$_2$CO.
The gas temperatures determined from NH$_3$ (2,2)/(1,1)
agree with the HiGal dust temperatures considering the uncertainties,
which agrees with previous results found in the active dense
clumps of Galactic disk clouds
\citep{Dunham2010,Giannetti2013,Battersby2014}.

%%%%%%%%%%%%%%%%%%%%%%%%%%%%%%%%%%%%%%%%%%%%%%%%%%%%%%%%%%%%%%%%%%%%%%%%%%%%%%%%%%%%
%\centering
\begin{table*}[t]
\caption{Thermal and non-thermal parameters.}
\label{table:velocity-dispersion}
\centering
\begin{tabular}
%{p{2cm}p{2cm}p{2cm}p{1.4cm}p{2cm}}
{cccccccccc}
\hline\hline % inserts double horizontal lines
& \multicolumn{4}{c}{NH$_3$} & & \multicolumn{4}{c}{H$_2$CO}\\
\cline{2-5} \cline{7-10}
Sources & $\sigma_{T}$ & $\sigma_{NT}$ & $a_s$ & $R_p$ & & $\sigma_{T}$ & $\sigma_{NT}$ & $a_s$ & $R_p$\\
&  km s$^{-1}$ & km s$^{-1}$ & km s$^{-1}$ & & & km s$^{-1}$ & km s$^{-1}$ & km s$^{-1}$ & \\
\hline % inserts single horizontal line
G5.89-0.39   & 0.13 & 1.68 & 0.36 & 0.046 &  & 0.10 & 3.17 & 0.38 & 0.014 \\
G5.90-0.44   & 0.12 & 0.92 & 0.32 & 0.121 &  & 0.08 & 2.09 & 0.34 & 0.027 \\
G5.97-1.36   & 0.13 & 0.87 & 0.35 & 0.160 &  & ...  & 0.90 & ...  & ...   \\
G6.91-0.22   & 0.08 & 0.75 & 0.22 & 0.090 &  & ...  & 2.20 & ...  & ...   \\
G9.04-0.52   & 0.08 & 0.57 & 0.22 & 0.144 &  & ...  & 0.82 & ...  & ...   \\
G9.21-0.20   & 0.09 & 0.70 & 0.24 & 0.121 &  & ...  & 1.18 & ...  & ...   \\
G9.88-0.75   & 0.10 & 0.80 & 0.27 & 0.115 &  & ...  & 1.57 & ...  & ...   \\
G10.99-0.08  & 0.08 & 0.74 & 0.20 & 0.074 &  & ...  & ...  & ...  & ...   \\
G11.92-0.61  & 0.10 & 0.70 & 0.26 & 0.142 &  & ...  & 2.41 & ...  & ...   \\
G12.43-1.11  & 0.11 & 0.82 & 0.31 & 0.141 &  & ...  & 1.44 & ...  & ...   \\
G12.68-0.18  & 0.12 & 1.02 & 0.32 & 0.099 &  & ...  & 2.74 & ...  & ...   \\
G12.91-0.26  & 0.11 & 1.27 & 0.30 & 0.056 &  & 0.11 & 2.40 & 0.39 & 0.026 \\
G13.28-0.30  & 0.08 & 0.53 & 0.21 & 0.153 &  & ...  & ...  & ...  & ...   \\
G14.20-0.19  & 0.10 & 1.14 & 0.26 & 0.051 &  & 0.12 & 1.96 & 0.44 & 0.051 \\
G14.33-0.64  & 0.11 & 1.25 & 0.29 & 0.053 &  & 0.11 & 1.72 & 0.39 & 0.052 \\
G15.66-0.50  & 0.11 & 1.08 & 0.28 & 0.068 &  & ...  & 2.03 & ...  & ...   \\
G17.10+1.02  & 0.17 & 0.72 & 0.44 & 0.372 &  & ...  & 1.25 & ...  & ...   \\
G18.21-0.34  & 0.09 & 0.68 & 0.24 & 0.120 &  & ...  & 2.13 & ...  & ...   \\
G19.01-0.03  & 0.09 & 0.63 & 0.25 & 0.161 &  & ...  & 2.57 & ...  & ...   \\
G22.55-0.52  & 0.09 & 0.58 & 0.23 & 0.161 &  & ...  & 1.82 & ...  & ...   \\
G28.61-0.03  & 0.14 & 1.06 & 0.36 & 0.119 &  & ...  & 2.63 & ...  & ...   \\
G28.86+0.07  & 0.15 & 1.38 & 0.39 & 0.082 &  & 0.10 & 2.59 & 0.41 & 0.025 \\
G30.24+0.57  & 0.13 & 1.39 & 0.35 & 0.063 &  & ...  & ...  & ...  & ...   \\
G30.70-0.07  & 0.11 & 1.61 & 0.29 & 0.033 &  & 0.10 & 2.07 & 0.36 & 0.030 \\
G31.40-0.26  & 0.13 & 1.44 & 0.34 & 0.055 &  & 0.10 & 2.12 & 0.40 & 0.035 \\
G31.70-0.49  & 0.07 & 0.44 & 0.19 & 0.191 &  & ...  & ...  & ...  & ...   \\
G34.37-0.66  & 0.07 & 0.32 & 0.18 & 0.323 &  & ...  & ...  & ...  & ...   \\
G35.03+0.35  & 0.12 & 1.20 & 0.33 & 0.075 &  & 0.11 & 2.27 & 0.39 & 0.030 \\
G35.19-0.74  & 0.11 & 1.33 & 0.30 & 0.049 &  & 0.10 & 2.86 & 0.31 & 0.012 \\
G37.87-0.40  & 0.12 & 2.11 & 0.33 & 0.025 &  & ...  & 2.67 & ...  & ...   \\
\hline %inserts single line
\end{tabular}
\tablefoot{Columns 2--5 and 6--9 are thermal linewidth, non-thermal
velocity dispersion, thermal sound speed, and the ratio of thermal
to nonthermal pressure obtained from NH$_3$ (1,1) and
para-H$_2$CO (3$_{03}$-2$_{02}$) with kinetic temperatures derived
from the NH$_3$ (2,2)/(1,1) and
para-H$_2$CO (3$_{21}$-2$_{20}$/3$_{03}$-2$_{02}$) line intensity ratios,
respectively.
Para-H$_2$CO non-thermal velocity dispersions without kinetic
temperatures are deduced from $\sigma_{NT}$ = $\Delta V$/2.355
\citep{Pan2009}, where $\Delta V$ is the measured FWHM linewidth.}
\end{table*}

\subsection{Non-thermal motion and turbulence}
We computed thermal linewidth ($\sigma_T$), non-thermal velocity
dispersion ($\sigma_{NT}$), thermal sound speed ($a_s$), and the
ratio of thermal to non-thermal pressure ($R_p$) (Lada 2003).
The $\sigma_T$, $\sigma_{NT}$, $a_s$, and $R_p$ are given by\\
\begin{equation}
\sigma_T = \sqrt{\frac{kT_{\rm kin}}{m_x}},
\end{equation}
\begin{equation}
\sigma_{NT} = \sqrt{\frac{\Delta V^2}{8\ln2}-\sigma_T^2},
\end{equation}
\begin{equation}
a_s = \sqrt{\frac{kT_{\rm kin}}{\mu m_H}},
\end{equation}
\begin{equation}
R_p = \frac{a_s^2}{\sigma_{NT}^2},
\end{equation}
where $k$ is the Boltzmann constant, $T_{\rm kin}$ is the kinetic
temperature of the gas, $m_x$ is the mass of the relevant
molecule, $\Delta V$ is the measured FWHM linewidth of  either the
para-H$_2$CO 3$_{03}$-2$_{02}$ or the NH$_3$ (1,1) transitions,
$\mu$ = 2.37 is the mean molecular weight for molecular clouds,
and $m_H$ is the mass of the hydrogen atom. The derived values
of $\sigma_T$, $\sigma_{NT}$, $a_s$, and $R_p$ are listed in
Table \ref{table:velocity-dispersion}.

Comparisons of velocity dispersion, thermal sound speed, and
kinetic temperature are shown in Figure \ref{figure:velocity-dispersion}
and Table \ref{table:velocity-dispersion}. The derived non-thermal
velocity dispersions of para-H$_2$CO and NH$_3$ are higher than
the thermal linewidths. This indicates
that the line broadening of para-H$_2$CO and NH$_3$ is dominated
by non-thermal motions in these massive clumps.  Para-H$_2$CO
traces higher non-thermal motions (average $\sigma_{NT}$ = 2.1 $\pm$ 0.6 km s$^{-1}$)
than those traced by NH$_3$ (for our selected sample, the average
becomes $\sigma_{NT}$ = 1.1 $\pm$ 0.4 km s$^{-1}$; for all NH$_3$
samples observed by \cite{Wienen2012}, the average becomes
$\sigma_{NT}$ = 0.9 $\pm$ 0.4 km s$^{-1}$). Para-H$_2$CO linewidths
appear to be affected strongly by non-thermal motions.

The average values of the Mach number (given as $M$ = $\sigma_{NT}$/$a_s$)
for para-H$_2$CO and NH$_3$ are 6.2 $\pm$ 1.5 and 3.4 $\pm$ 1.1,
which indicates that the velocity distributions within these massive
clumps are significantly influenced by supersonic non-thermal
components (e.g., turbulent motions, infall, outflow, rotation,
shocks, and/or magnetic fields; \citealt{Urquhart2015}). The mean
value of the Mach number derived from NH$_3$ agrees with the
result ($\sim$ 3.2) of the Bolocam Galactic Plane Survey (BGPS)
sources \citep{Dunham2011}. The determined ratio of thermal to
non-thermal pressure, $R_p$ (see Eq.(6)),
ranges from 0.01 to 0.05 and the average becomes 0.03 $\pm$ 0.01
for para-H$_2$CO. For NH$_3$, we find values between 0.02 and 0.37
and the average becomes 0.11 $\pm$ 0.08 (all NH$_3$ samples
observed by \cite{Wienen2012} yield 0.01--0.47 and the average
becomes 0.10 $\pm$ 0.06). The low $R_p$ values indicate that
non-thermal pressure is dominant in these massive clumps.

It is expected that the correlation between kinetic temperature
and line width is due to a conversion of turbulent energy into
heat \citep{Gusten1985,Molinari1996,Ginsburg2016,Immer2016}.
Recent para-H$_2$CO observations of the CMZ have shown that
the warm dense gas is most likely heated by turbulence \citep{Ao2013,Ginsburg2016,Immer2016}.
Clumps formed in turbulent molecular clouds are significantly
affected by the temperature of the cloud material \citep{Bethell2004}.
We examine whether there is a relationship between turbulence
and temperature in our massive clumps. We adopt the non-thermal
velocity dispersion of NH$_3$ and para-H$_2$CO as proxy for
the turbulence, and the kinetic temperatures of NH$_3$ (2,2)/(1,1)
and para-H$_2$CO (3$_{21}$-2$_{20}$/3$_{03}$-2$_{02}$)
as the gas kinetic temperature. Figure \ref{figure:velocity-dispersion}
shows that the non-thermal velocity dispersion of NH$_3$ and
para-H$_2$CO are significantly positively correlated with the
gas kinetic temperature. This implies that those massive clumps
are turbulent and the gas may be heated by turbulent heating.

\section{Summary}
We have measured the kinetic temperature with
para-H$_2$CO ($J$$_{K_AK_C}$ = 3$_{03}$-2$_{02}$, 3$_{22}$-2$_{21}$,
and 3$_{21}$-2$_{20}$) and compare the kinetic temperature derived
from this formaldehyde 218 GHz line triplet with those obtained
from ammonia for 30 massive star forming clumps using the 15m JCMT.
The main results are the following:
\begin{enumerate}
\item
The integrated intensity distributions of para-H$_2$CO, CH$_3$OH,
and NH$_3$ agree well with the 870 $\mu$m intensity distributions.
The integrated intensities and linewidths of H$_2$CO and CH$_3$OH are
also consistent in our clumps. They may trace similar regions
and/or be chemically linked, while the correlation with NH$_3$
is less pronounced.

\item
Using the RADEX non-LTE model, we derive gas kinetic temperatures
by modeling the measured para-H$_2$CO 3$_{22}$-2$_{21}$/3$_{03}$-2$_{02}$
and 3$_{21}$-2$_{20}$/3$_{03}$-2$_{02}$ line ratios. We find that
the two ratios are good thermometers to trace kinetic temperatures
in dense regions ($n$(H$_2$) $\gtrsim$ 10$^5$ cm$^{-3}$) of the
massive clumps, while for lower densities the
3$_{21}$-2$_{20}$/3$_{03}$-2$_{02}$ line ratio should be preferred.

\item
The gas kinetic temperature of the massive clumps measured by
NH$_3$ (2,2)/(1,1) line ratios \citep{Wienen2012} ranges from 11 to
61 K (average 27 $\pm$ 12 K). The derived dust temperature range from Herschel
HiGal data is 11 -- 54 K with an average of 26 $\pm$ 8 K.
The gas kinetic temperature derived from para-H$_2$CO (3$_{21}$-2$_{20}$/3$_{03}$-2$_{02}$)
line ratios of the massive clumps ranges from 30 to 61 K with an
average of 46 $\pm$ 9 K, which is higher than that measured by the
NH$_3$ (2,2)/(1,1) transitions and the dust emission.

\item
A comparison of kinetic temperatures derived from para-H$_2$CO,
NH$_3$ (2,2)/(1,1), and the dust emission indicates that in many
cases para-H$_2$CO traces a similar kinetic temperature to the
NH$_3$ (2,2)/(1,1) transitions and the dust associated with the
HII regions. Distinctly higher temperatures are probed by
para-H$_2$CO in the clumps associated with outflows/shocks.

\item
Kinetic temperatures obtained by para-H$_2$CO trace turbulence to a higher
degree than NH$_3$ (2,2)/(1,1) in the massive clumps. The non-thermal
velocity dispersions of para-H$_2$CO and, to a lesser degree, NH$_3$
are positively
correlated with the gas kinetic temperature. The massive clumps are
significantly influenced by supersonic non-thermal motions.
\end{enumerate}

\begin{acknowledgements}
We thank the staff of the JCMT telescope for their assistance in observations.
The authors are grateful for the helpful comments of the referee.
We thank Jens Kauffmann and Yan Gong for valuable comments.
This work was funded by The
National Natural Science Foundation of China under grant 11433008 and The
Program of the Light in China's Western Region (LCRW) under grant
Nos.XBBS201424 and The National Natural Science Foundation of China under
grant 11373062, and partly supported by the National Basic Research Program
of China (973 program, 2012CB821802). C.H acknowledges support by a visiting
professorship for senior international scientists of the Chinese Academy of
Sciences (2013T2J0057). This research has used NASA's Astrophysical Data System (ADS).
\end{acknowledgements}

\onecolumn

\Online
\begin{appendix} %First online appendix

\section{Spectra of para-H$_2$CO and CH$_3$OH.}

%%%%%%%%%%%%%%%%%%%Fig.1-S870um-NH3-H2CO-CH3OH-intensities%%%%%%%%
\begin{figure*}[h]
\vspace*{0.2mm}
\begin{center}
\includegraphics[width=4.3cm]{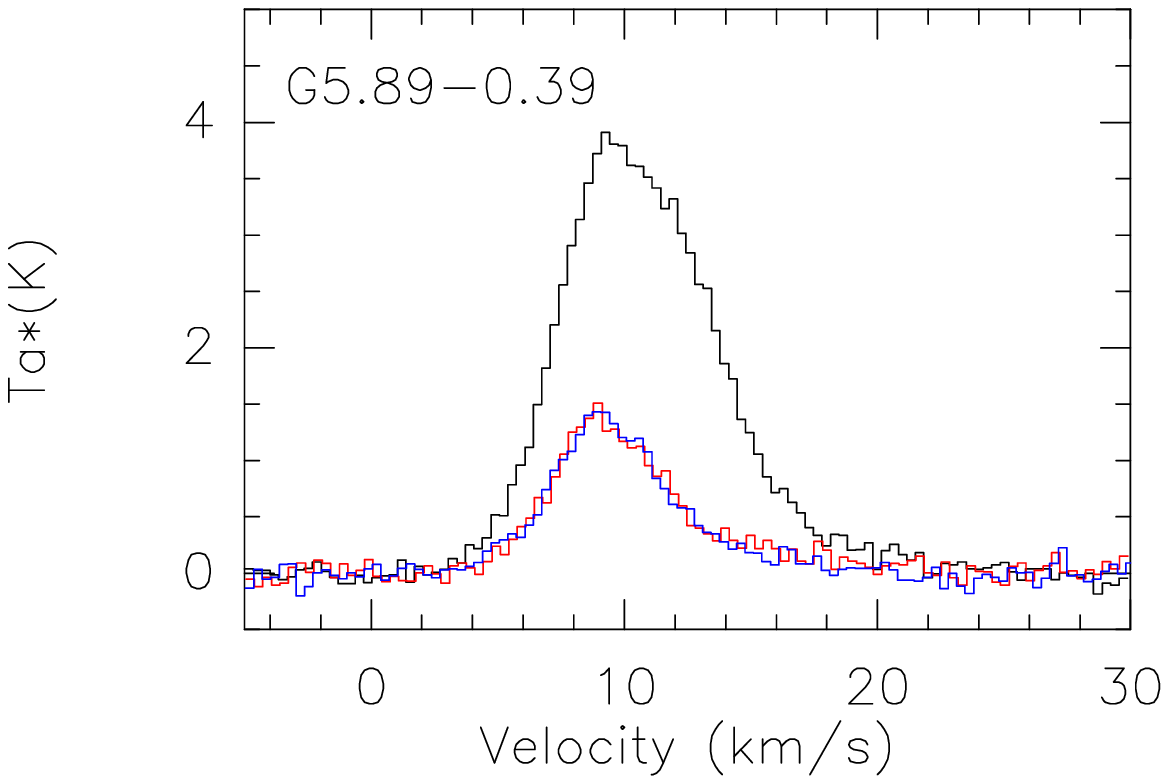}
\includegraphics[width=4.3cm]{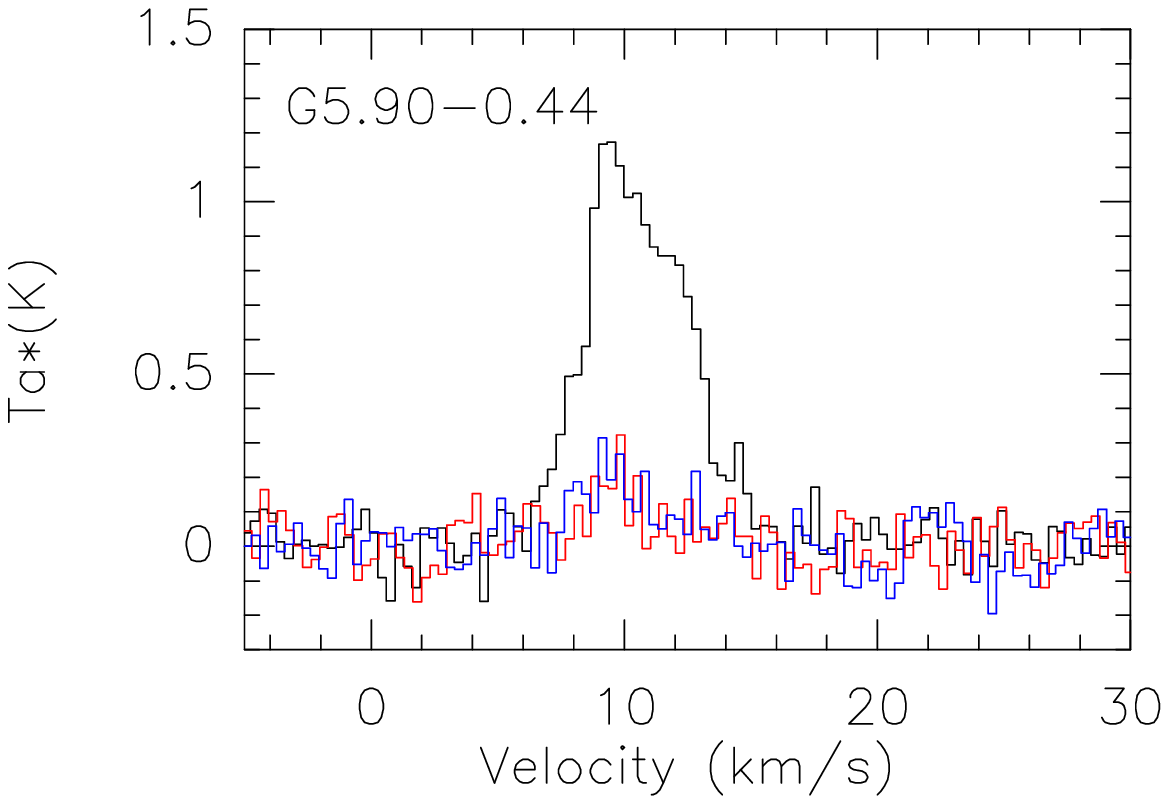}
\includegraphics[width=4.3cm]{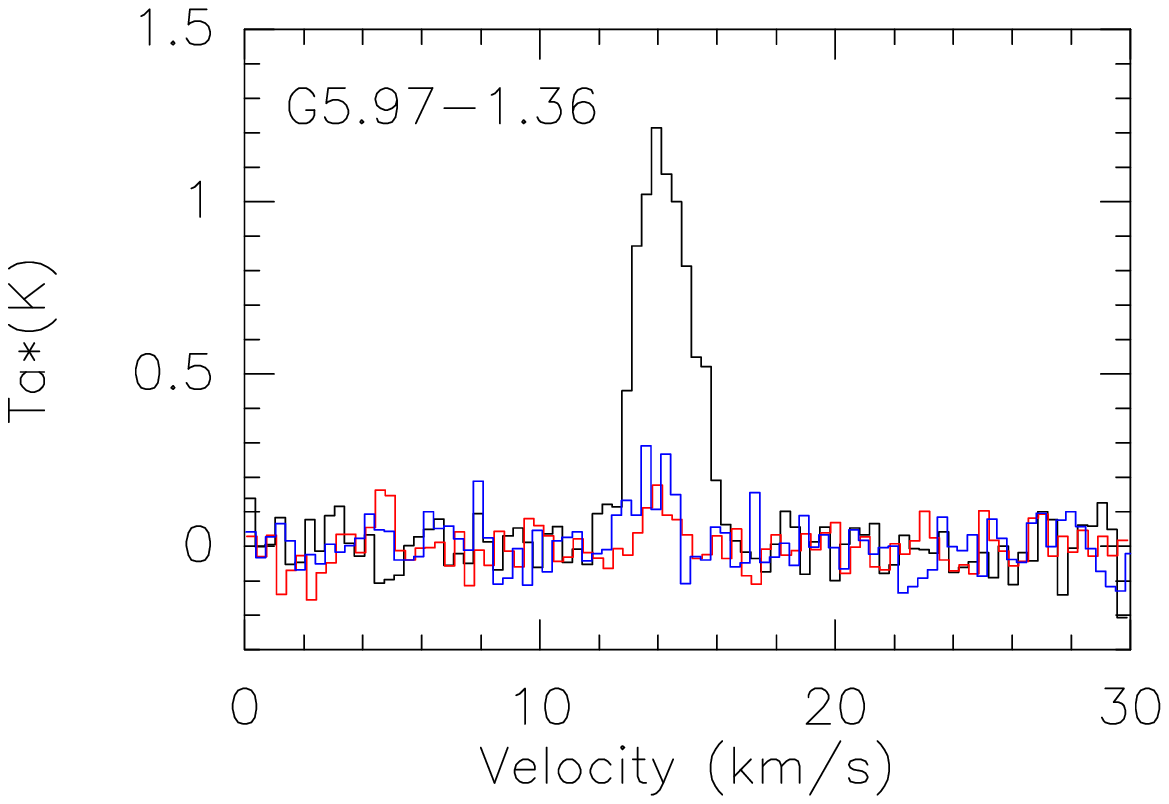}
\includegraphics[width=4.3cm]{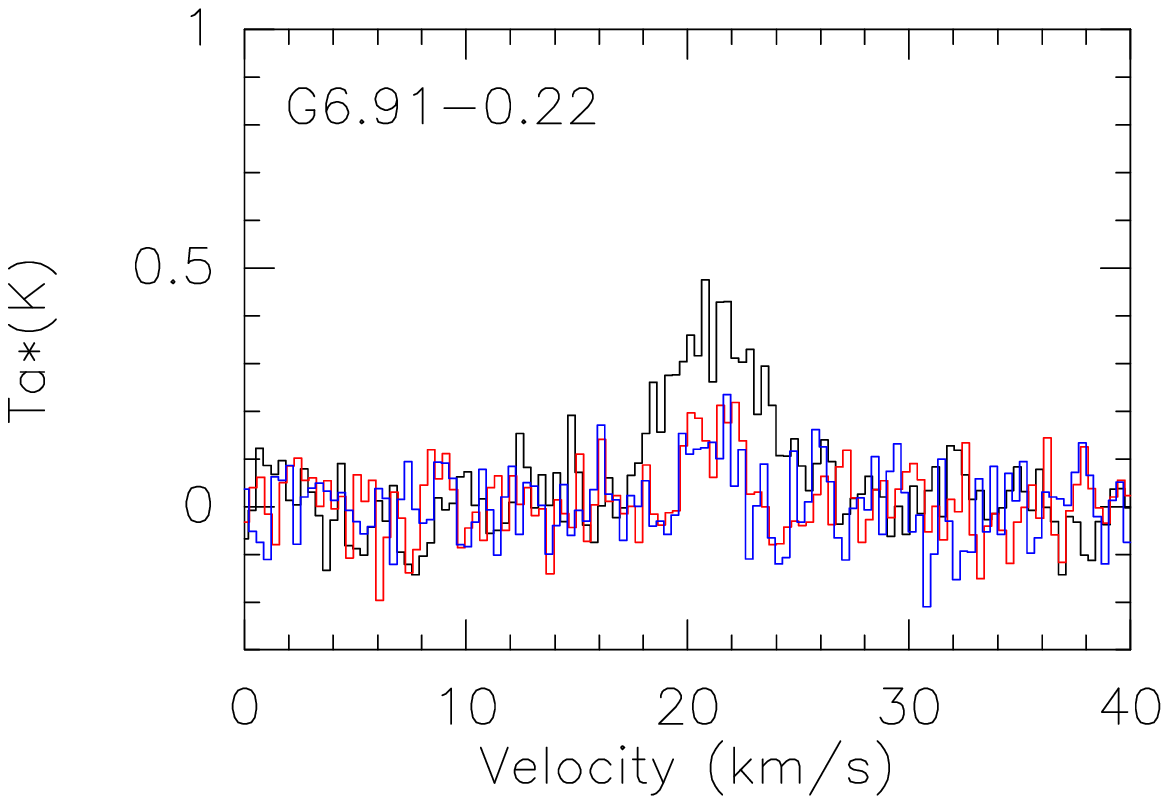}
\includegraphics[width=4.3cm]{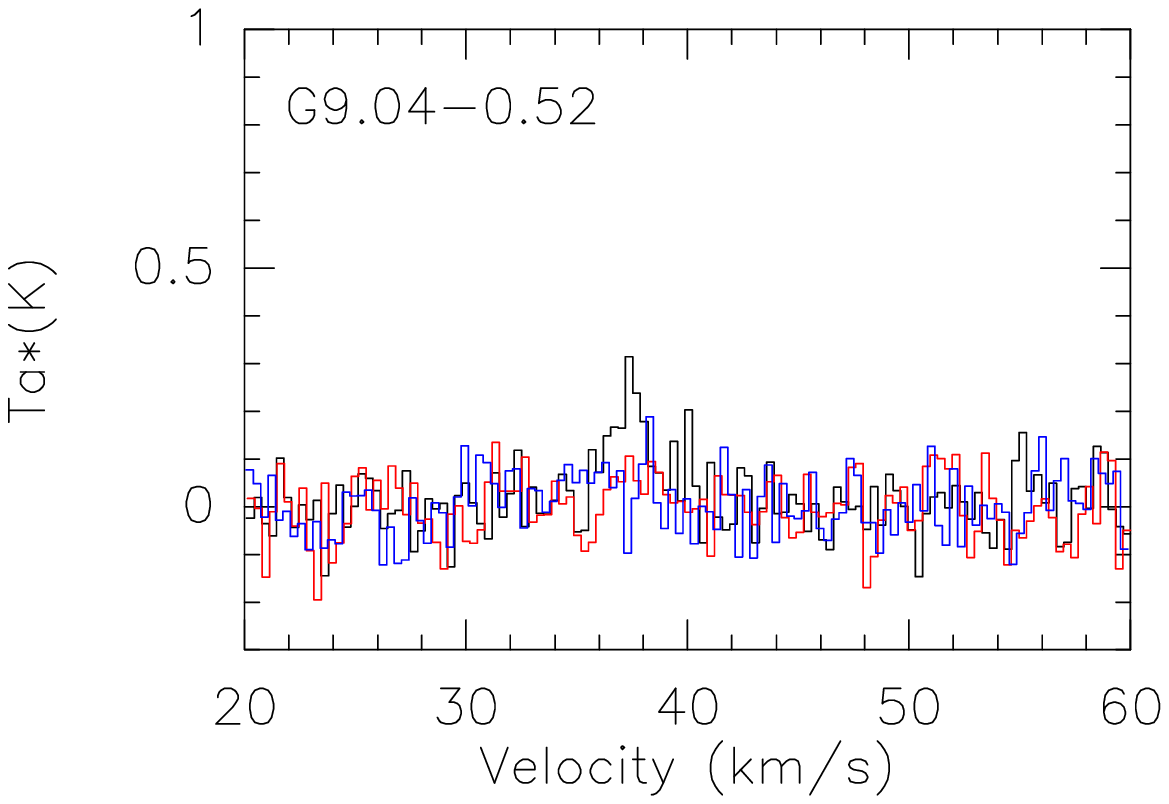}
\includegraphics[width=4.3cm]{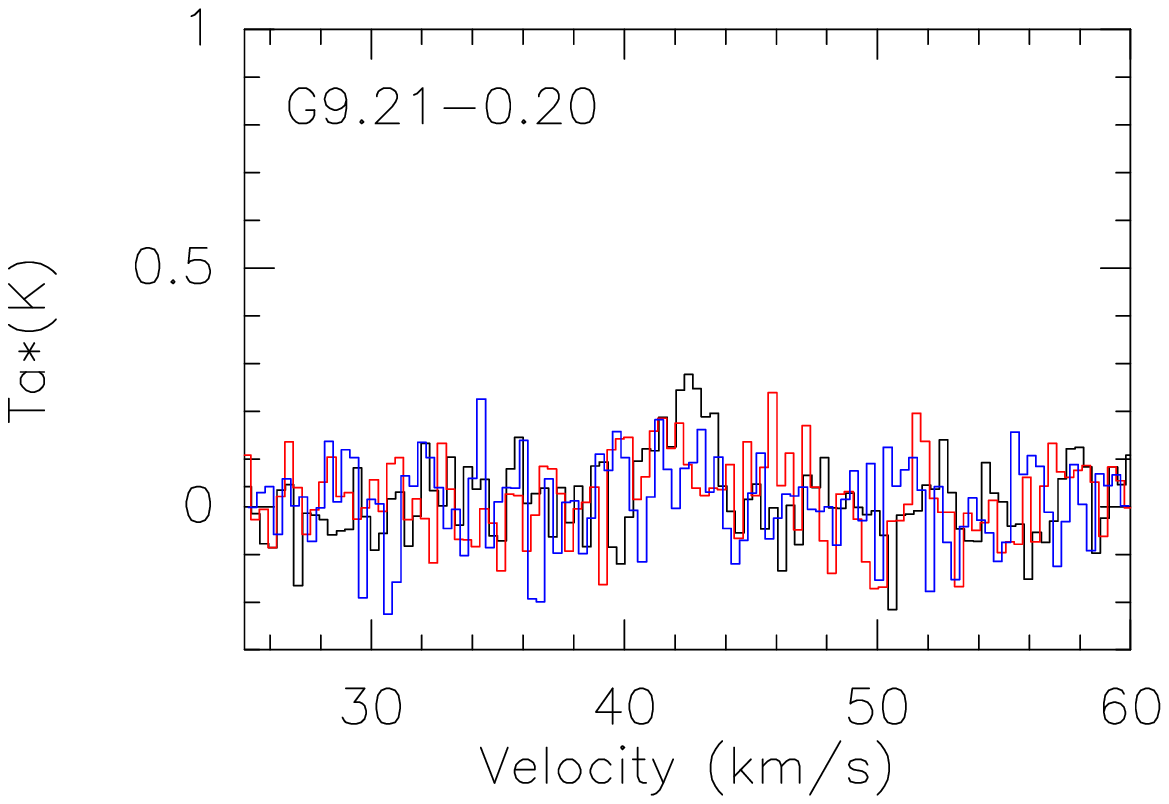}
\includegraphics[width=4.3cm]{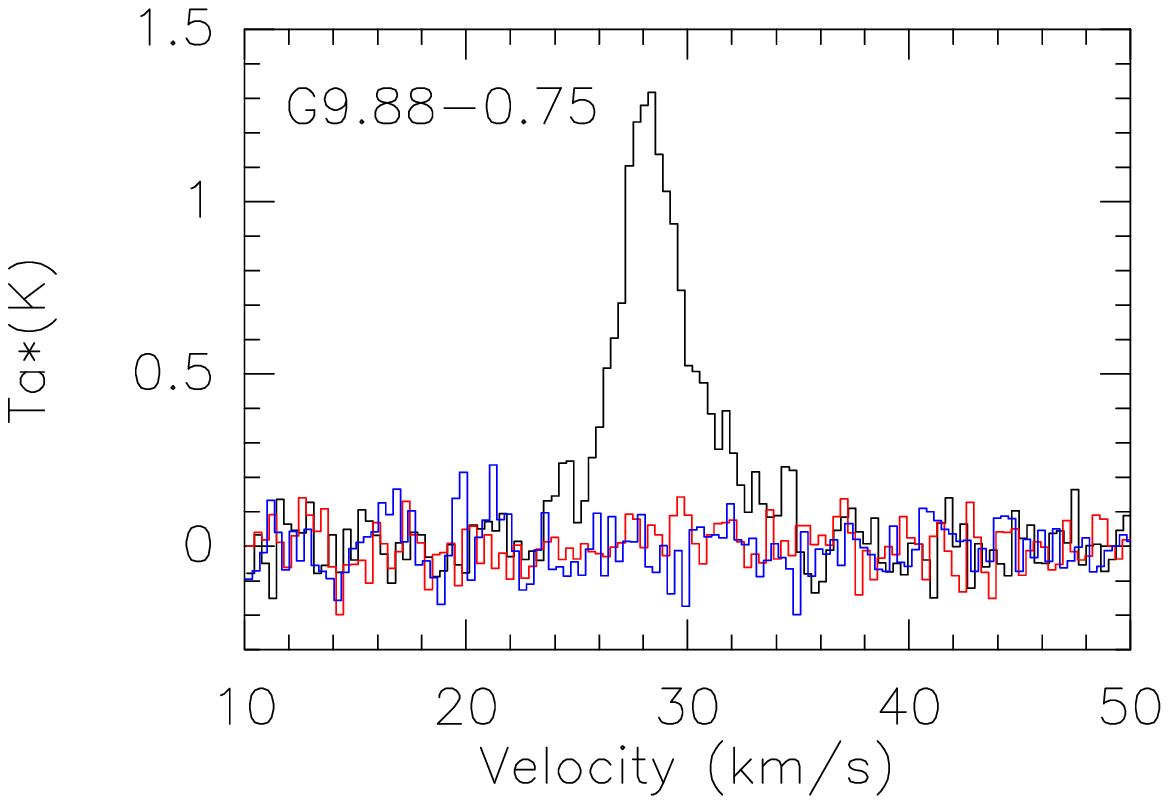}
\includegraphics[width=4.3cm]{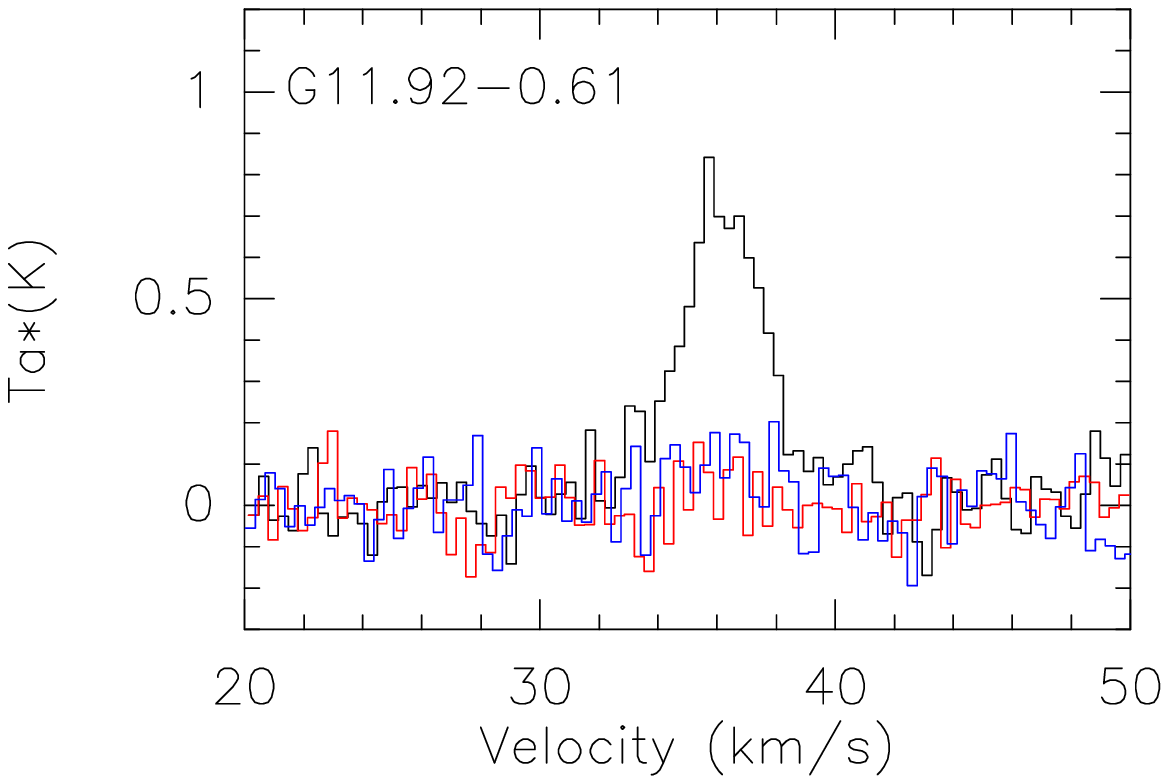}
\includegraphics[width=4.3cm]{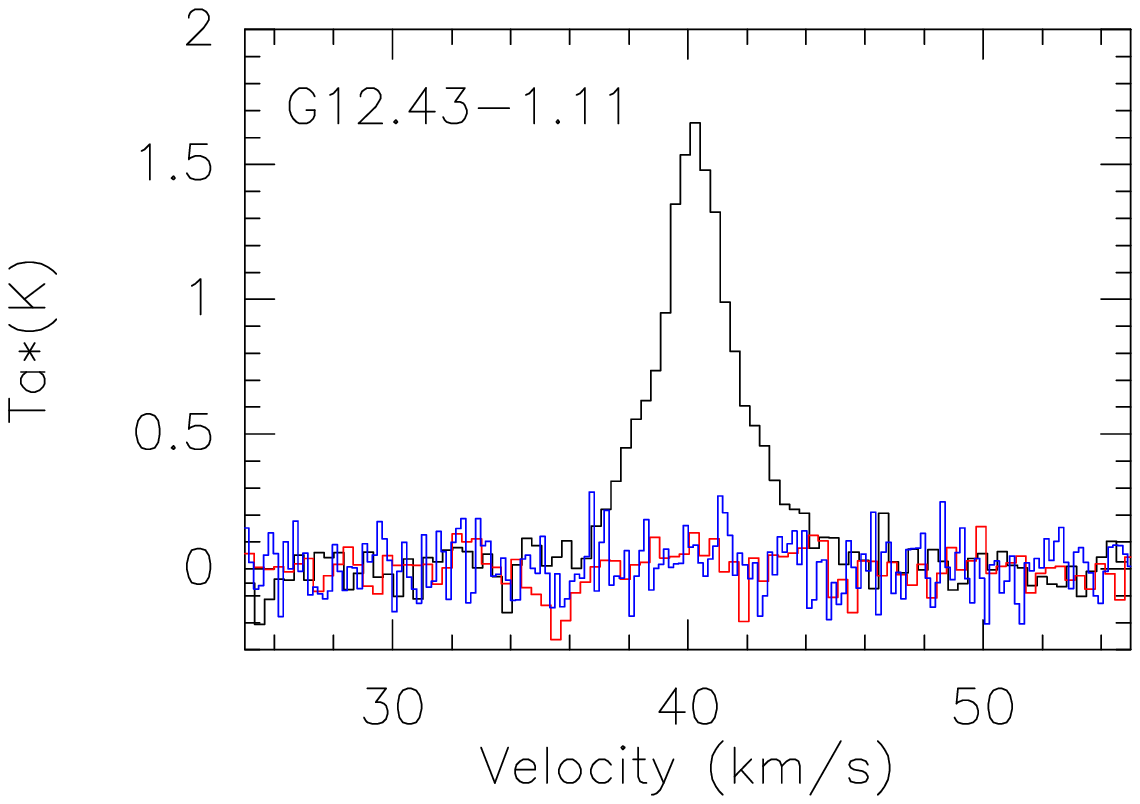}
\includegraphics[width=4.3cm]{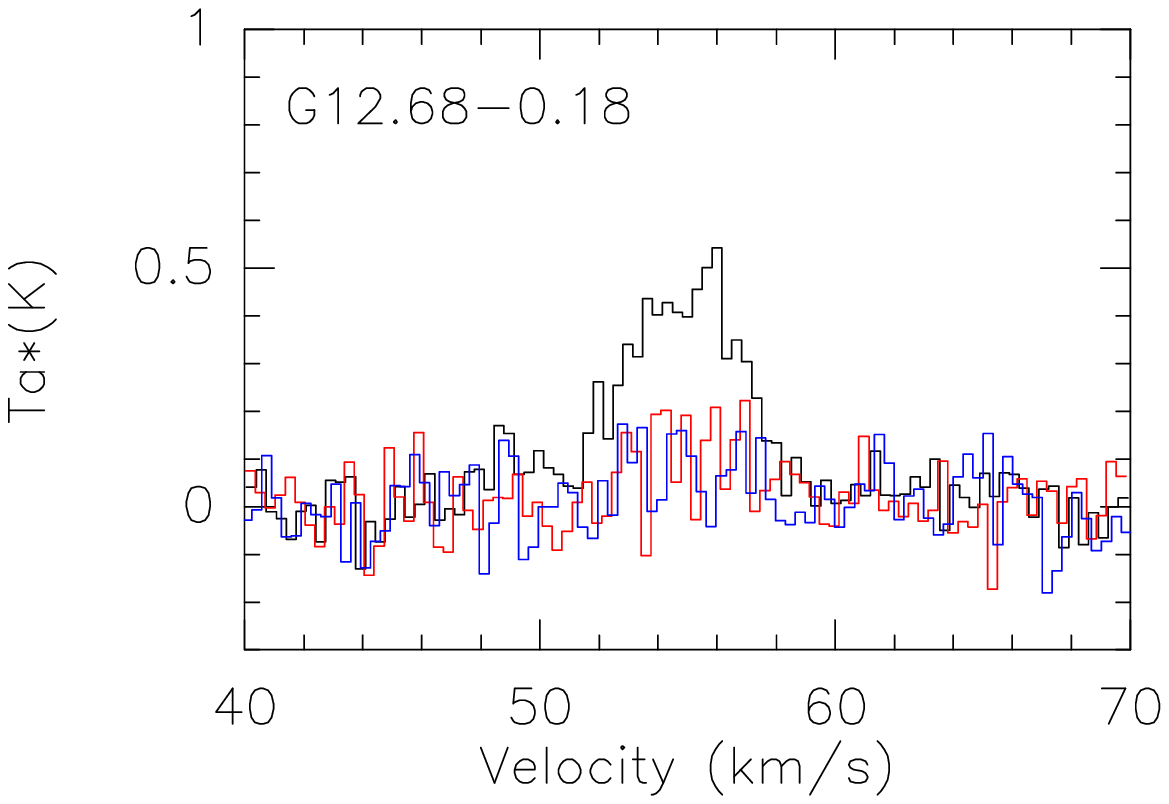}
\includegraphics[width=4.3cm]{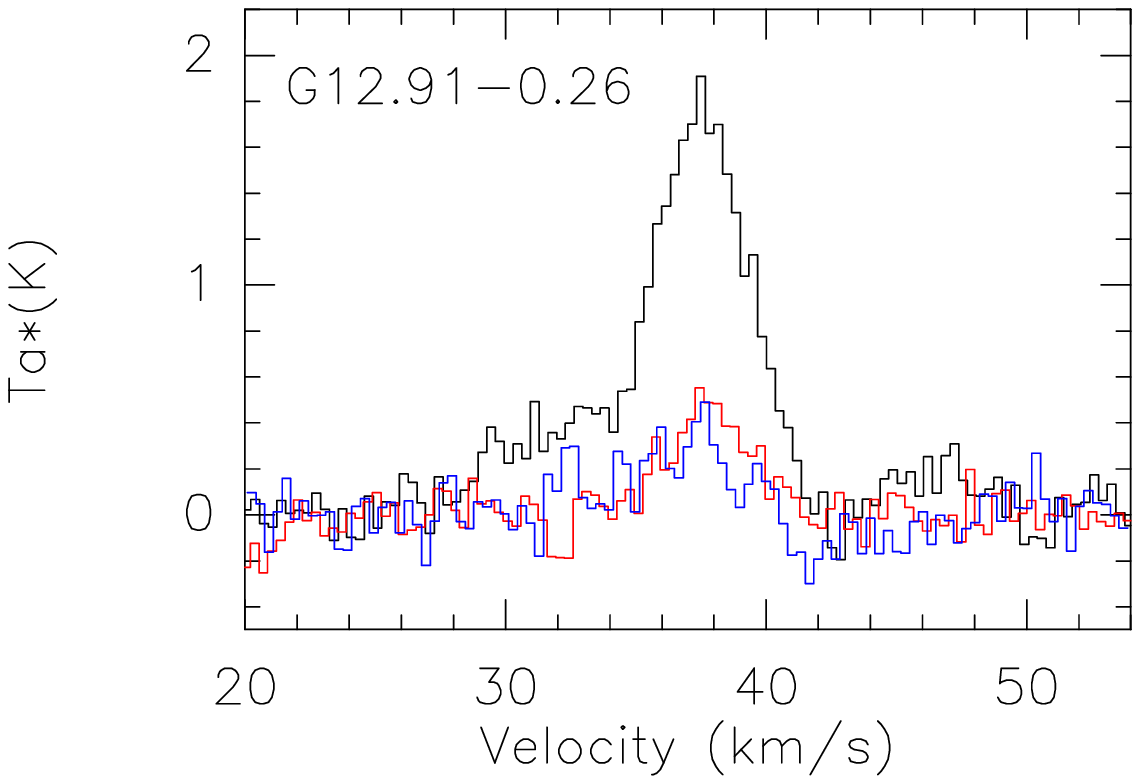}
\includegraphics[width=4.3cm]{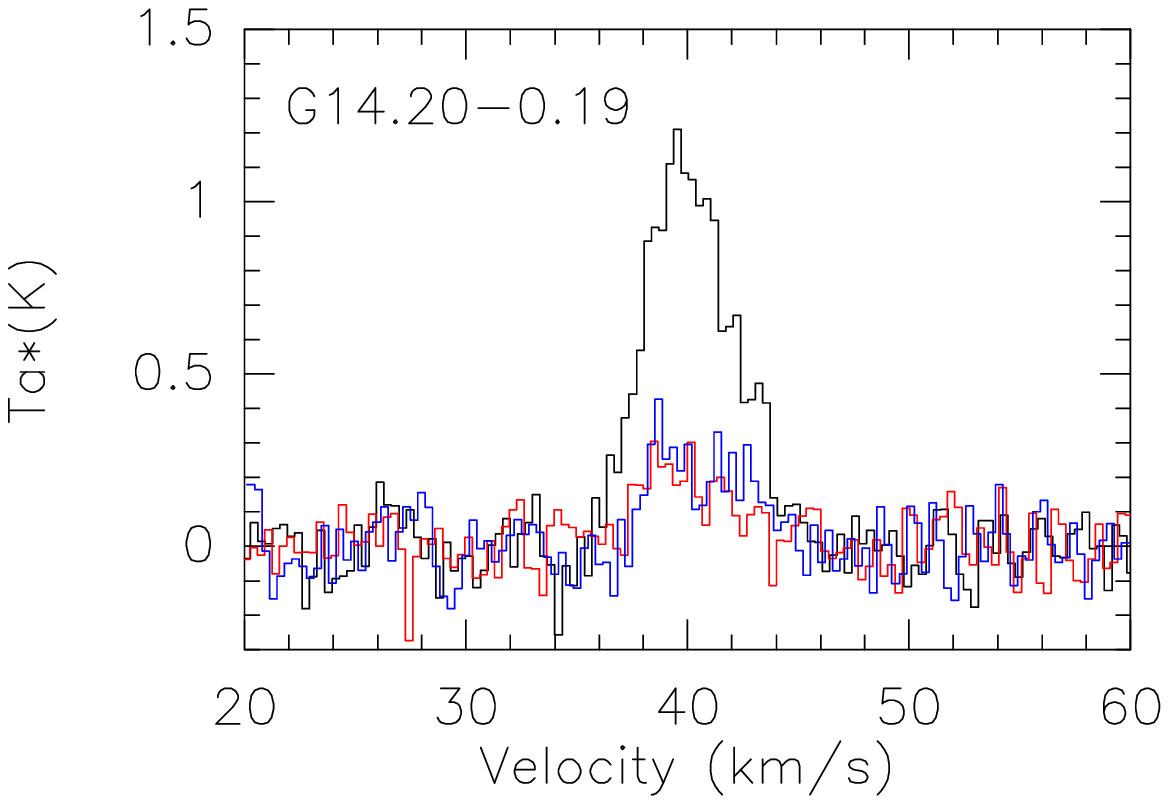}
\includegraphics[width=4.3cm]{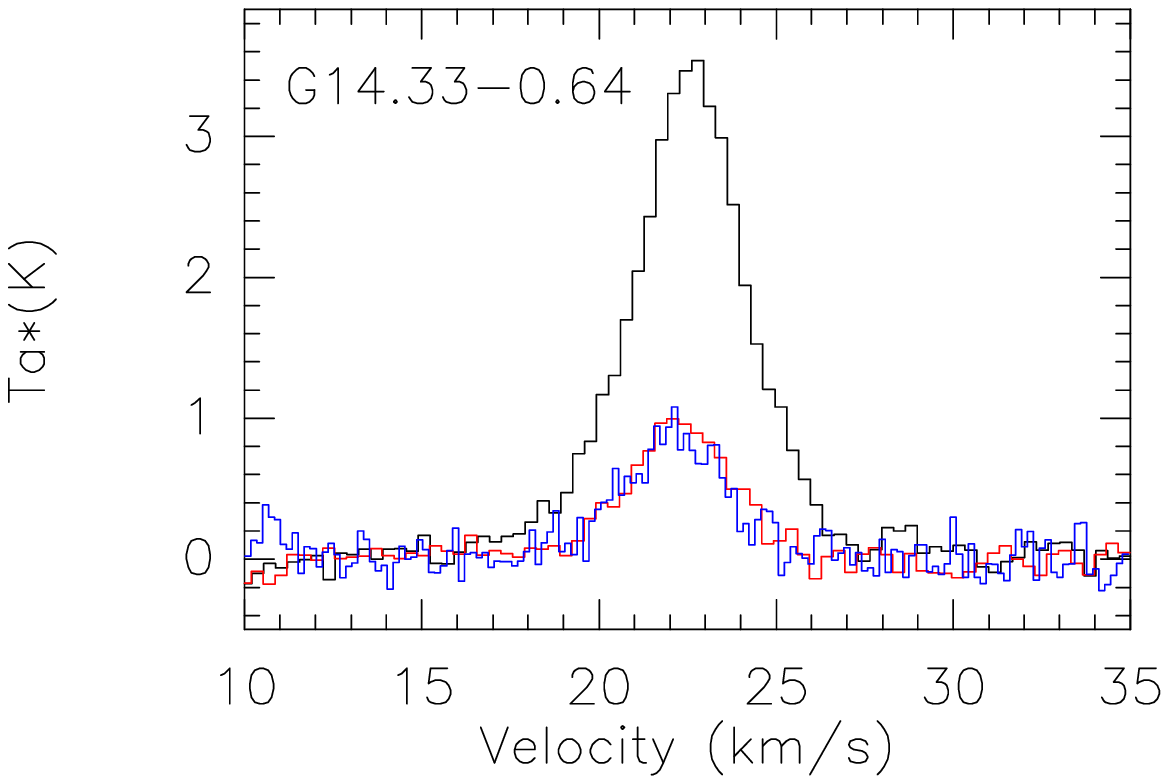}
\includegraphics[width=4.3cm]{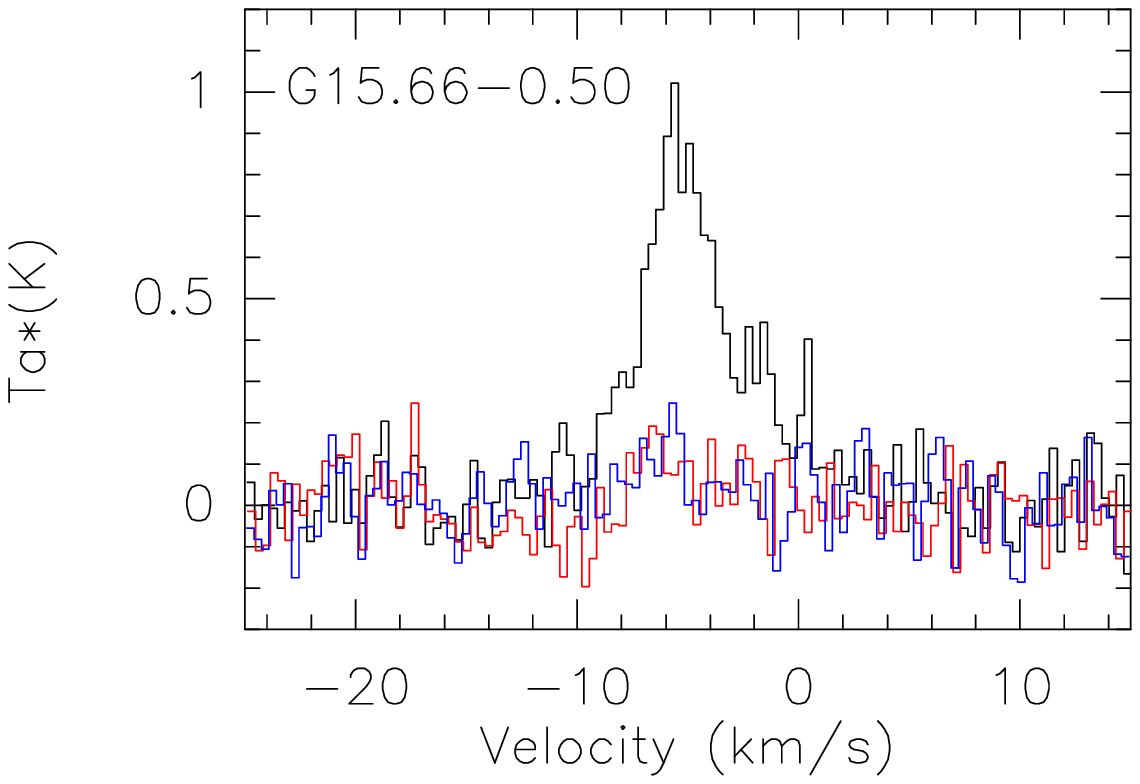}
\includegraphics[width=4.3cm]{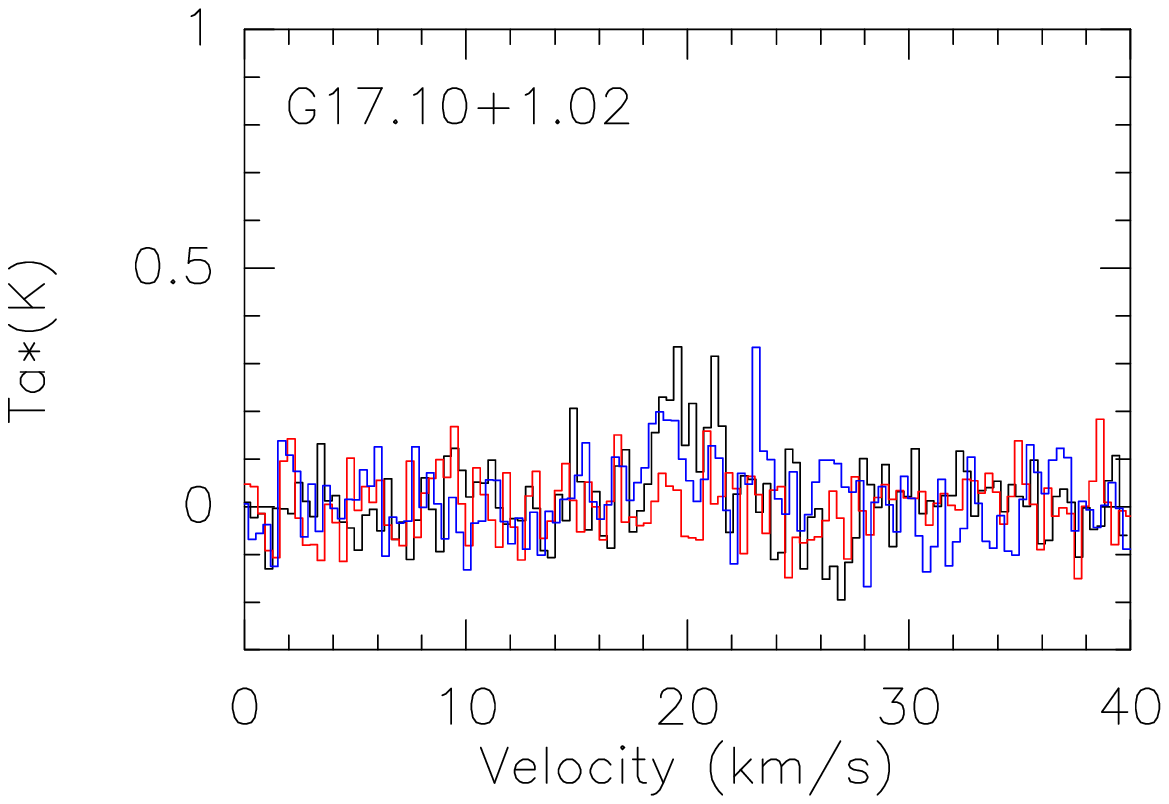}
\includegraphics[width=4.3cm]{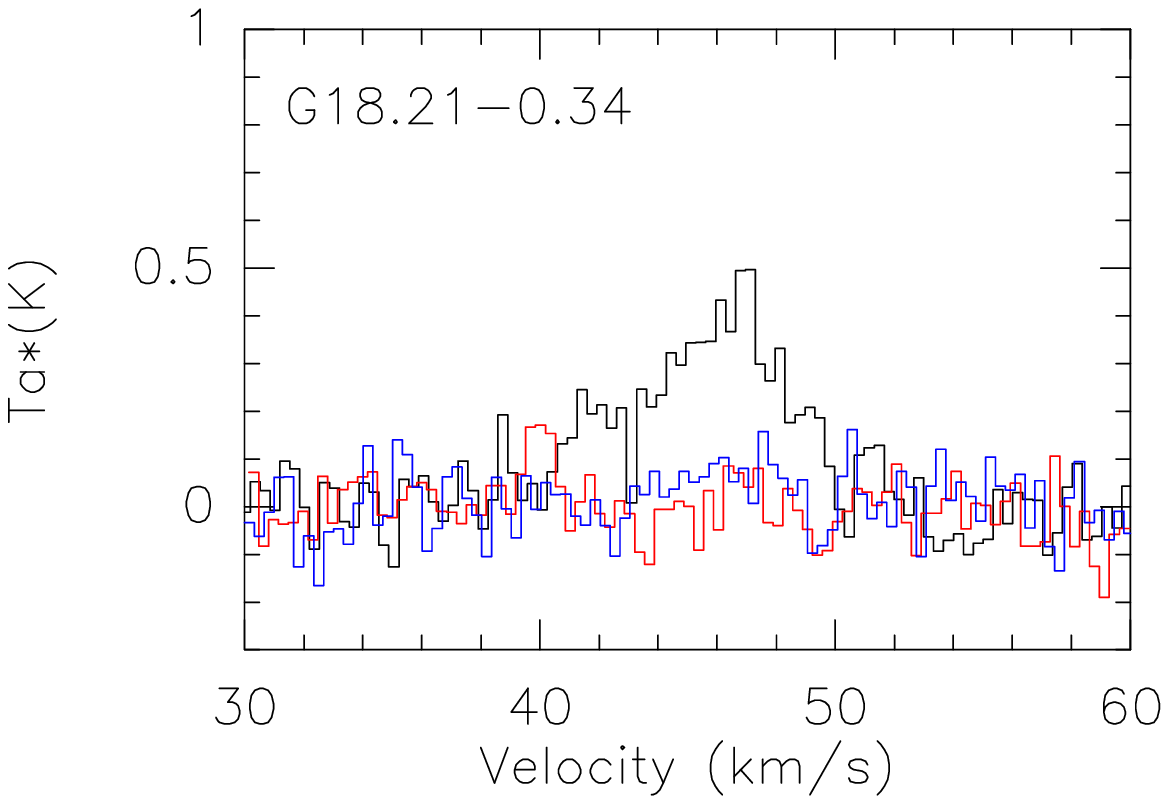}
\includegraphics[width=4.3cm]{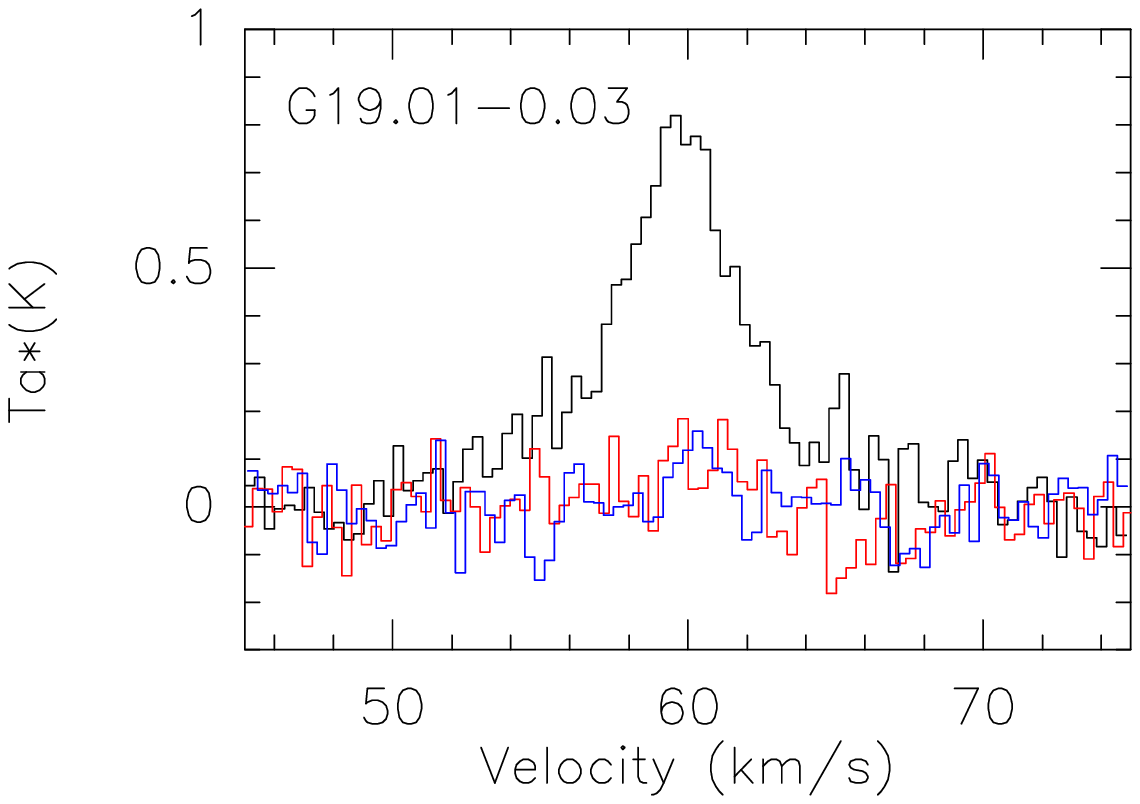}
\includegraphics[width=4.3cm]{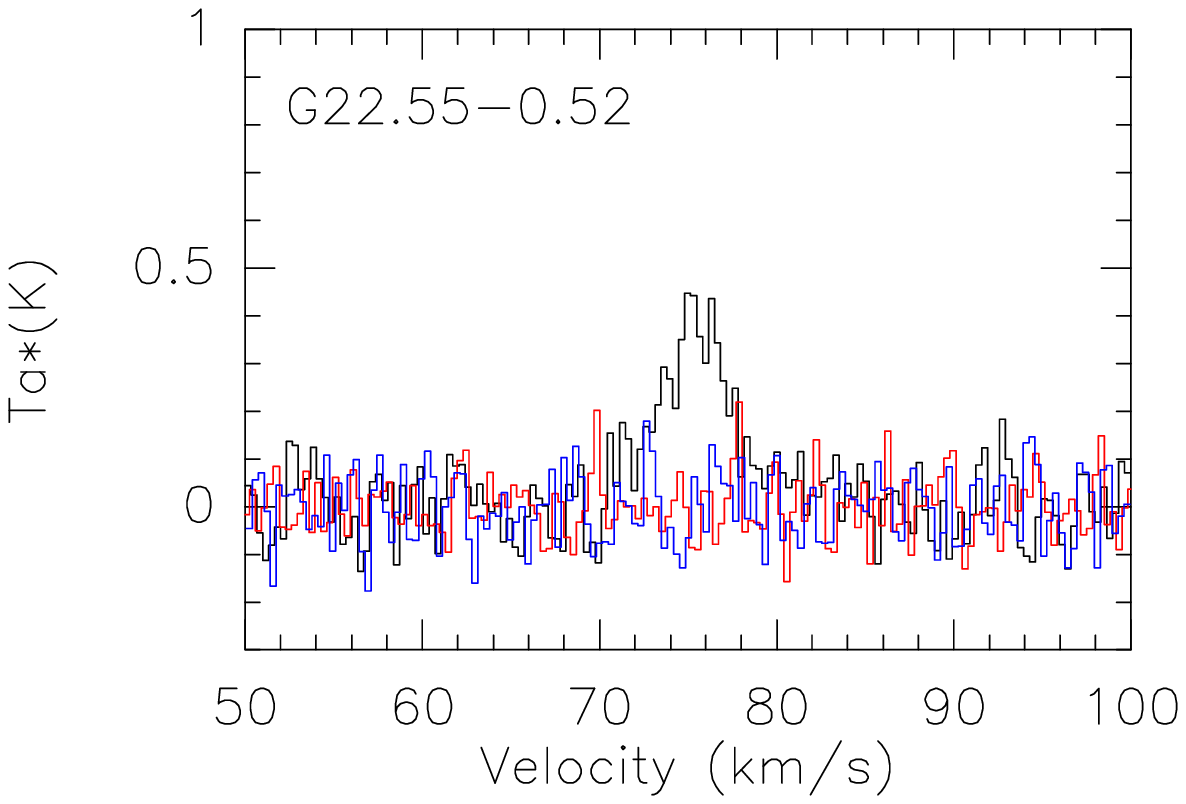}
\includegraphics[width=4.3cm]{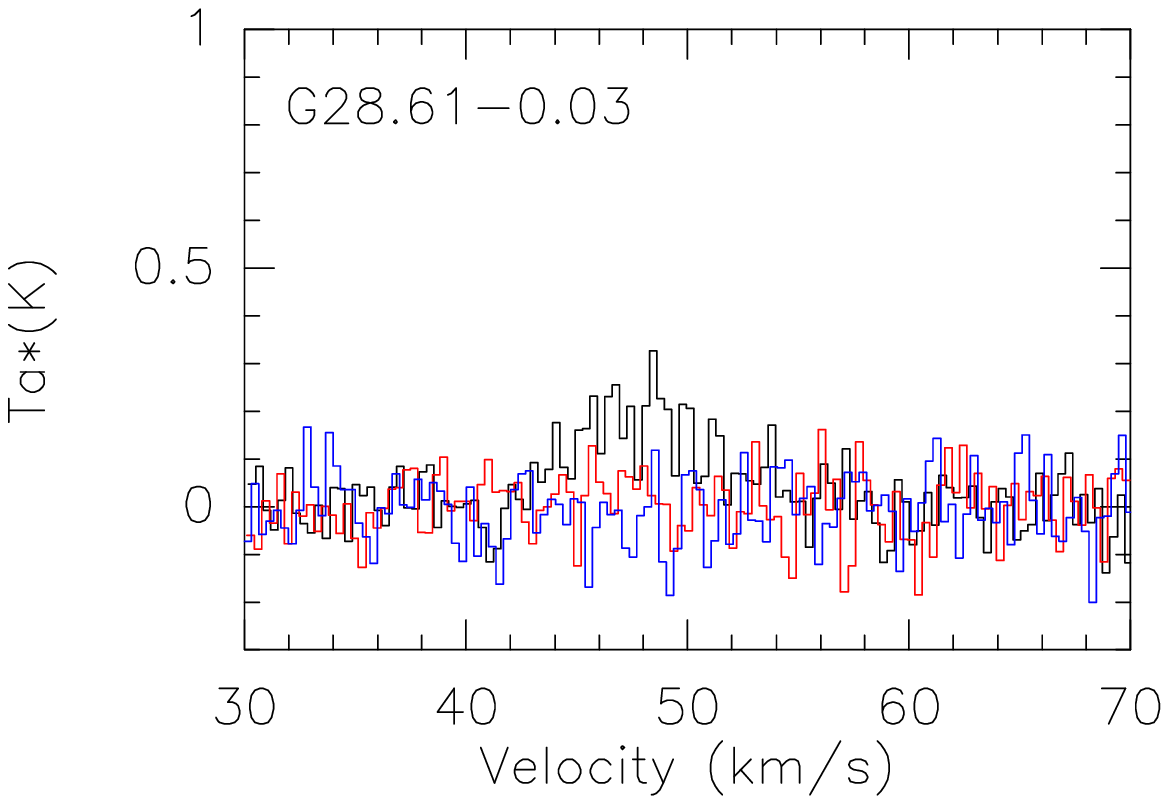}
\includegraphics[width=4.3cm]{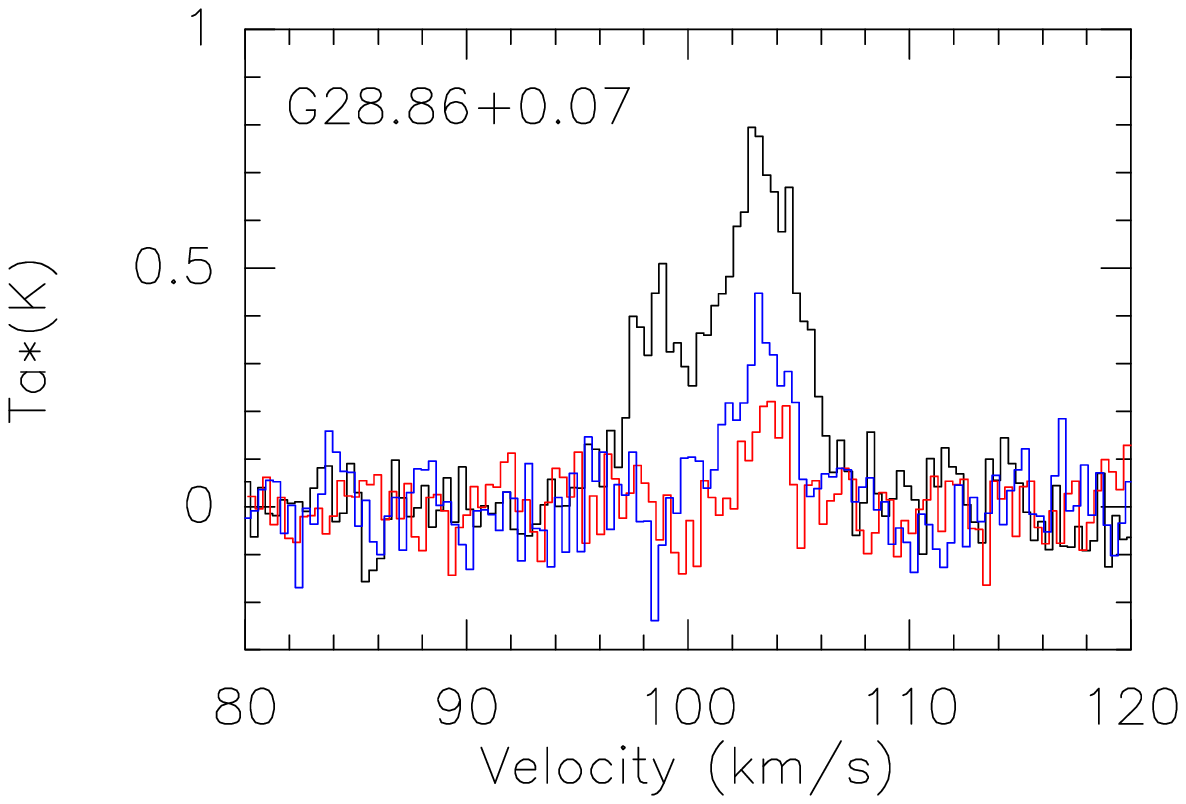}
\includegraphics[width=4.3cm]{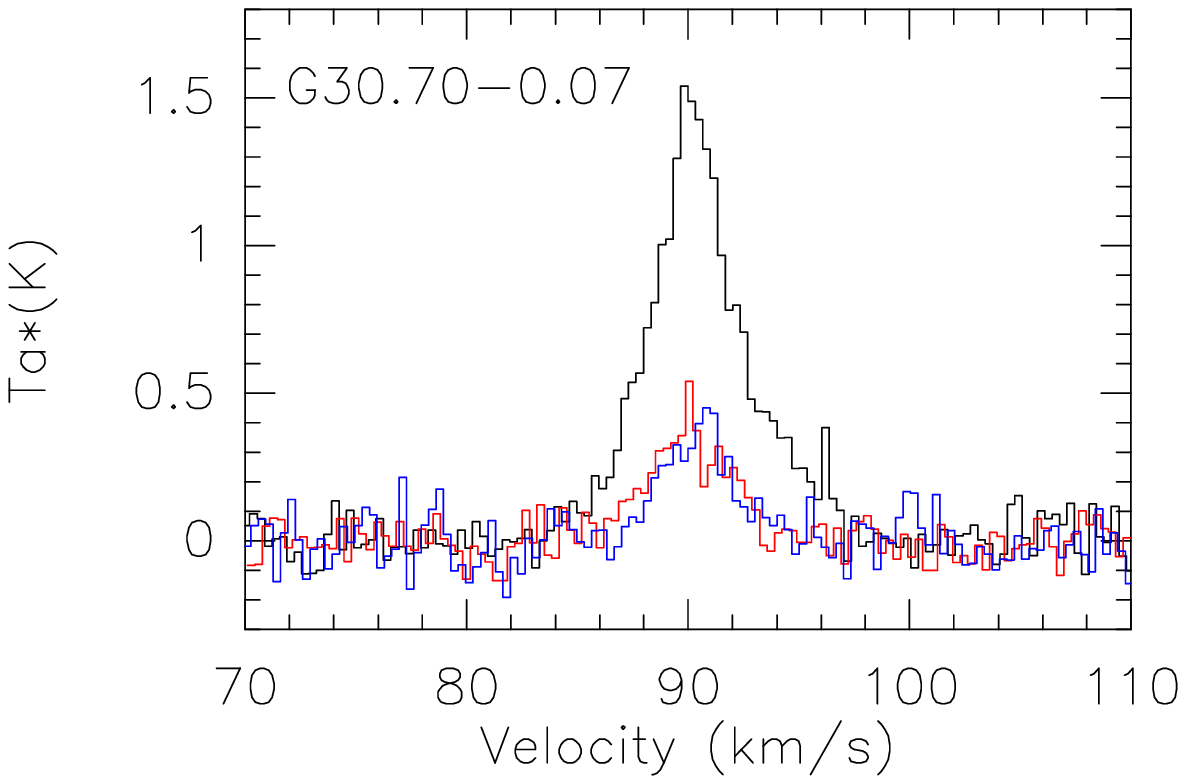}
\includegraphics[width=4.3cm]{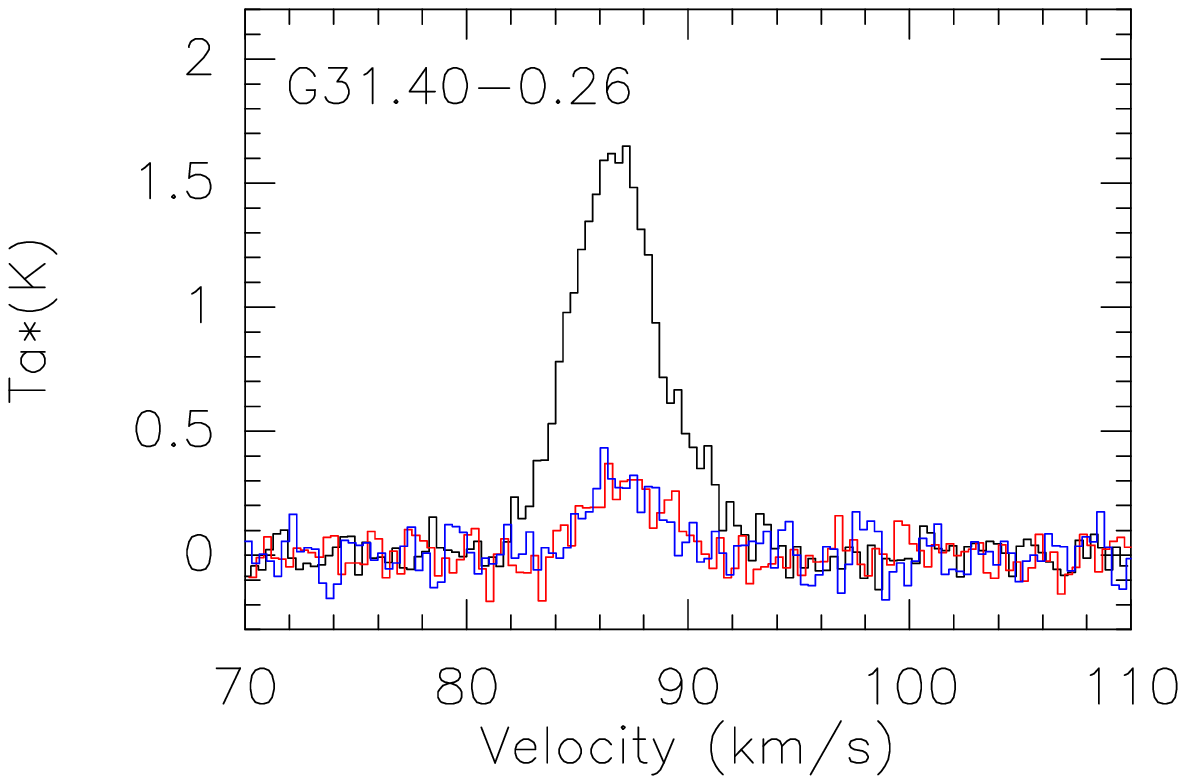}
\includegraphics[width=4.3cm]{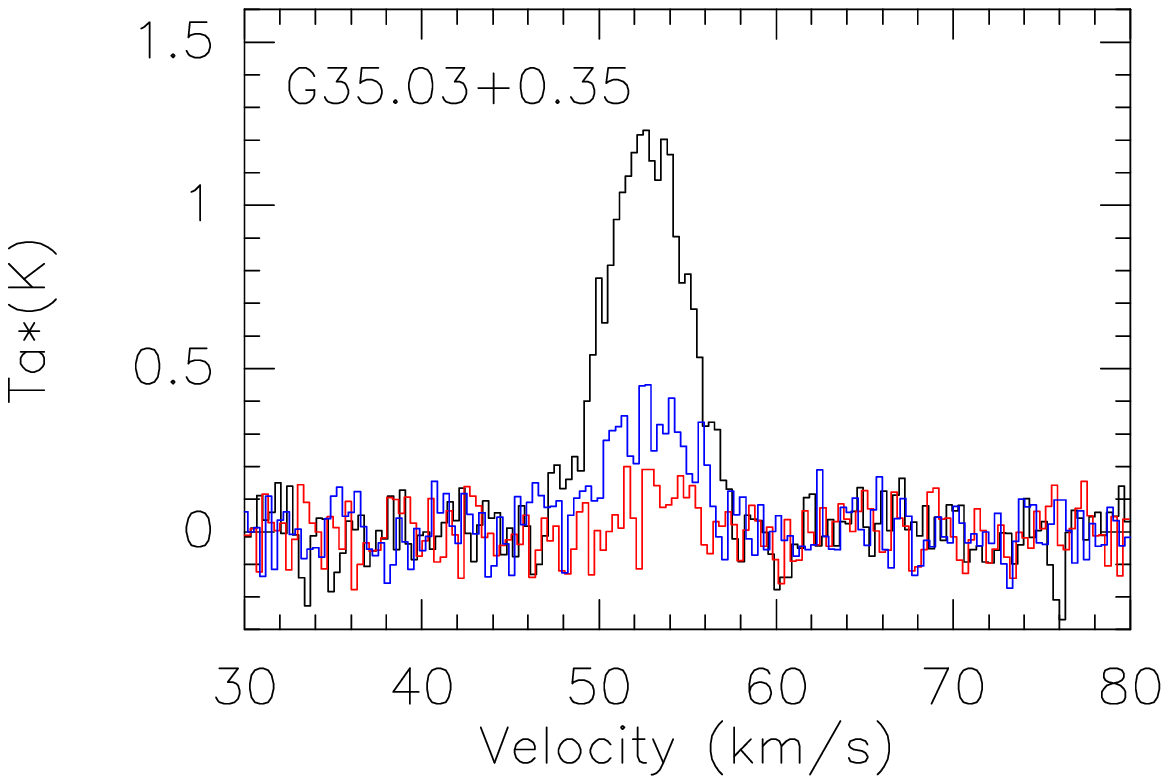}
\includegraphics[width=4.3cm]{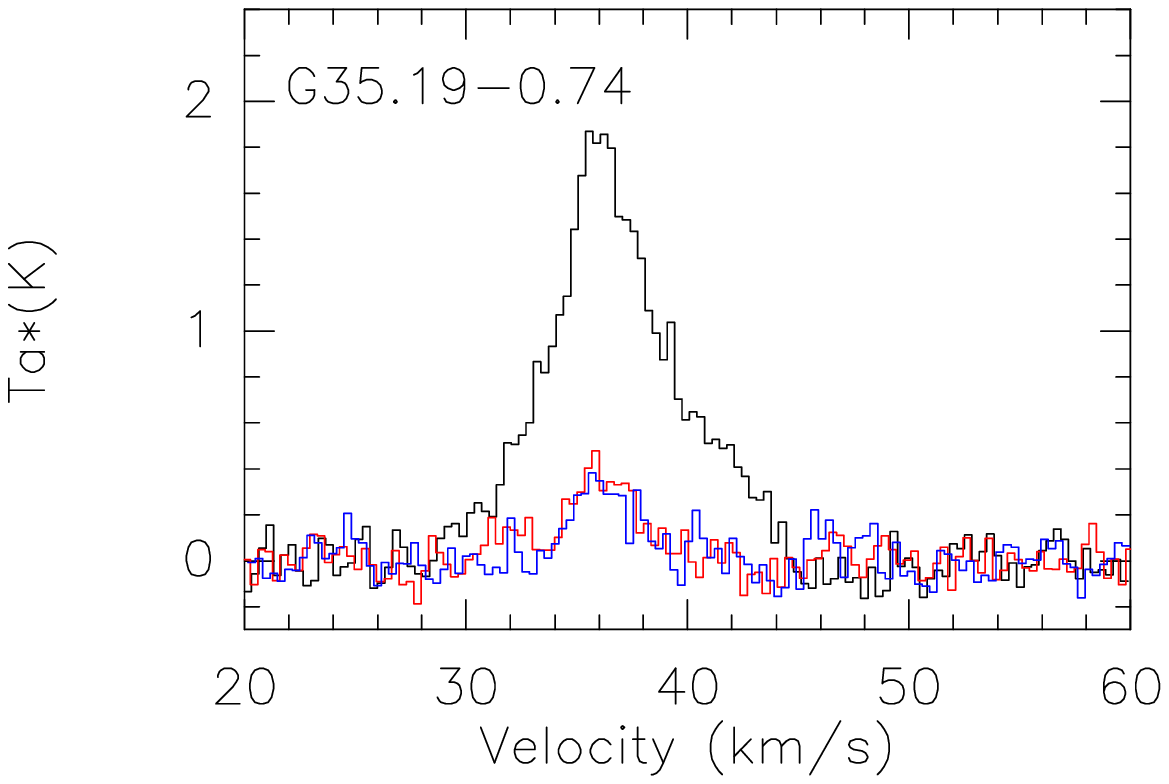}
\includegraphics[width=4.3cm]{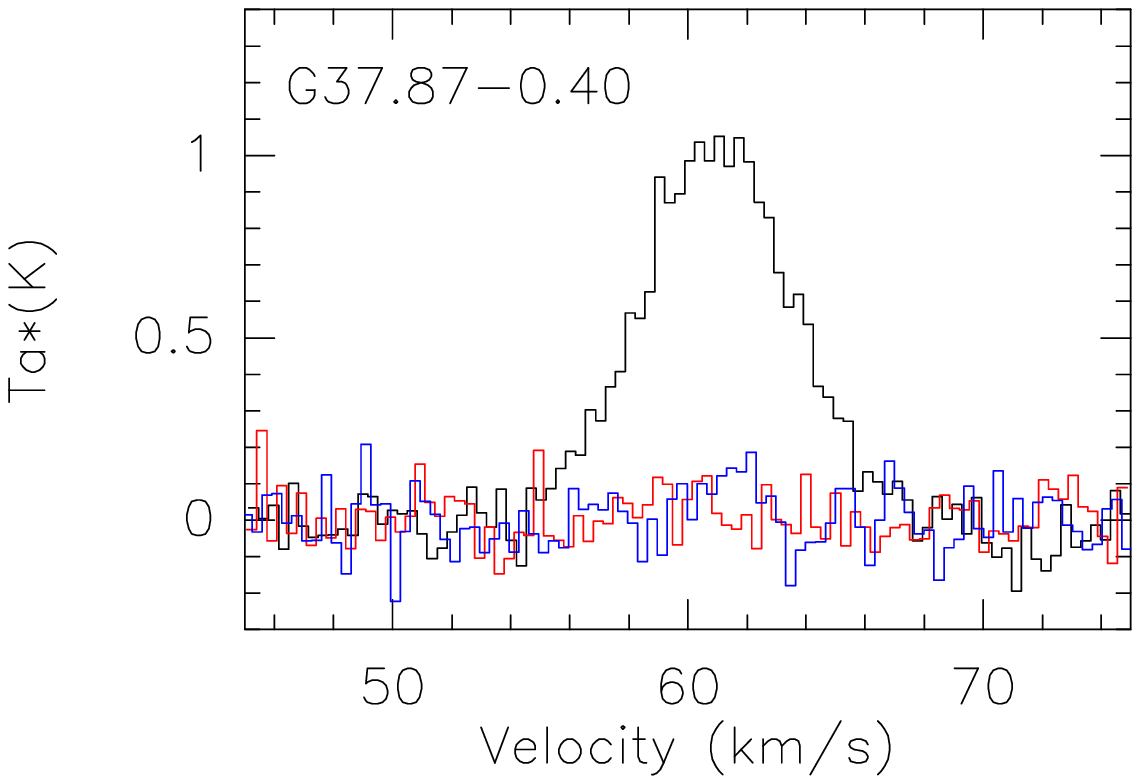}
\end{center}
\caption{Spectra of para-H$_2$CO. Black: para-H$_2$CO 3$_{03}$ -- 2$_{02}$,
Red: para-H$_2$CO 3$_{22}$ -- 2$_{21}$, and Blue: para-H$_2$CO 3$_{21}$ -- 2$_{20}$.}
\label{figure:H2CO-spectrum}
\end{figure*}

%%%%%%%%%%%%%%%%%%%Fig.1-S870um-NH3-H2CO-CH3OH-intensities%%%%%%%%
\begin{figure*}[t]
\vspace*{0.2mm}
\begin{center}
\includegraphics[width=4.3cm]{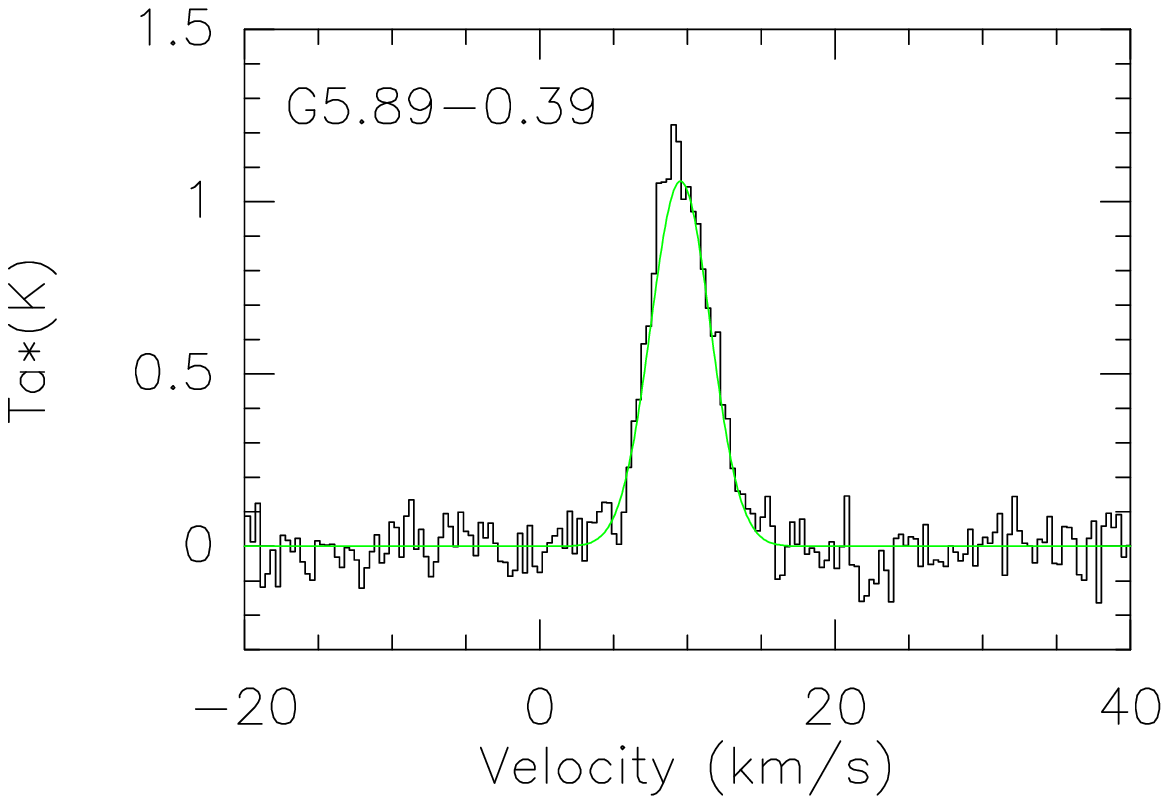}
\includegraphics[width=4.3cm]{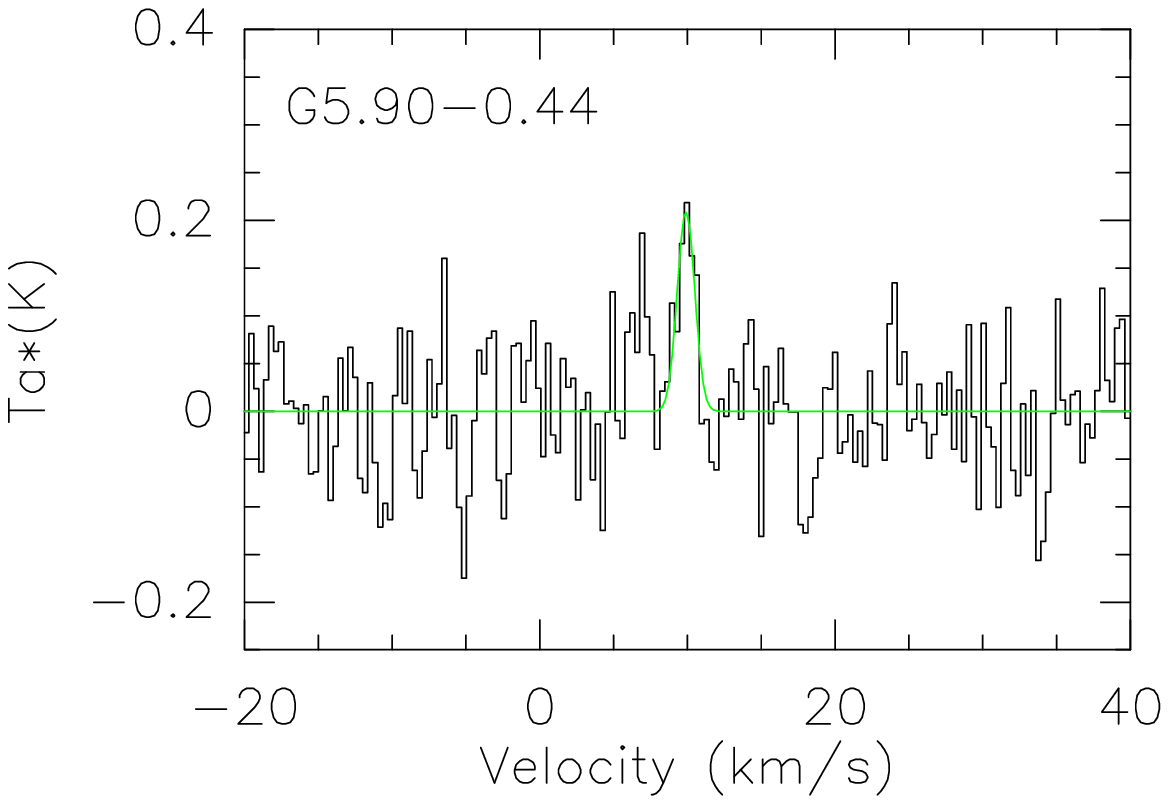}
\includegraphics[width=4.3cm]{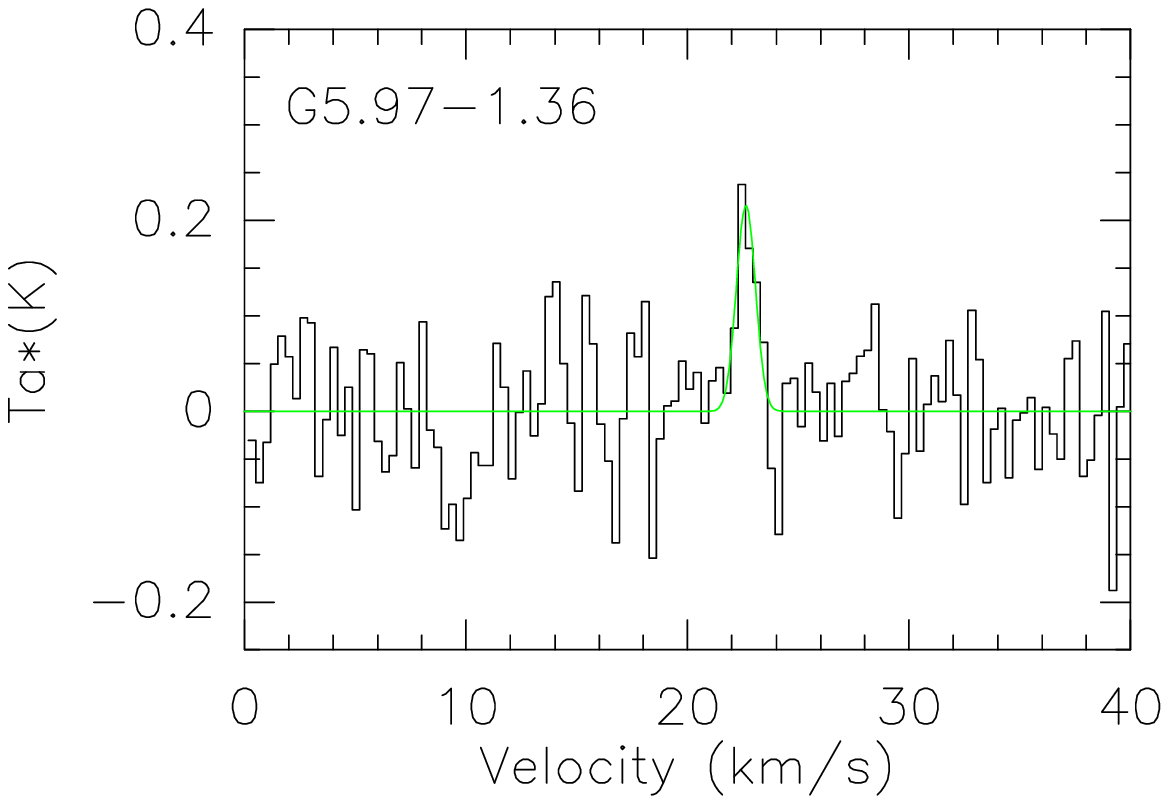}
\includegraphics[width=4.3cm]{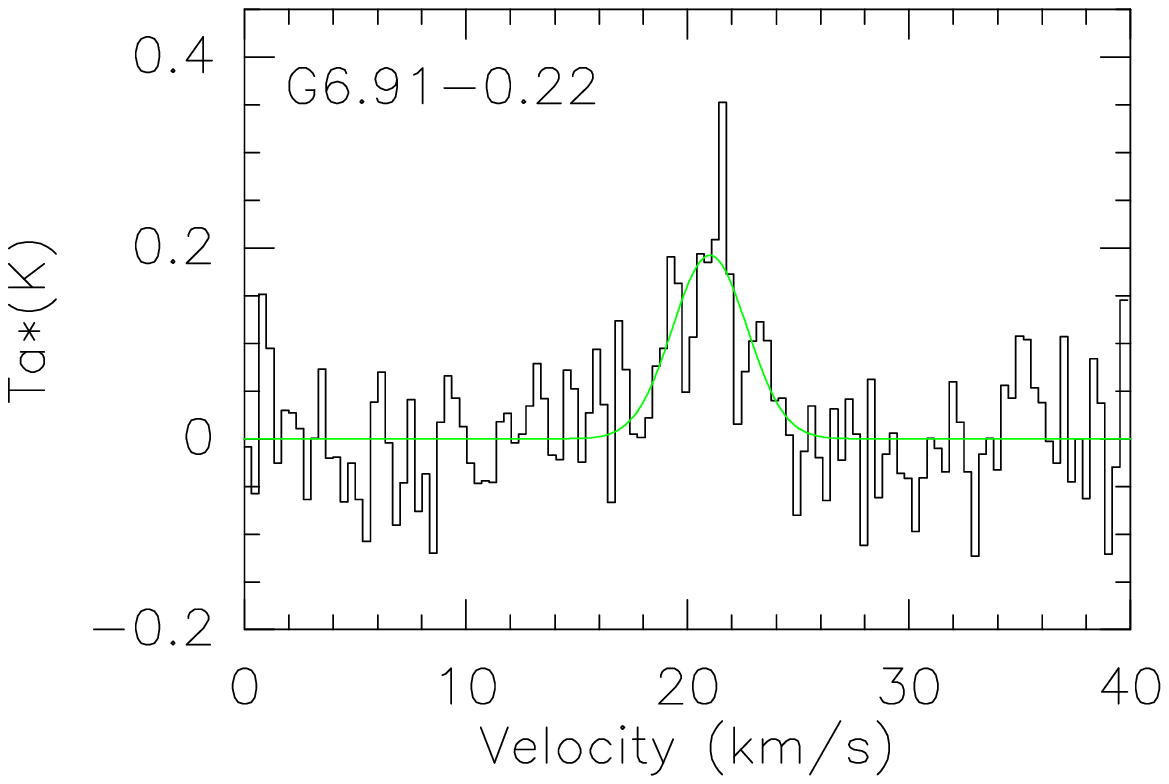}
\includegraphics[width=4.3cm]{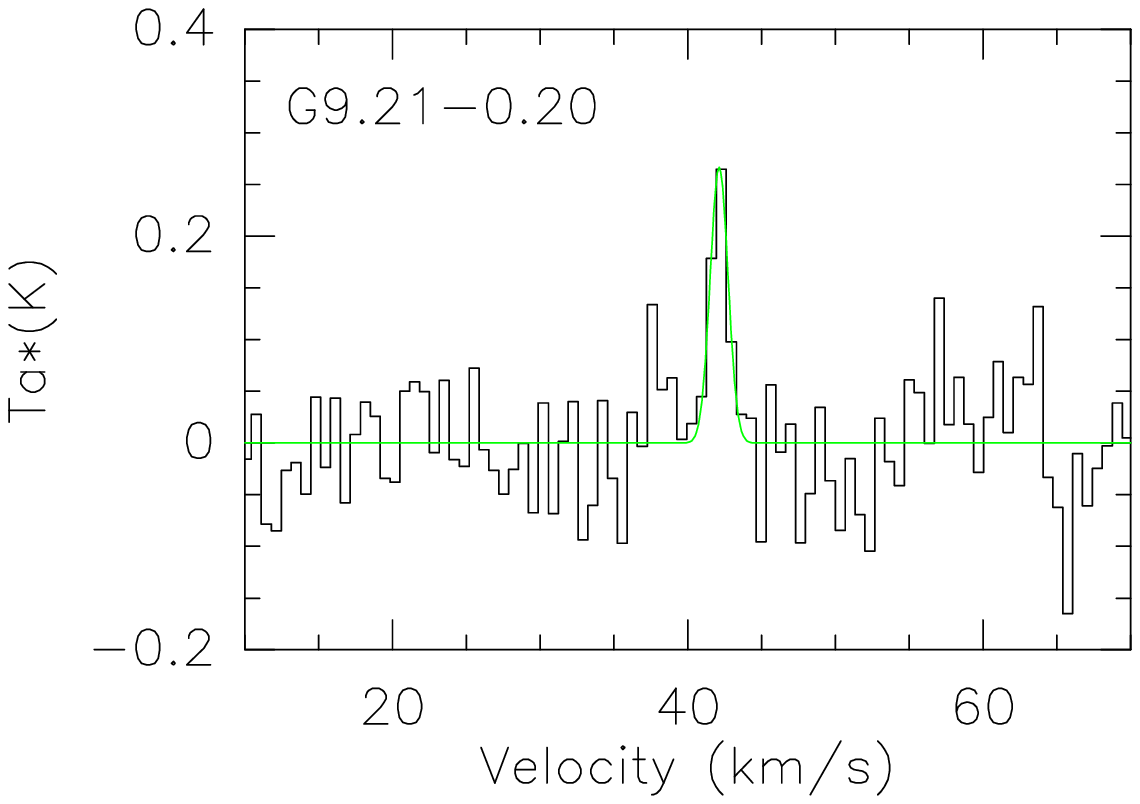}
\includegraphics[width=4.3cm]{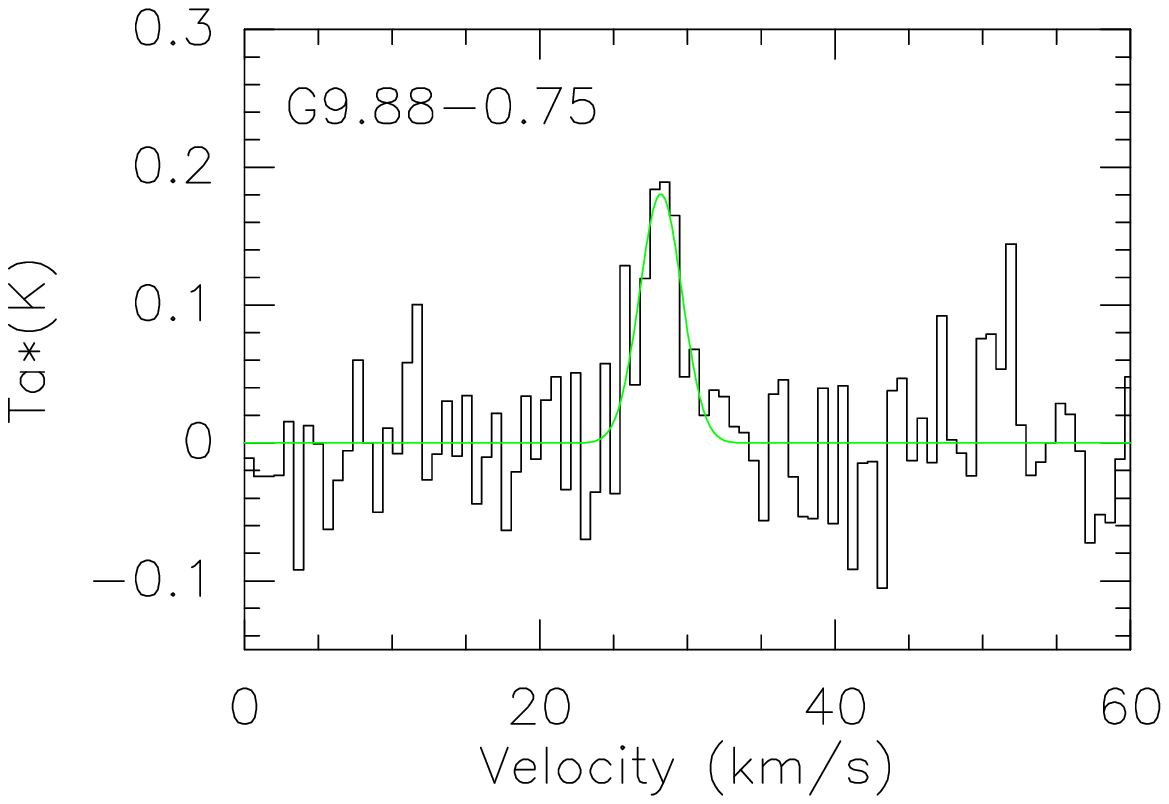}
\includegraphics[width=4.3cm]{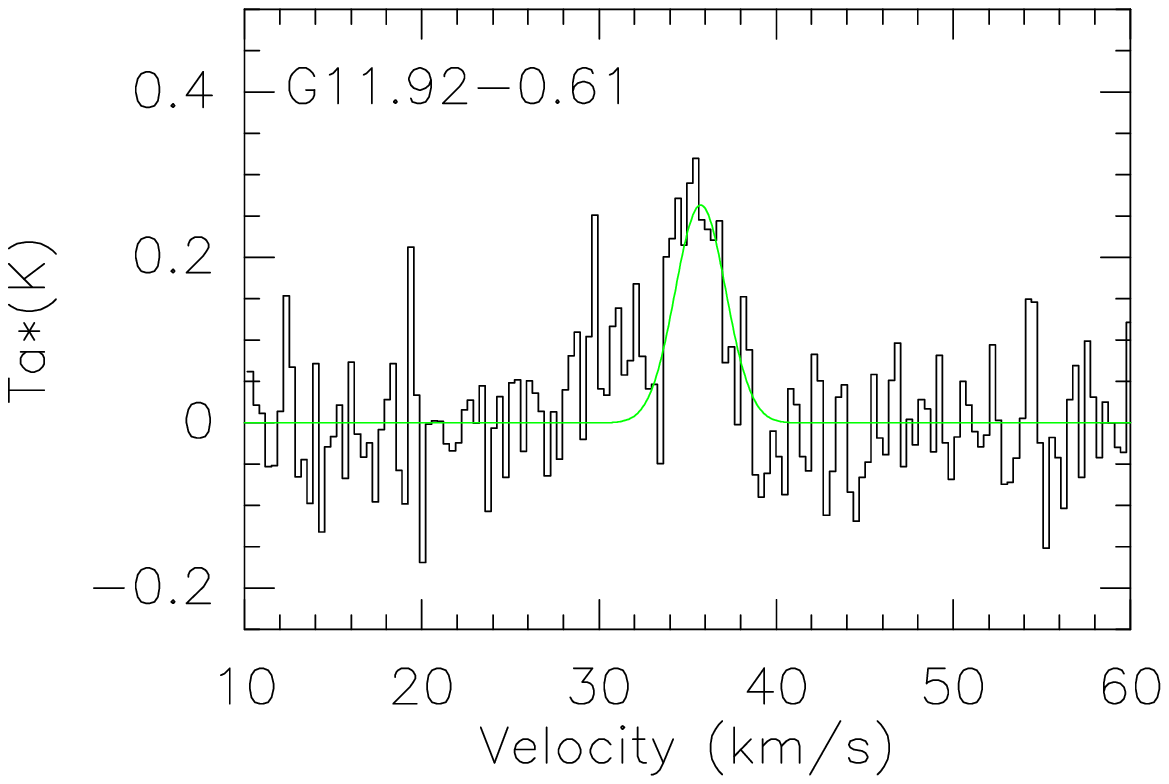}
\includegraphics[width=4.3cm]{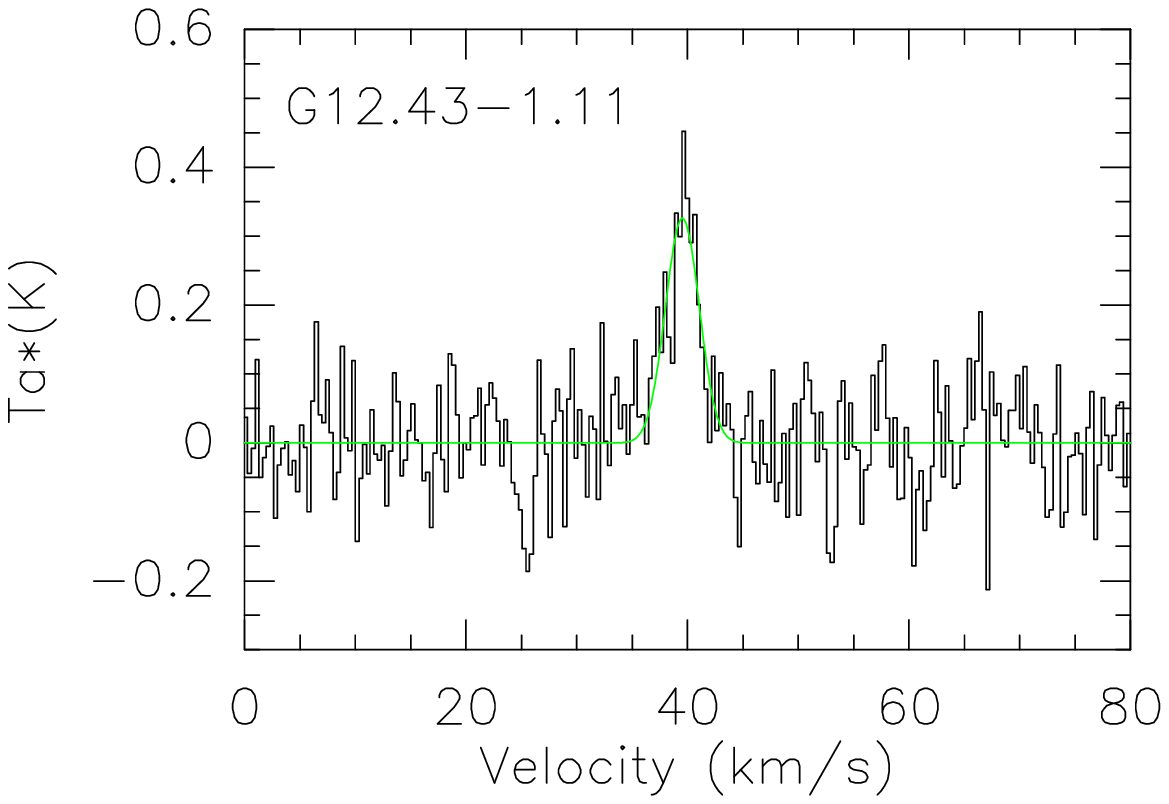}
\includegraphics[width=4.3cm]{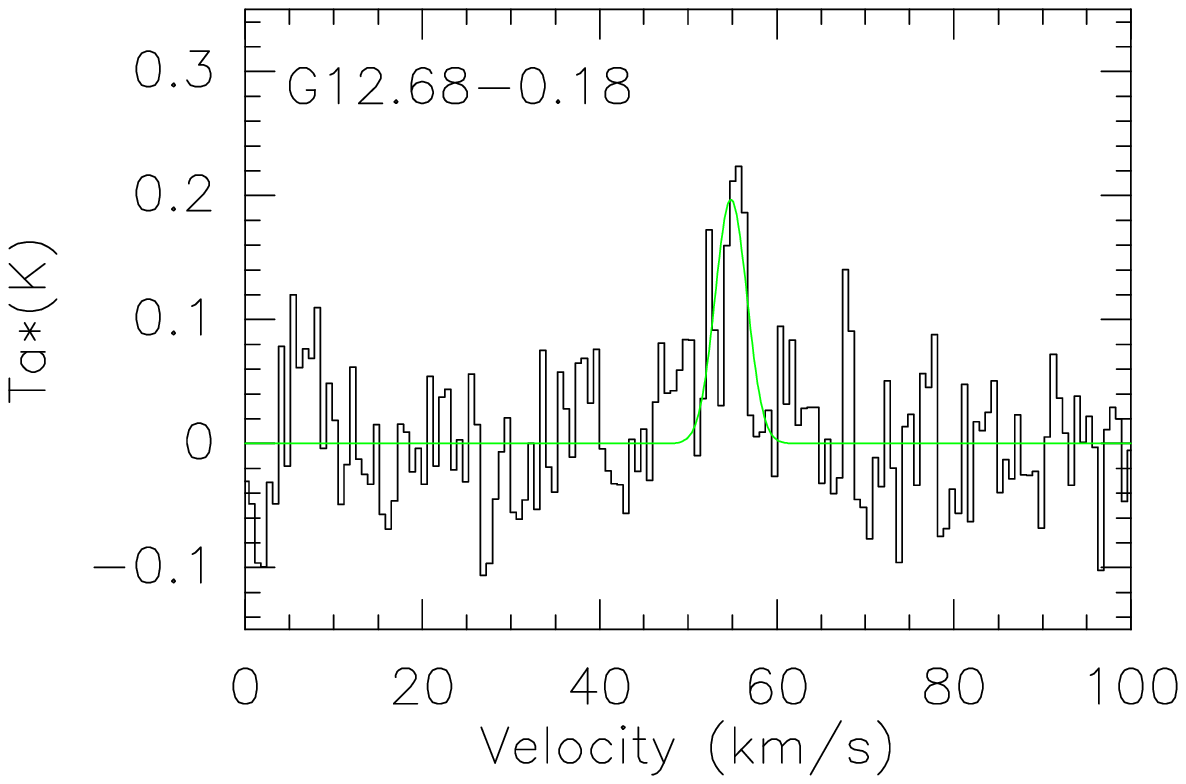}
\includegraphics[width=4.3cm]{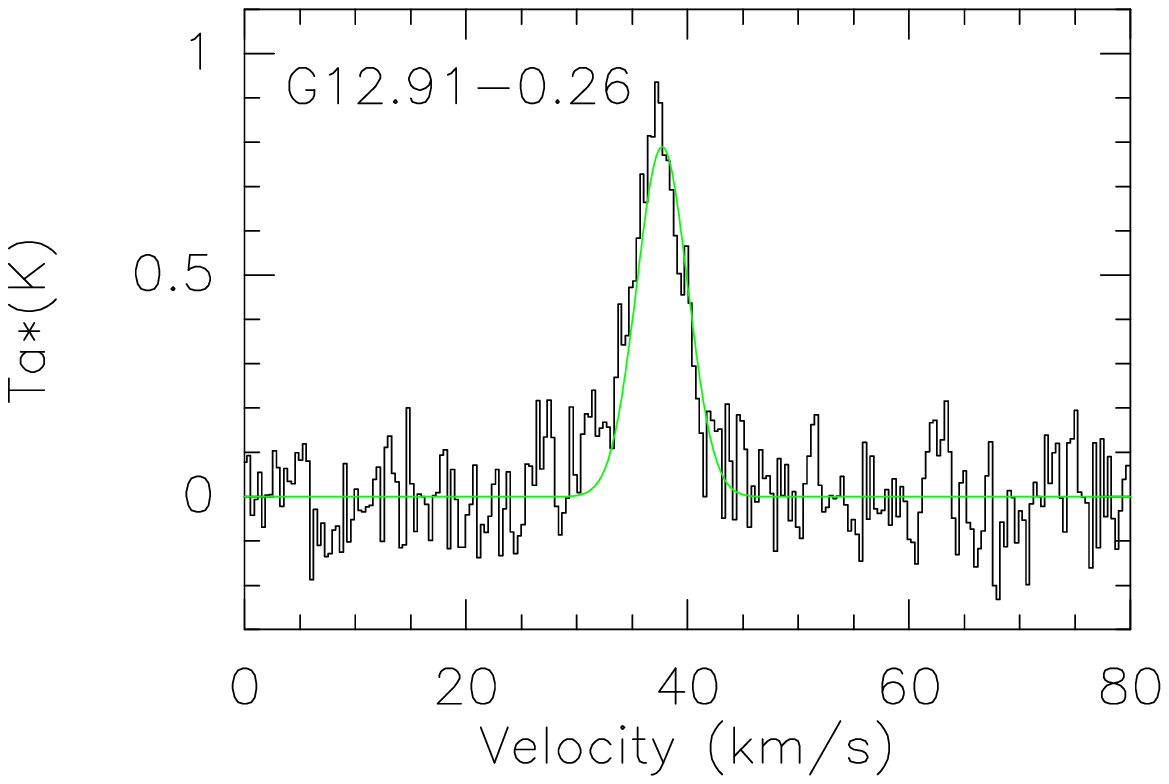}
\includegraphics[width=4.3cm]{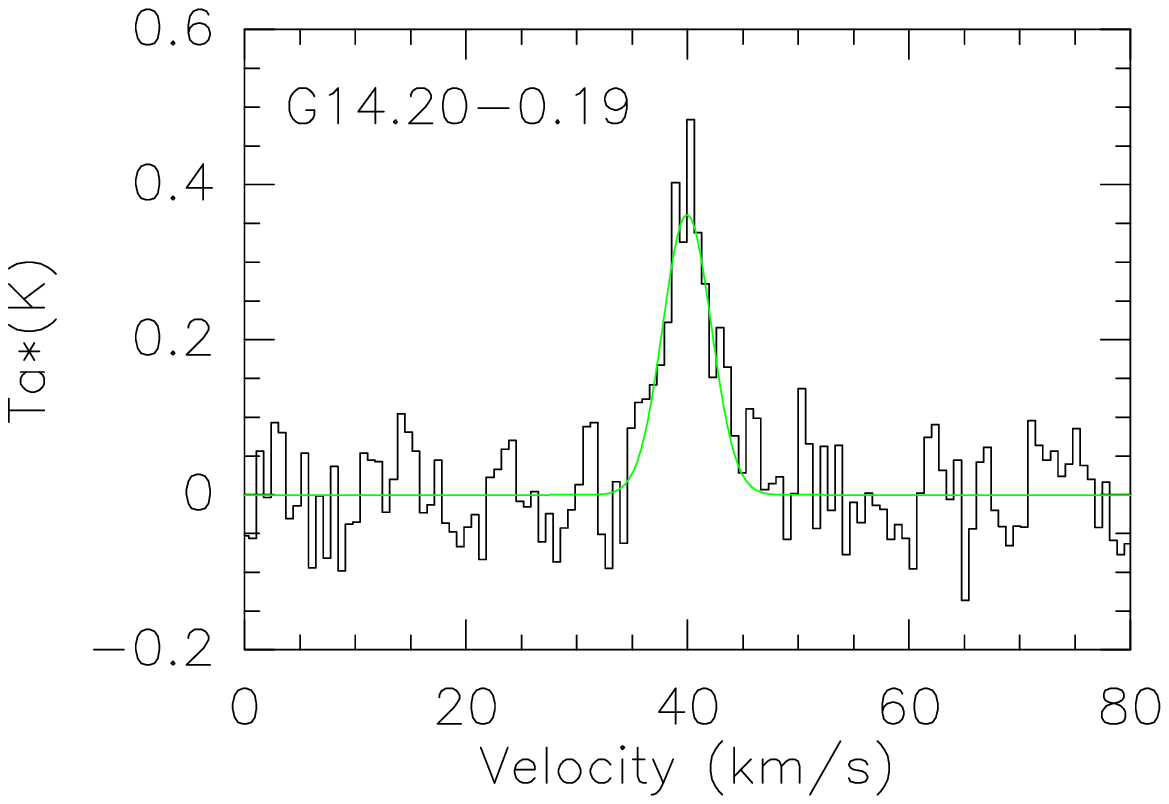}
\includegraphics[width=4.3cm]{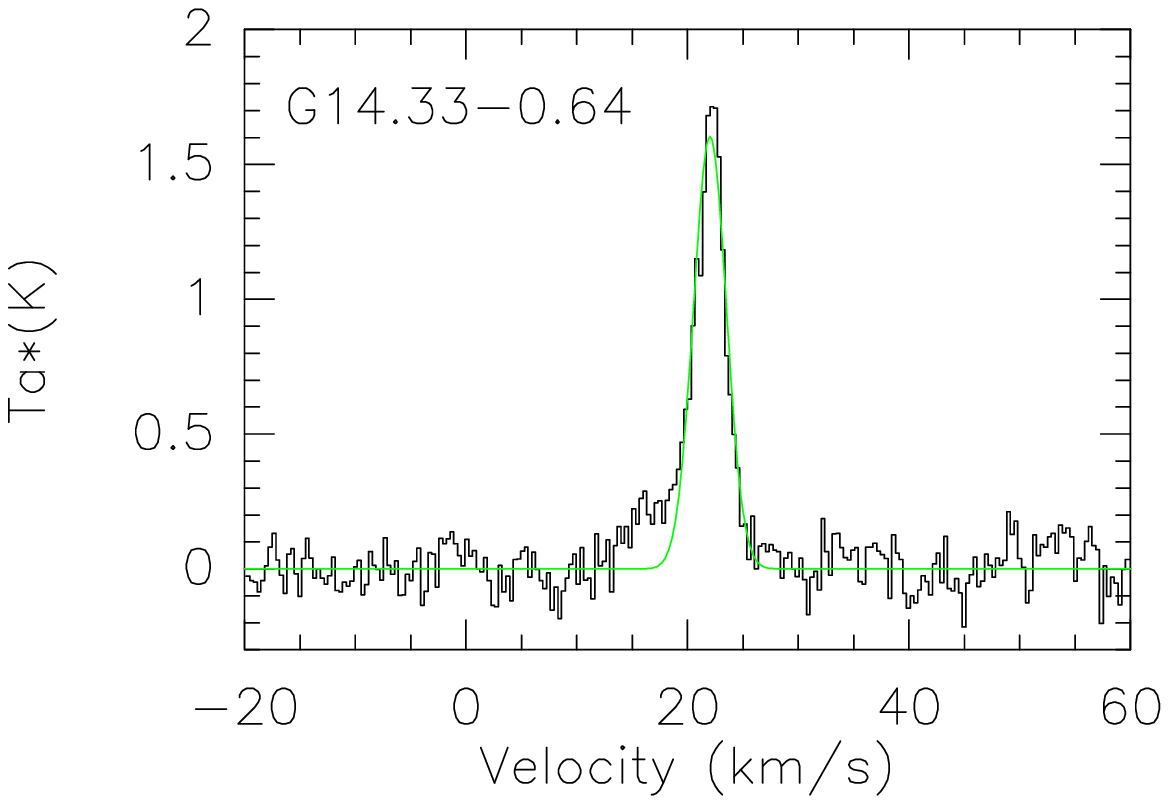}
\includegraphics[width=4.3cm]{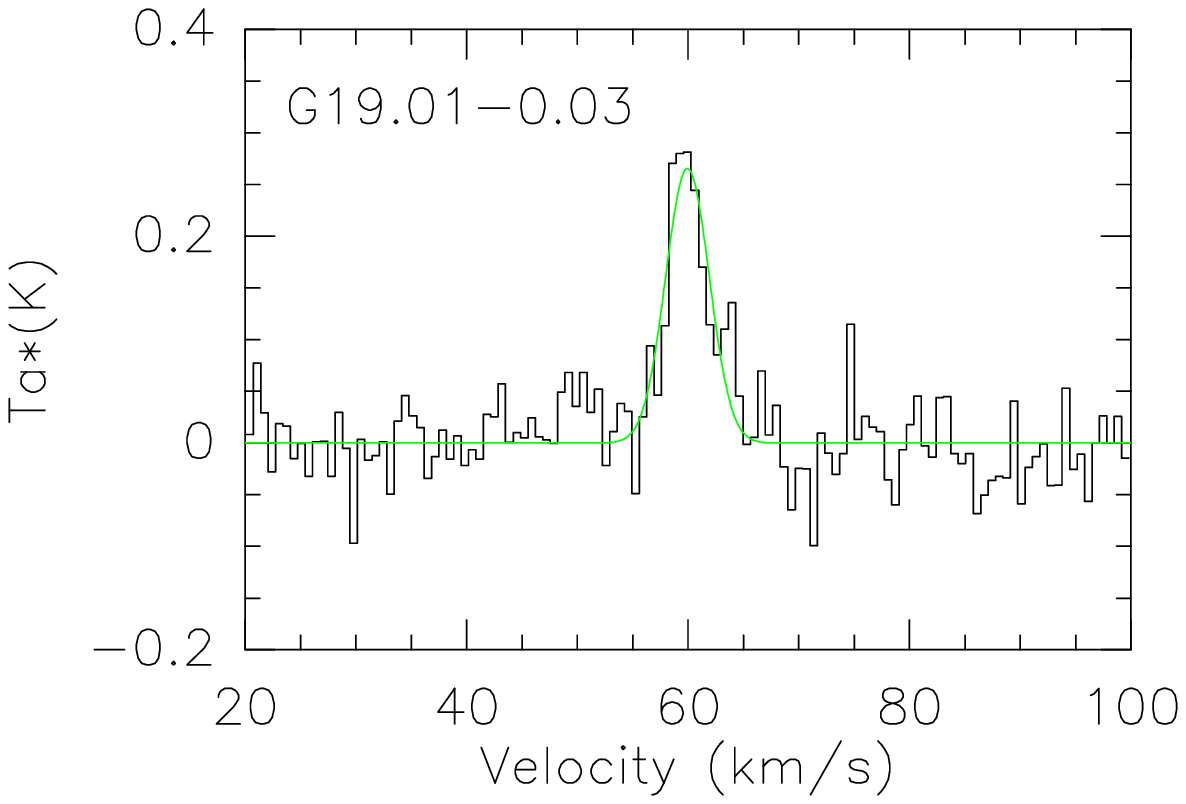}
\includegraphics[width=4.3cm]{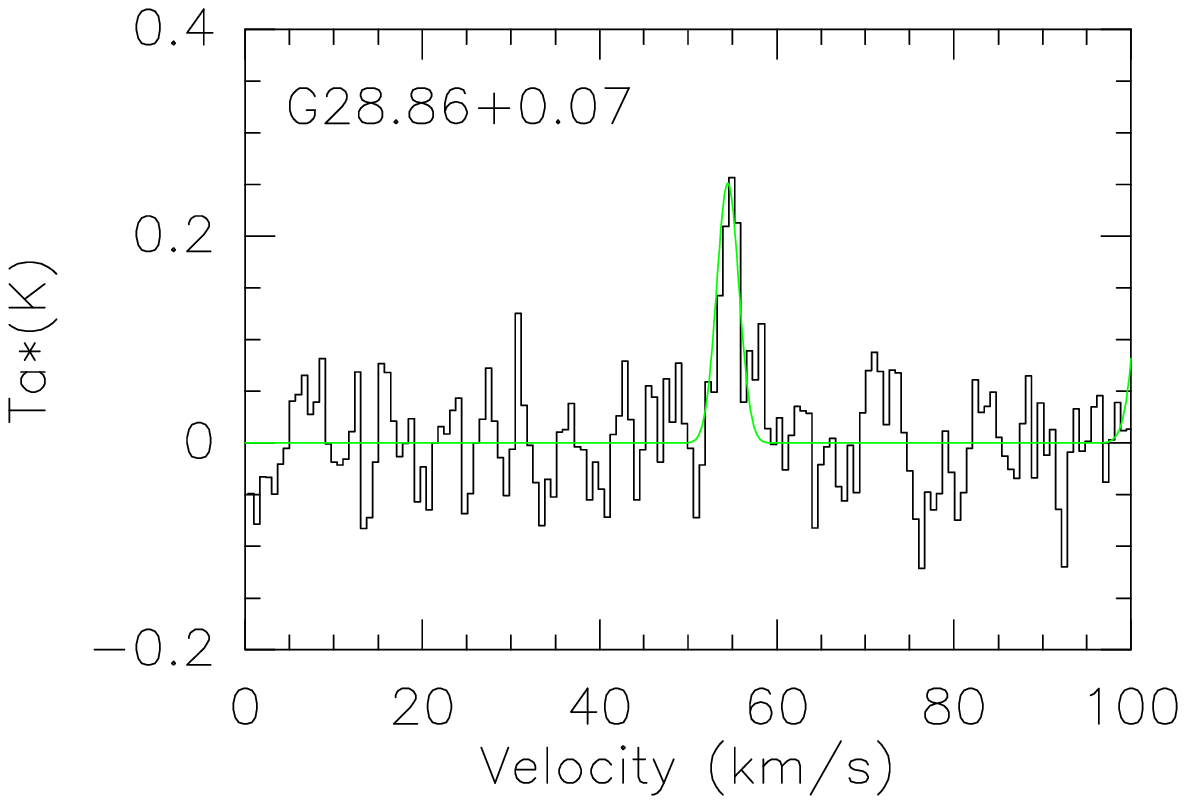}
\includegraphics[width=4.3cm]{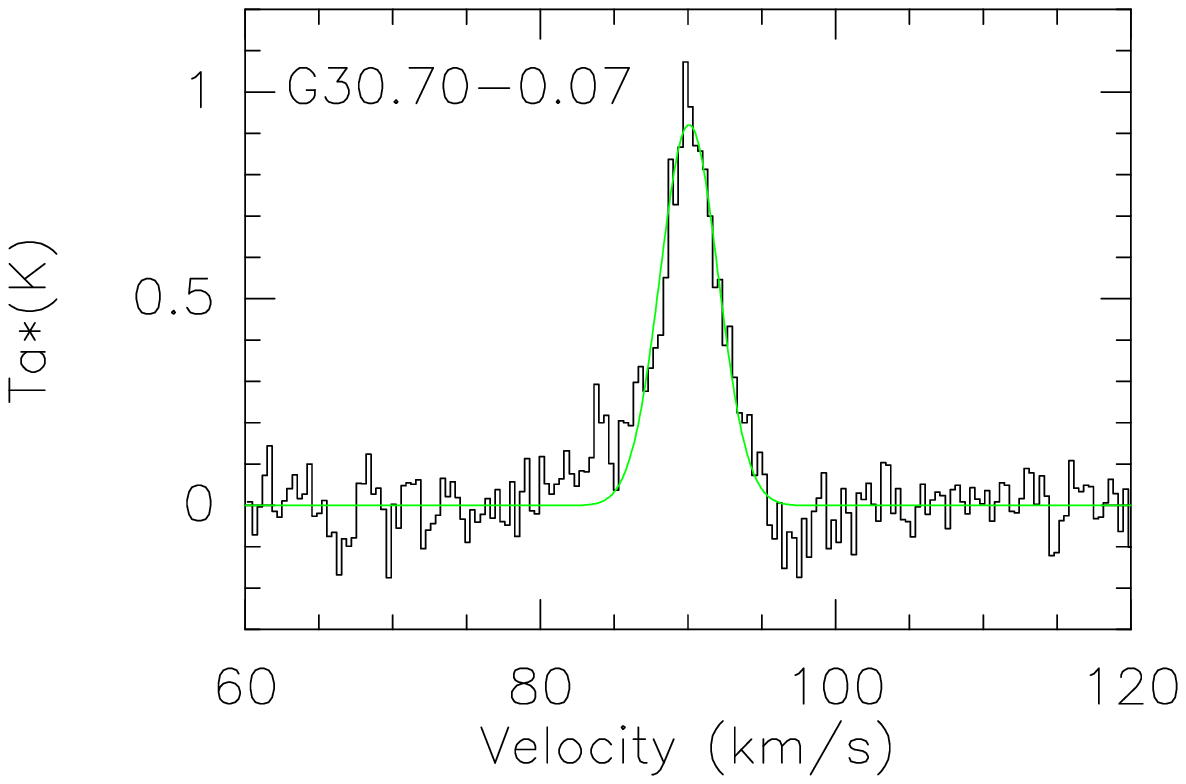}
\includegraphics[width=4.3cm]{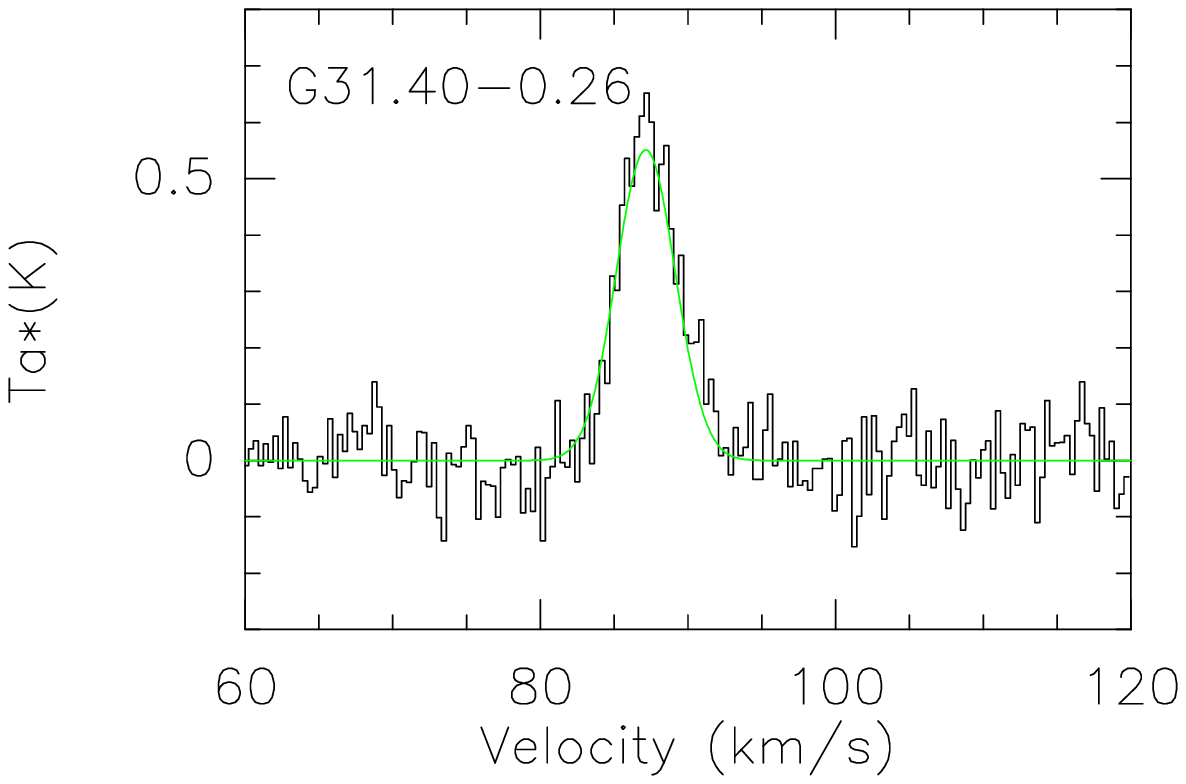}
\includegraphics[width=4.3cm]{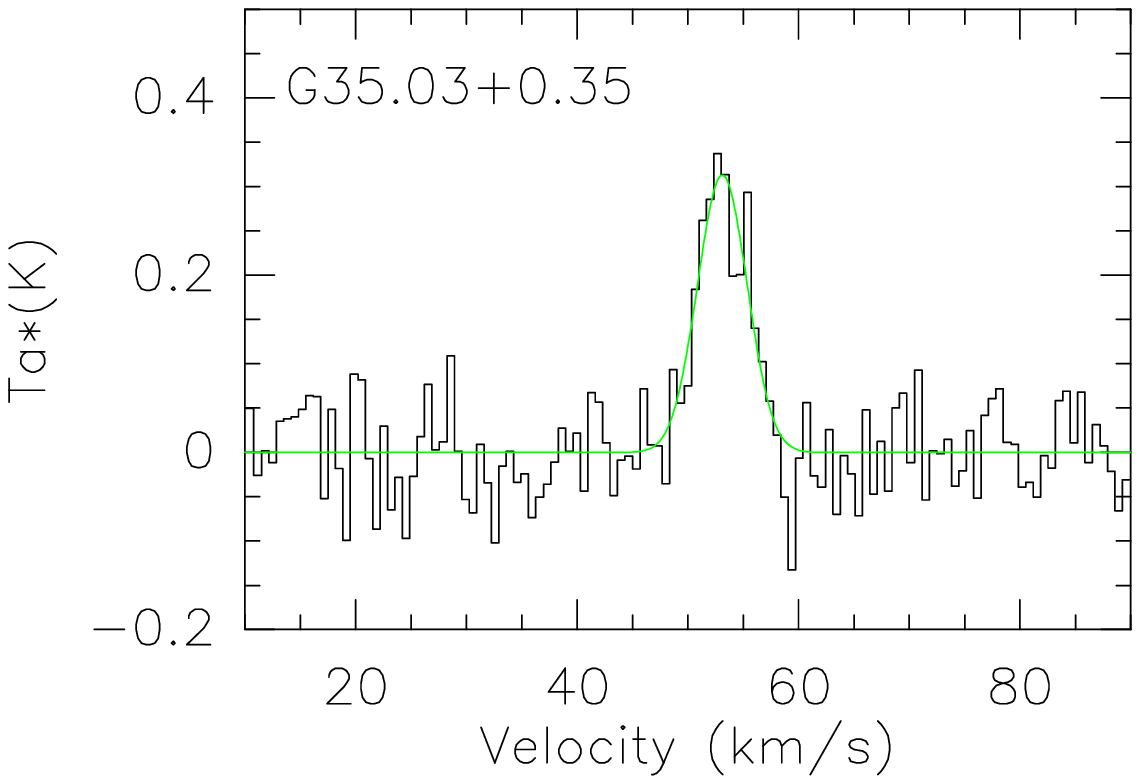}
\includegraphics[width=4.3cm]{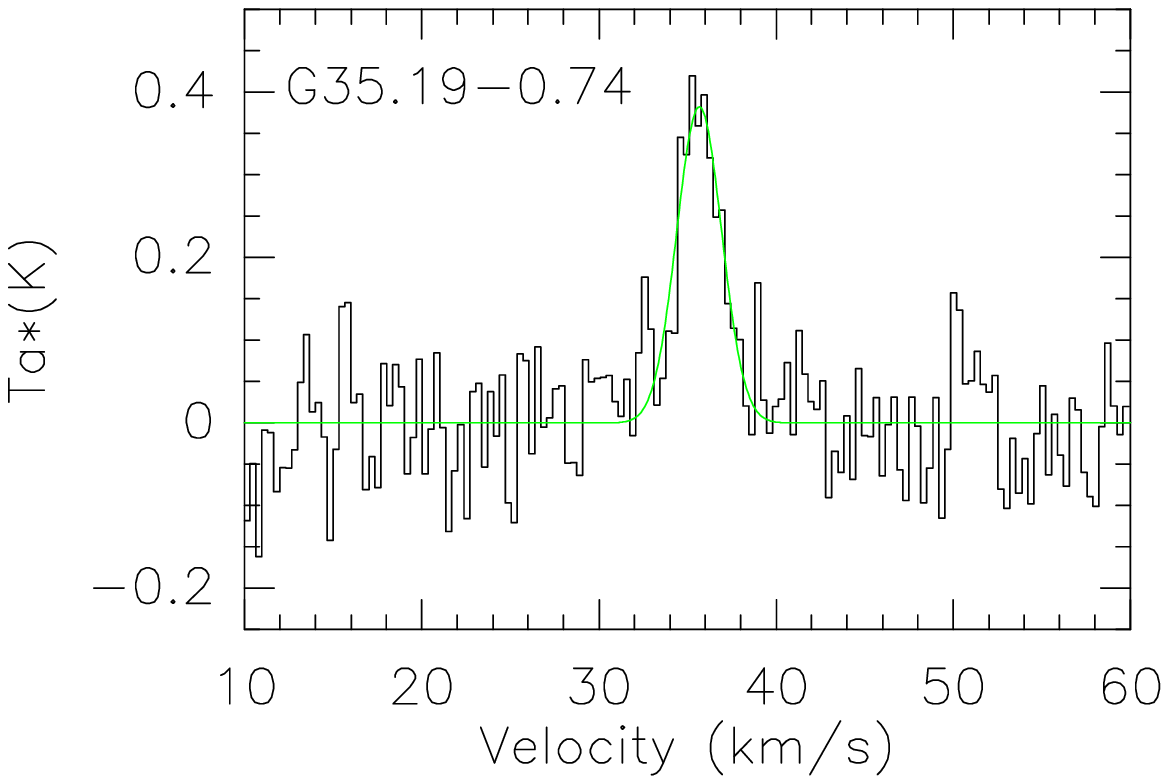}
\end{center}
\caption{CH$_3$OH (4$_{22}$-3$_{12}$) spectra.}
\label{figure:CH3OH-spectrum}
\end{figure*}

\end{appendix}

\end{document}